%% file: OpticalSignatures.tex
\documentclass[aps,superscriptaddress,showpacs,pra,twocolumn]{revtex4-2}
\usepackage[english]{babel}
\usepackage[dvipsnames]{xcolor}
\usepackage[normalem]{ulem}
\usepackage[colorlinks]{hyperref}
\hypersetup{colorlinks,linkcolor=red,citecolor=blue,urlcolor=blue,final}
\usepackage[T1]{fontenc}
\usepackage[latin9]{inputenc}
\usepackage{geometry}
\usepackage{array}
\usepackage{verbatim}
\usepackage{float}
\usepackage{booktabs}
\usepackage{bm}
\usepackage{amsmath}
\usepackage{amssymb}
\usepackage{graphicx}

\makeatletter

\providecommand{\tabularnewline}{\\}

\usepackage{lipsum}
\makeatother

\begin{document}

\title{Optical signatures of the coupled spin-mechanics of a levitated magnetic microparticle}

\author{Vanessa Wachter}
\affiliation{Max Planck Institute for the Science of Light, Staudtstra\ss{}e 2, 91058 Erlangen, Germany}
\affiliation{Department of Physics, University of Erlangen-N\"urnberg, Staudtstra\ss{}e 7, 91058 Erlangen, Germany}
\author{Victor A. S. V. Bittencourt}
\affiliation{Max Planck Institute for the Science of Light, Staudtstra\ss{}e 2, 91058 Erlangen, Germany}
\author{Shangran Xie}
\thanks{Currently at: School of Optics and Photonics, Beijing Institute of Technology, 100081 Beijing, China}
\affiliation{Max Planck Institute for the Science of Light, Staudtstra\ss{}e 2, 91058 Erlangen, Germany}
\author{Sanchar Sharma}
\affiliation{Max Planck Institute for the Science of Light, Staudtstra\ss{}e 2, 91058 Erlangen, Germany}
\author{Nicolas Joly}
\affiliation{Department of Physics, University of Erlangen-N\"urnberg, Staudtstra\ss{}e 7, 91058 Erlangen, Germany}
\affiliation{Max Planck Institute for the Science of Light, Staudtstra\ss{}e 2, 91058 Erlangen, Germany}
\author{Philip Russell}
\affiliation{Max Planck Institute for the Science of Light, Staudtstra\ss{}e 2, 91058 Erlangen, Germany}
\author{Florian Marquardt}
\affiliation{Max Planck Institute for the Science of Light, Staudtstra\ss{}e 2, 91058 Erlangen, Germany}
\affiliation{Department of Physics, University of Erlangen-N\"urnberg, Staudtstra\ss{}e 7, 91058 Erlangen, Germany}
\author{Silvia Viola Kusminskiy}
\affiliation{Max Planck Institute for the Science of Light, Staudtstra\ss{}e 2, 91058 Erlangen, Germany}
\affiliation{Department of Physics, University of Erlangen-N\"urnberg, Staudtstra\ss{}e 7, 91058 Erlangen, Germany}
\begin{abstract}
We propose a platform that combines the fields of cavity optomagnonics and levitated optomechanics in order to control and probe the coupled spin-mechanics of magnetic
dielectric particles. We theoretically study the dynamics of a levitated
Faraday-active dielectric microsphere serving as an optomagnonic cavity, placed in an external magnetic field  and driven by an external laser. We find that the optically driven magnetization
dynamics induces angular oscillations of the particle with low associated damping.
Further, we show that the magnetization and angular motion dynamics can
be probed via the power spectrum of the outgoing light. Namely, the
characteristic frequencies attributed to the angular oscillations
and the spin dynamics are imprinted in the light spectrum by two main
resonance peaks. Additionally, we demonstrate that a ferromagnetic
resonance setup with an oscillatory perpendicular magnetic field
can enhance the resonance peak corresponding to the spin oscillations
and induce fast rotations of the particle around its anisotropy axis.
\end{abstract}
\maketitle

\section{Introduction}

Since the pioneering work of Ashkin \citep{ashkin2006optical}, trapping
dielectric particles by optical forces has found a wide range of applications
\citep{grier2003revolution}, ranging from sensing ultra weak forces
\citep{hebestreit2018sensing,ranjit2015attonewton,ranjit2016zeptonewton,geraci2010short,geraci2015sensing,hempston2017force,blakemore2019three,monteiro2020force,monteiro2017optical}
and temperature gradients \citep{bykov2015flying,zeltner2018flyingparticle,zhang2014novel}
to manipulating biological systems \citep{ashkin1987cells,ashkin1987bacteria,lang2003resource,neuman2004optical,marago2013optical,grier2003revolution}.
Furthermore, dielectric particles levitated in vacuum are highly isolated
from environmental noise and decoherence, offering an ideal playground
to control the dynamics of micro- and nanoresonators \citep{gieseler2013thermal,gieseler2014dynamic,millen2014nanoscale,ricci2017optically,arita2013laserinduced,hebestreit2018measuring,hoang2016torsional,rashid2018precessionmotion}.
This has brought forth unprecedented performance in cavity and feedback
cooling of  levitated particle motion \citep{delic2019cavity,windey2019cavitybased,tabbenjohanns2019colddamping,conangla2019optimalfeedback,gieseler2012subkelvin,li2011millikelvin}
and the realization of fast rotation in high vacuum \citep{reinmann2018ghzrotation,ahn2018opticallylevitated}.
The high degree of isolation also makes levitated particles a key
component in many proposals for exploring quantum mechanics with mesoscopic
objects \citep{chang2010cavity,romero2011optically,romero2010toward,bateman2014near,scala2013matter,magrini2021real,tebbenjohanns2021quantum}.

Magnetically ordered dielectrics introduce magnetization as an additional
degree of freedom to levitated particles. For example, due to conservation of angular momentum, a change in the magnetization
of a freely moving body can
lead to its mechanical rotation and vice versa -- the so called Einstein-de
Haas effect \citep{einstein_de_haas} and its reciprocal, the Barnett
effect \citep{barnett1915magnetization}. Furthermore, the magnetization
couples to the particle orientation, defined by the specific magneto-crystalline
anisotropy of the material \citep{chikazumi2009physics}. These effects
are prominent in magnetically levitated systems and can be harnessed
for probing mesoscopic quantum mechanics including rotational degrees of freedom, as well as for sensing \citep{romero2012quantum,wang2019dynamics,timberlake2019acceleration,lewandowski2021high,slezak2018cooling,hsu2016cooling,prat2017ultrasensitive,gieseler2020single,o2019magneto,houlton2018axisymmetric,johnsson2016macroscopic,cirio2012quantum,vinante2020ultralow,perdriat2021spin}. 

The magnetization of dielectrics also can couple to light via magneto-optical
effects, the best known being the Faraday effect: the magnetization-induced
rotation of the polarization of light. Conversely, light can also
induce a small precession of the magnetization (the inverse Faraday
effect) \citep{landau1984electrodynamics, stancil2009spin}. This generally weak coupling can be enhanced by confinement
of light inside of the dielectric, which has given rise to the field of  (cavity optomagnonics) \citep{violakusminskiyCoupledSpinlightDynamics2016a,liu2016optomagnonics,sharma2017light,almpanis2018dielectric,almpanis2020spherical,osada2018orbital,kusminskiy2019cavity,PhysRevLett.117.133602,haigh2018selection,PhysRevLett.116.223601,osada2018brillouin,zhang2016optomagnonic, rameshti2021cavity}.
The combination of optomagnonic effects with levitation provides a
novel system within which light, magnetism and angular motion are intertwined.

In this work, we propose a new scheme for optomagnonically controlling and probing the coupled dynamics of levitated magnetic dielectric microparticles.
We consider the system depicted in Fig.$~$\ref{fig:schematic_model}:
a spherical dielectric magnet is levitated in the presence of an external static magnetic field. A pump laser drives two internal electromagnetic field modes
of the particle, which couple to the magnetization dynamics via the
cavity-enhanced optomagnonic coupling. In turn, the magnetization
of the particle couples to its angular motion via both magneto-crystalline
anisotropy and dissipation of magnons into the lattice. We show that
the optically driven magnetization dynamics induce angular oscillations,
which exhibit low damping due to the lack of direct dissipation of
the angular motion. The driven angular motion is limited in amplitude
by the optical quality factor of the particle and the driving
power that the system can support. The coupling between magnetization and angular motion can be
probed via the power spectrum of the cavity modes, which exhibits two
main resonance peaks that can be linked to the angular oscillations and to the
spin dynamics. Finally, we study the case of a driven ferromagnetic
resonance (FMR) by adding an oscillatory magnetic field perpendicular
to the static magnetic field. We show that this configuration can enhance the visibility of the characteristic peaks in the spectrum and can  induce, via the magnetization, fast rotations
of the particle around its anisotropy axis. 

The paper is structured as follows. In Sec. \ref{sec:Model} we present
the model and derive the corresponding Hamiltonian. In Sec. \ref{sec:Equations-of-motion}
we derive the coupled Heisenberg equations of motion with their conserved
quantities and discuss the torque generated by the applied laser via
the optomagnonic coupling. Sec. \ref{sec:SteadyState-1} presents
the steady state configurations of the system. In Sec. \ref{sec:Force-and-dynamical}
we investigate the nonlinear dynamics of the system numerically. Sec.
\ref{sec:Power-Spectrum} is devoted to the computation of the power
spectrum of the cavity field fluctuations and the resonance peaks.
In Sec.$~$\ref{sec:Driven-System} we show the effects of an additional
driving magnetic field and we discuss some generalizations to the
model in Sec.$~$\ref{sec:Generalisations}. We draw conclusions and
discuss further directions in Sec. \ref{sec:Conclusion}.

\section{Model\label{sec:Model}}

\begin{figure}[t]
\includegraphics[width=1\columnwidth]{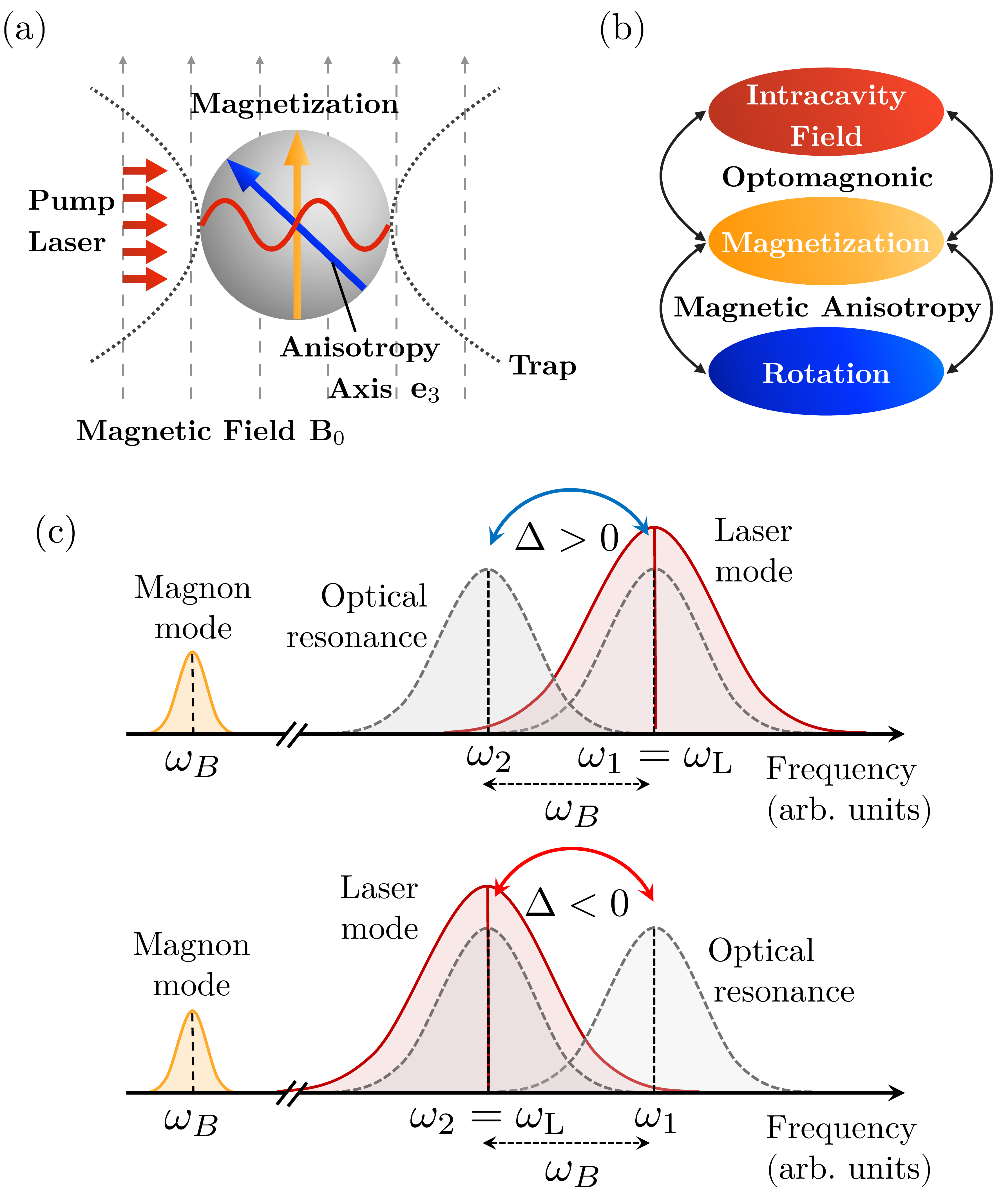}
\caption{Schematic of the  model, showing the couplings between the relevant
degrees of freedom. (a) A levitated ferrimagnetic microsphere serves
as an optomagnonic cavity. The microsphere exhibits uniaxial anisotropy
and we assume that its magnetization is homogeneous, pointing initially along
the direction of an external magnetic field. (b) The intracavity field
couples via optomagnonic effects to the magnetization which in turn
couples to the rotation through the magnetic anisotropy. (c) Schematic
illustration of the relevant frequencies of the system and the pumping
scheme. Two optical modes have frequencies $\omega_{1}$ and $\omega_{2}$
separated by the frequency of the magnon mode, i.e. $\omega_{1}-\omega_{2}=\omega_{B}$.
One of the two modes is pumped in resonance with the laser frequency
$\omega_{\mathrm{L}}$.}
\label{fig:schematic_model}
\end{figure}

We consider a Faraday-active dielectric microparticle that is levitated
in the presence of an external static magnetic field $\mathbf{B}_0$
as depicted in Fig.$~$\ref{fig:schematic_model}. The magnetization
of the particle is saturated along $\mathbf{B}_0$. The particle supports
internal electromagnetic modes. If the radius of the particle is much
larger than the driving laser wavelength, light is trapped by total
internal reflection, forming optical whispering gallery modes \citep{heebner2008optical}.
Otherwise, for radii comparable to the driving light wavelength, such
modes correspond to Mie resonances \citep{bohren2008absorption,kuznetsov2016optically}.
In case of an unmagnetized particle the underlying Mie mode frequencies are solely
determined by the polarization and the angular mode number, and the modes
are degenerate with respect to the azimuthal number. The magnetization
of the particle breaks this degeneracy and causes a splitting of the
Mie modes into modes with different azimuthal numbers \citep{almpanis2020spherical,ford1978scattering}.

Driving the particle with light can induce a precession of the magnetization
around its equilibrium orientation. The associated quantized magnetic excitations
are denoted magnons. The mean number of magnons in a coherent state is directly related to the amplitude of the magnetization precession. In the following we focus on the homogeneous
magnon mode, known as the Kittel mode, consisting of a uniformly precessing
magnetization. Its frequency
can be tuned by the external magnetic field and usually lies in the GHz regime. For a driving electromagnetic field at optical frequencies, the resonant dipolar coupling between the magnetization and the magnetic field is suppressed due to the mismatch of frequencies.
The coupling to the electromagnetic field takes place instead through a  Raman scattering
type of process involving two photons, and it is dominated by the
dipolar coupling to the \emph{electric component} of the field \citep{fleury1968scattering}.

In this work, we consider that a pump laser of suitable wavelength can drive simultaneously two internal
optical modes of the levitated particle which satisfy the so-called \emph{triple-resonance condition} where the frequency
difference between the cavity modes equals the magnetization's precession
frequency \citep{PhysRevLett.117.133602,zhang2016optomagnonic}. In
other words, transitions between the modes can occur through emission
or absorption of a magnon. The triple-resonance condition can be fulfilled
in the visible/near-infrared part of the electromagnetic spectrum,
where the frequency splitting between neighboring Mie modes for micrometer-sized particles is in
the range of a few GHz, the typical frequency of magnons \citep{almpanis2020spherical}. We consider the only source of birefringence (and thus
of torque) to be magnetic. This is valid for example for an Yttrium Iron Garnet (YIG) sphere. For magnetic dielectric
particles which also exhibit optical or geometric anisotropy \citep{nieminen2001optical,friese1998optical,simpson2007polarization,donato2016light}, the model can be generalized accordingly. Our model takes optical losses and magnetic damping into consideration. The latter corresponds to damping of the magnetization precession,
i.e., the relaxation to its equilibrium configuration \citep{gilbert2004phenomenological}.
For the coupling of the confined light to the particle magnetization we
focus on the Faraday effect (circular birefringence), relegating consideration
of the Cotton-Mouton effect (linear birefringence) to Sec.$~$\ref{sec:Cotton-Mouton-Effect}.
The magnetization and the orientation of the particle are coupled
by a magnetocrystalline anisotropy, such that a mechanical torque
can be induced. We assume that the material exhibits uniaxial crystalline
anisotropy, such that the magnetization, in the absence of a magnetic
field, is aligned either parallel to the anisotropy axis, in the case
of an easy-axis magnet, or perpendicular to it, in the case of a hard-axis
magnet. We assume that the translational degrees of freedom are completely
decoupled from the rest, namely, we neglect effects related to the
center of mass motion of the particle. We therefore do not explicitly
include the trapping potential in our model. We discuss briefly deviations
from these assumptions in Sec.$~$\ref{sec:Generalisations}. 

A suitable material for an experimental realization of this system
is YIG, since it exhibits ferrimagnetic properties at room temperature
and shows high transparency in the near-infrared. Further, it has
a large Faraday rotation per unit length $\theta_{\mathrm{F}}=240^{\circ}/\mathrm{cm}$
with a dielectric index of $\varepsilon\approx5$ \citep{stancil2009spin,antonov1969optical}
at these frequencies, making it the material of choice in the state-of-the-art
optomagnonic experiments \citep{PhysRevLett.117.133602,haigh2018selection,PhysRevLett.116.223601,osada2018brillouin,zhang2016optomagnonic}.
In the following we investigate the proposed model for parameter values
corresponding to YIG, although the framework applies in general for
other materials too.

\subsection{Hamiltonian}

The Hamiltonian describing the system reads
\begin{align}
\hat{H} & =\hat{H}_{\mathrm{Sp}}+\hat{H}_{\mathrm{Cav}}+\hat{H}_{\mathrm{OM}}\label{eq:initialHamiltonian}
\end{align}
where $\hat{H}_{\mathrm{Sp}}$ is the Hamiltonian of the freely rotating
magnetic sphere in the presence of the magnetic field and $\hat{H}_{\mathrm{Cav}}$
that of the optical modes confined in the dielectric sphere. $\hat{H}_{\mathrm{OM}}$
describes the coupling between the optical modes and the magnetization
of the sphere. The operator character of the quantities is represented
by the ``hat'' notation. In the following we present each term in detail.

The first term in Eq.$~$\eqref{eq:initialHamiltonian} is given by
\begin{align}
\hat{H}_{\mathrm{Sp}} & =\dfrac{\hbar^{2}\hat{\mathbf{L}}^{2}}{2I}-\hbar\gamma\hat{\mathbf{S}}\cdot\mathbf{B}_0-\hbar^{2}D(\hat{\mathbf{S}}\cdot\mathbf{e}_{{\rm 3}}(\hat{\Omega}))^{2}\,,\label{eq:HSp}
\end{align}
where the first term is the kinetic energy of the rotational motion
with dimensionless angular momentum $\hat{\mathbf{L}}=(\hat{L}_{x},\hat{L}_{y},\hat{L}_{z})$
and $I=2mR^{2}/5$ the moment of inertia of a sphere of radius $R$
and mass $m$. In the following we will express the inertia in terms
of a rotation frequency of the rigid solid $\omega_{I}=\hbar/I$.

The second term in Eq.$~$\eqref{eq:HSp} corresponds to the Zeeman energy describing the interaction
between the particle's magnetization and the homogeneous magnetic field
$\mathbf{B}_0$. For the Kittel mode, all spins precess in phase and
can be described by the dimensionless macrospin $\hat{\mathbf{S}}=(\hat{S_{x}},\hat{S}_{y},\hat{S}_{z})$,
which relates to the magnetization as $\hat{\mathbf{M}}=\hat{\mathbf{S}}M_{\mathrm{S}}/S$
with $S=|\mathbf{S}|$ the total spin of the sample and $M_{\mathrm{S}}$ the saturation magnetization. $\gamma=1.76\cdot10^{11}\,\mathrm{s}^{-1}\mathrm{T}^{-1}$
denotes the absolute value of the gyromagnetic ratio. We consider
a magnetic field parallel to the $z$ axis, namely $\mathbf{B}_0=B_0\mathbf{e}_{z}$,
such that the Zeeman energy of the Hamiltonian reduces to $-\hbar\gamma\hat{S}_{z}B_0$
and we denote $\gamma B_0=\omega_{B}$ as the Larmor precession frequency.
For a sphere, this corresponds to the frequency of the Kittel mode. A magnon corresponding to this mode has energy $\hbar \omega_{B}$.

The third term in Eq.$~$\eqref{eq:HSp} accounts for uniaxial magnetocrystalline anisotropy (cubic anisotropy will be discussed in Sec.$~$\ref{subsec:Cubic-anisotropy}).
The anisotropy axis is fixed to the particle and labeled as $\mathbf{e}_{3}(\hat{\Omega})$.
As $\mathbf{e}_{3}(\hat{\Omega})$ depends on the Euler angles $\hat{\Omega}$,
this term couples the spin with the orientation of the sphere with
coupling constant $\omega_{D}=\hbar D$. Here $|D|=K_{u}V/(\hbar S)^{2}$
\citep{gatteschi2006molecular, rusconi2017quantum} denotes
the anisotropy strength with $K_{u}$ the anisotropy energy density
and $V$ the volume of the particle. For $D>0$, $\mathbf{e}_{3}(\hat{\Omega})$
is called an ``easy axis'', i.e. the interaction energy is minimized
if the magnetization and $\mathbf{e}_{3}$ are parallel. For $D<0$,
$\mathbf{e}_{3}(\hat{\Omega})$ is a ``hard axis'', and in this case
it is energetically favorable for the magnetization to lie in the
plane perpendicular to $\mathbf{e}_{3}$. 

\begin{figure}[t]
\includegraphics[width=1\columnwidth]{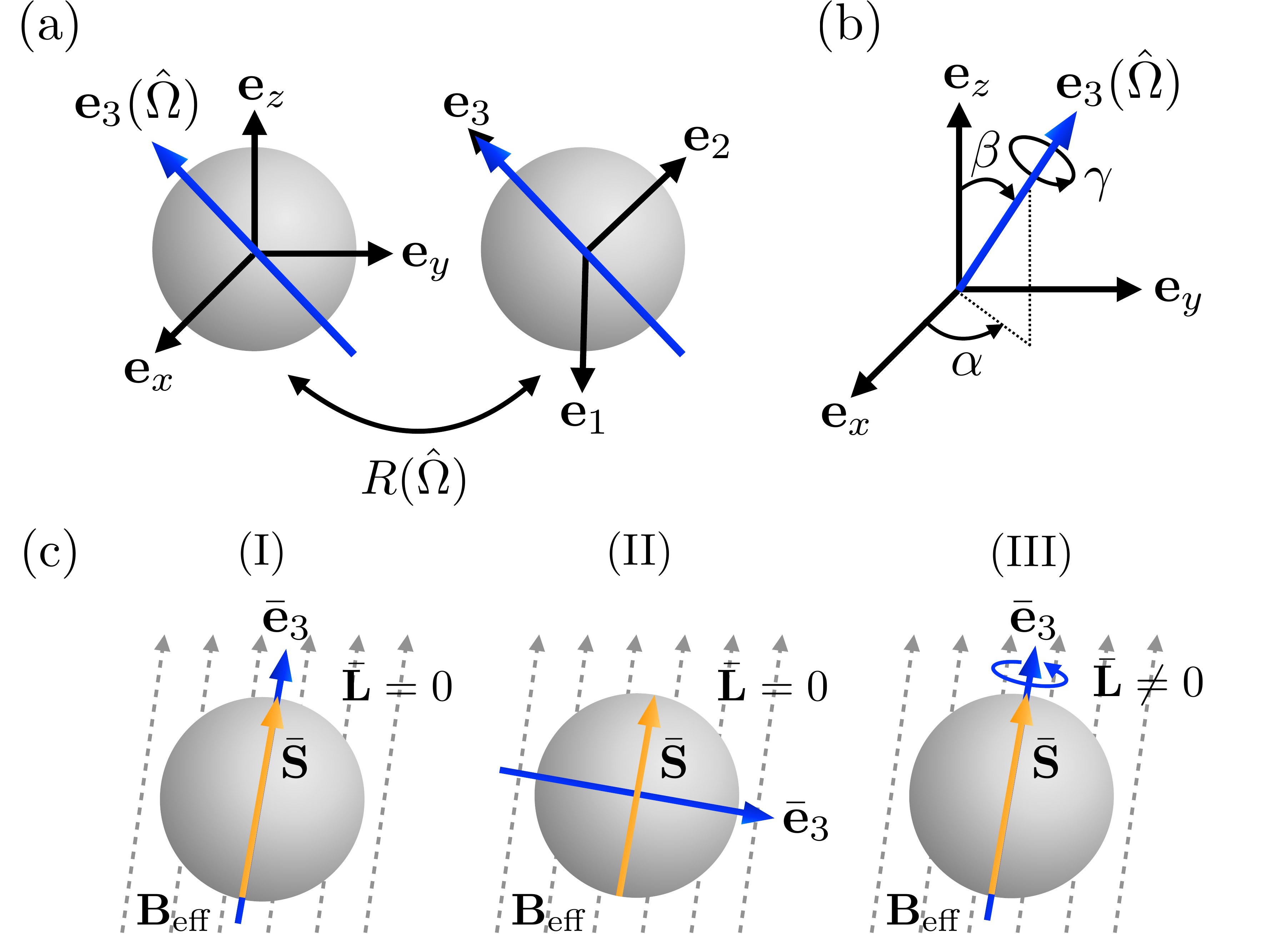}\caption{Definition of the Euler angles. (a) The rotation matrix $R(\hat{\Omega})$
aligns the fixed laboratory frame $O\mathbf{e}_{x}\mathbf{e}_{y}\mathbf{e}_{z}$
with the body frame $O\mathbf{e}_{1}\mathbf{e}_{2}\mathbf{e}_{3}$
where $\mathbf{e}_{3}(\hat{\Omega})$ denotes the anisotropy direction.
(b) The Euler angles correspond to three successive rotations that
are necessary for aligning the lab frame with the frame in which $\mathbf{e}_{3}$
is fixed. (c) Steady state configurations of the system. }
\label{Fig:euler_angles}
\end{figure}

The orientation of $\mathbf{e}_{3}(\hat{\Omega})$ is defined by an
operator specified by the Euler angles $\hat{\Omega}=\{\hat{\alpha},\hat{\beta},\hat{\gamma}\}$,
such that $\mathbf{e}_{3}(\hat{\Omega})$ is given in terms of elements
of the rotation matrix $R(\hat{\Omega})$ (Eq.$~$\eqref{eq:rotationMatrix-1}
and \eqref{eq:e3-1-1}) connecting the fixed laboratory frame $O\mathbf{e}_{x}\mathbf{e}_{y}\mathbf{e}_{z}$
to the rotating frame $O\mathbf{e}_{1}\mathbf{e}_{2}\mathbf{e}_{3}$
(which we call the body frame): $(\mathbf{e}_{1},\mathbf{e}_{2},\mathbf{e}_{3})^{T}=R(\hat{\Omega})(\mathbf{e}_{x},\mathbf{e}_{y},\mathbf{e}_{z})^{T}$
(see Fig.$~$\ref{Fig:euler_angles}). Note that from now on for simplicity
we write $\mathbf{e}_{3}$ without the explicit dependence on $\hat{\Omega}$.

The cavity Hamiltonian $\hat{H}_{\mathrm{Cav}}$ describes the electromagnetic
field inside the particle and is given by 
\begin{equation}
\hat{H}_{{\rm Cav}}=\dfrac{1}{2}\int\mathrm{d}^{3}\mathbf{r}\left(\varepsilon_{0}\mathbf{\hat{E}}^{\ast}(\mathbf{r},t)\bar{\mathbf{\boldsymbol{\varepsilon}}}\mathbf{\hat{E}}(\mathbf{r},t)+\dfrac{1}{\mu_{0}}\hat{\mathbf{B}}^{2}(\mathbf{r},t)\right)\label{eq:H_cav}
\end{equation}
with $\varepsilon_{0}$ and $\mu_{0}$ denoting, respectively, the
vacuum permittivity and permeability, and $\bar{\boldsymbol{\varepsilon}}$
the static part of the effective relative permittivity tensor that depends
on the magnetization $\mathbf{M}$. Considering only the Faraday effect and a saturation magnetization
along the $z$ axis, the permittivity tensor is given by 
\begin{equation}
\boldsymbol{\varepsilon}(\mathbf{M})=\bar{\boldsymbol{\varepsilon}}+\delta\boldsymbol{\varepsilon},\label{eq:permittivity_tensor_Faraday}
\end{equation}
with the static
\begin{equation}
\bar{\boldsymbol{\varepsilon}}=\begin{pmatrix}\varepsilon & -\mathrm{i}fM_{z} & 0\\
\mathrm{i}fM_{z} & \varepsilon & 0\\
0 & 0 & \varepsilon
\end{pmatrix},\label{eq:permittivity_static}
\end{equation}
and dynamical part
\begin{equation}
\delta\boldsymbol{\varepsilon}=\begin{pmatrix}0 & 0 & \mathrm{i}fM_{y}\\
0 & 0 & -\mathrm{i}fM_{x}\\
-\mathrm{i}fM_{y} & \mathrm{i}fM_{x} & 0
\end{pmatrix}.\label{eq:permittivity_dynamic}
\end{equation}
Here $\varepsilon$ denotes the relative isotropic permittivity of
the unmagnetized material, and $f$ a material-dependent constant.
It is related to the Faraday rotation per unit length $\theta_{\mathrm{F}}$, 
\begin{equation}
\theta_{\mathrm{F}}=\dfrac{\omega fM_{\mathrm{S}}}{2c\sqrt{\varepsilon}},\label{eq:faraday_rot}
\end{equation}
for optical frequency $\omega$ and vacuum speed of light $c$. The decomposition into a static and dynamical part as given in Eqs. \eqref{eq:permittivity_tensor_Faraday}-\eqref{eq:permittivity_dynamic} is valid for small deviations of the magnetization with respect to the $z$-axis, such that $M_z$ can be considered approximately constant $M_z\approx M_{\mathrm{S}}$ ($M_{x,y}/M_{\mathrm{S}}\ll 1$). The static magnetization causes a Zeeman-like splitting of the optical Mie modes \citep{almpanis2020spherical},
i.e. it lifts the $(2l+1)$- degeneracy with respect
to the azimuthal number $-l\leq m\leq l$, where $l$ is the angular mode number. This spliting is in the GHz range \citep{almpanis2020spherical}.

The electric and magnetic intracavity field, $\hat{\mathbf{E}}$ and
$\hat{\mathbf{B}}$, are quantized in the usual way such that $\hat{H}_{\mathrm{Cav}}$
is written in terms of the cavity operators $\hat{a}_{k}$ and the
corresponding frequencies $\omega_{k}$ (with $k$ labelling the mode
indices) as
\begin{equation}
\hat{H}_{\mathrm{Cav}}=\sum_{k}\hbar\omega_{k}\bigg(\hat{a}_{k}^{\dagger}\hat{a}_{k}+\dfrac{1}{2}\bigg).\label{eq:H_cav_quant}
\end{equation}

The optomagnonic Hamiltonian $\hat{H}_{\mathrm{OM}}$ is derived from
the time-averaged electromagnetic energy \citep{landau1984electrodynamics,violakusminskiyCoupledSpinlightDynamics2016a} 
\begin{equation}
\bar{U}=\dfrac{1}{4}\varepsilon_{0}\int\mathrm{d^{3}\mathbf{r}}\sum_{ij}E_{i}^{\ast}(\mathbf{r},t)\delta\varepsilon_{ij}(\mathbf{M})E_{j}(\mathbf{r},t).\label{eq:em_energy}
\end{equation}
We are interested in the interaction between the Kittel mode and a single pair of WGMs for
which the optomagnonic interaction Hamiltonian, after quantizing the
electric field $\mathbf{E}(\mathbf{r},t)\rightarrow\hat{\mathbf{E}}^{+}(\mathbf{r},t)=\sum_{k}\mathbf{E}_{k}(\mathbf{r})\hat{a}_{k}(t)$
(with $\mathbf{E}_{k}(\mathbf{r})$ indicating the electric field of the $k$-th eigenmode), can be expressed as \citep{PhysRevLett.117.133602}
\begin{equation}
\hat{H}_{\mathrm{OM}}=\hbar g(\hat{S}_{+}\hat{a}_{1}^{\dagger}\hat{a}_{2}+\hat{S}_{-}\hat{a}_{2}^{\dagger}\hat{a}_{1}),\label{eq:H_om_quant}
\end{equation}
where $\hat{S}_{\pm}=\hat{S}_{x}\pm\mathrm{i}\hat{S}_{y}$. The optomagnonic
coupling constant $g$ is defined as 
\begin{equation}
g=-\dfrac{\varepsilon_{0}fM_{\mathrm{S}}}{4\hbar S}\tilde{g}\label{eq:g}
\end{equation}
with $\tilde{g}$ (given in Appendix$~$\ref{sec:Transition-Amplitude})
denoting the transition amplitude between two Zeeman-split Mie modes. Here the selection rule for the initial and final azimuthal mode
numbers $m_{\mathrm{f}}-m_{\mathrm{i}}=1$ is fulfilled, and we assume
that the frequency splitting between two neighboring modes satisfies
the triple-resonance condition. The triple resonance can be achieved by tuning the frequency of the Kittel mode by the external magnetic field  $\mathbf{B}_0$. For a micron sized YIG sphere $g\sim0.03\,\mathrm{Hz}$.
This coupling is enhanced by driving one of the two optical modes, by a factor given by the square root of the number of circulating photons in the cavity \citep{kusminskiy2019cavity,kusminskiy2019quantum}. 

\begin{table}[t]
\caption{Definition and values of the relevant frequencies and couplings in
the system scaled by the total spin $S$. The values correspond to
the physical parameters given in Table$~$\ref{tab:parameters} and
are given in terms of the magnon frequency $\omega_{B}\equiv\gamma B_0=10^{10}\,\mathrm{s^{-1}}$,
corresponding to an applied magnetic field of $\sim60~\mathrm{mT}$.
Note that the full dependence of the optomagnonic coupling $g$ on $R$ cannot be obtained analytically, due to the non-trivial dependence of the transition amplitude $\tilde{g}$ on $R$ (see Fig.$~$\ref{fig:coupling_radius}). }
\begin{tabular}{>{\raggedright}p{0.5\columnwidth}>{\raggedright}p{0.5\columnwidth}}
\toprule 
Definition & Value (for $R$ in $\mu$m)\tabularnewline
\midrule
\multicolumn{1}{l}{$\omega_{I}S\equiv\hbar S/I=5\hbar S/(2mR^{2})$} & $2.15\cdot10^{-8}\omega_{B}/R^{2}$\tabularnewline
$\omega_{D}S\equiv\hbar DS=K_{u}V/(\hbar S)$ & $2.32\cdot10^{-1}\omega_{B}$\tabularnewline
$gS=\dfrac{\varepsilon_{0}fM_{\mathrm{S}}}{4\hbar}\tilde{g}$ & $5.596\cdot10^{-2}\omega_{B}/R$\tabularnewline
\bottomrule
\end{tabular}\label{tab:couplings}
\end{table}

\begin{table}[t]
\caption{Values of the physical parameters used in the calculations. The values
correspond to YIG \citep{stancil2009spin,wu2013recent}, however,
the framework holds generally. Note that we choose $m_{\mathrm{i}}=0$
such that the coupling strength is maximized (see Appendix \ref{sec:Transition-Amplitude}).}
\begin{tabular}{ll}
\toprule 
Parameter and values & Description \tabularnewline
\midrule
$\rho_{m}=5\cdot10^{3}\,\mathrm{kg}\,\mathrm{m^{-3}}$ & Mass density\tabularnewline
$S=\rho_{m}/738\,\mathrm{amu}$ & Total spin\tabularnewline
$K_{u}=10^{3}\:\mathrm{J/\mathrm{m^{3}}}$\footnote{For the cubic symmetry of YIG one finds two constants for the anisotropy
density in the literature. Here we approximate the uniaxial constant
with a value in the order of the first-order cubic anisotropy constant.
See Sec. \ref{subsec:Cubic-anisotropy} for details.} & Magnetic anisotropy constant \tabularnewline
$\varepsilon=5$ & Relative permittivity\tabularnewline
$n=2.2$ & Refractive index\tabularnewline
$\theta_{\mathrm{F}}=240^{\circ}/\mathrm{cm}$ & Faraday rotation\tabularnewline
$M_{\mathrm{S}}=140\,\mathrm{kA}/\mathrm{{m}}$ & Saturation magnetization\tabularnewline
$\lambda_{0}=1500\,\mu\mathrm{m}$ & Laser wavelength\tabularnewline
$m_{\mathrm{i}}=0$ & Initial azimuthal mode number\tabularnewline
$\eta_{\mathrm{G}}=10^{-4}$ & Gilbert damping parameter\tabularnewline
\bottomrule
\end{tabular}\label{tab:parameters}
\end{table}

The optical modes are driven externally by a laser, described by adding
a driving term $i\hbar\epsilon_{\mathrm{L}1,2}\left(\hat{a}_{1,2}^{\dagger}\mathrm{e}^{-\mathrm{i\omega_{Las}t}}-\hat{a}_{1,2}\mathrm{e}^{\mathrm{i}\omega_{\mathrm{Las}}t}\right)$
to the Hamiltonian. The parameter $\epsilon_{Li}=\sqrt{\dfrac{\kappa_{\mathrm{rad}}P_{i}}{\hbar\omega_{\mathrm{Las}}}}$
depends on the laser power $P_{i}$ and the radiative decay rate $\kappa_{\mathrm{rad}}$
(see Eq.$~$\eqref{eq:kappa}). Collecting all terms and working in
a frame rotating with the laser frequency $\omega_{\mathrm{Las}}$,
the total system Hamiltonian reads 
\begin{align}
\nonumber
\hat{H}= & \dfrac{1}{2}\hbar\omega_{I}\hat{\mathbf{L}}^{2}-\hbar\omega_{B}\hat{S}_{z}-\hbar\Delta_{1}\hat{a}_{1}^{\dagger}\hat{a}_{1}-\hbar\Delta_{2}\hat{a}_{2}^{\dagger}\hat{a}_{2}\\\nonumber
 & -\hbar\omega_{D}\left(\hat{\mathbf{S}}\cdot\hat{\mathbf{e}}_{3}\right)^{2}+\hbar g(\hat{S}_{+}\hat{a}_{1}^{\dagger}\hat{a}_{2}+\hat{S}_{-}\hat{a}_{2}^{\dagger}\hat{a}_{1})\\
 & +\mathrm{i}\hbar\epsilon_{\mathrm{L1}}\big(\hat{a}_{1}^{\dagger}-\hat{a}_{1}\big)+\mathrm{i}\hbar\epsilon_{\mathrm{L2}}\big(\hat{a}_{2}^{\dagger}-\hat{a}_{2}\big)
\label{eq:hamiltonian_reduced}
\end{align}
with detuning $\Delta_{i}=\omega_{\mathrm{Las}}-\omega_{i}$. The
assumption that the drive laser couples to both modes is a simplification
that can be refined \footnote{For instance, one could consider individual driving terms at different
frequencies for each mode, which would require additional unitary
transformations to a rotating frame in which $\hat{H}$ is time-independent.}.

\begin{figure}[t]
\includegraphics[width=0.95\columnwidth]{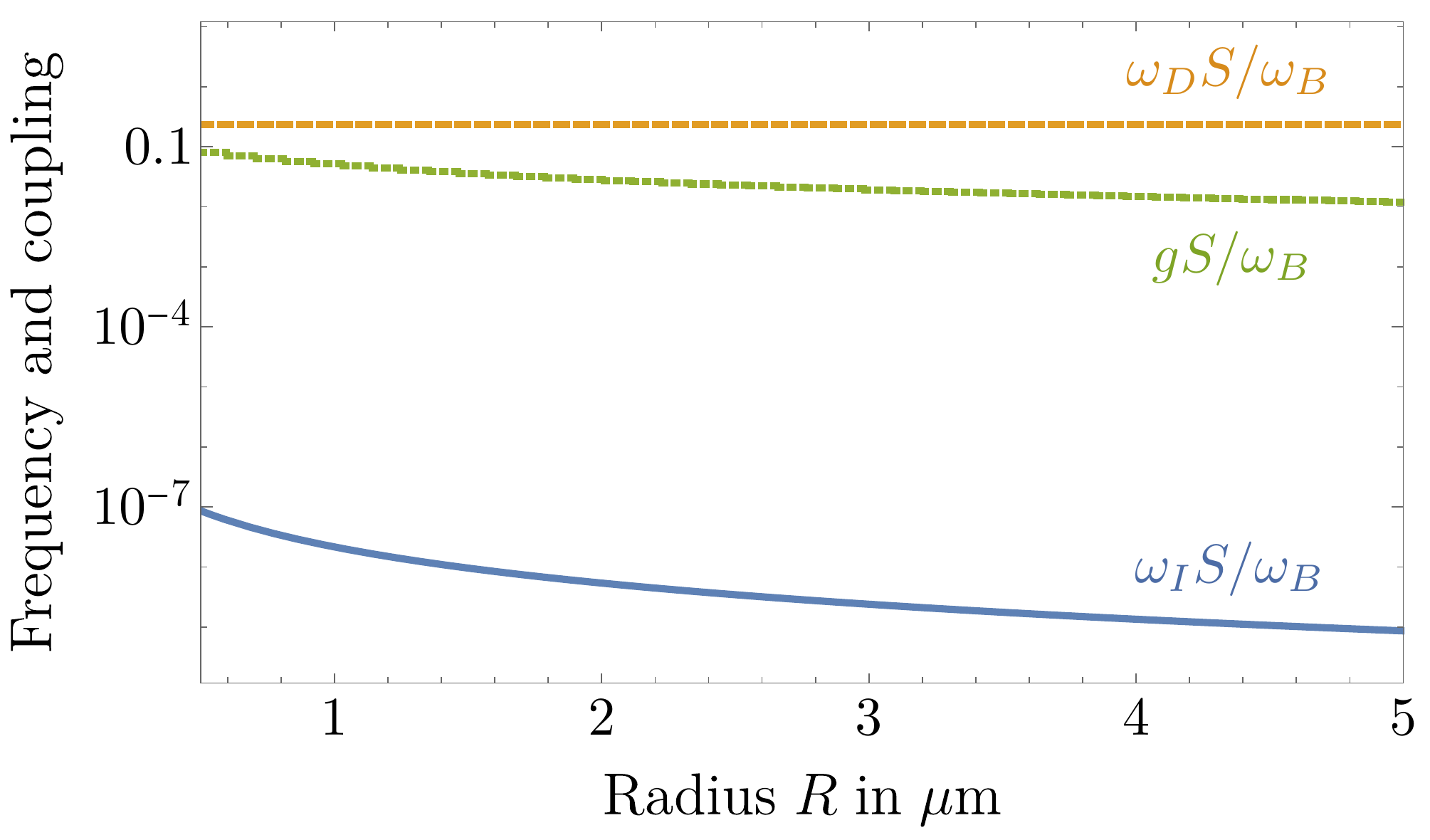}\caption{Frequencies and couplings defined in Table \ref{tab:couplings} as
a function of the radius $R$. The values are given in units of the
Larmor frequency $\omega_{B}=\gamma B_0$ which is controlled by the
applied magnetic field and is typically in the GHz range. Note that
the vertical axis is in logarithmic scale.}
\label{fig:coupling_radius}
\end{figure}

In Table$~$\ref{tab:couplings} we summarize the relevant frequencies
and couplings and show their dependence on the size of the particle
in Fig.$~$\ref{fig:coupling_radius}(a). The optomagnonic coupling
$g$ and the anisotropy strength $\omega_{D}$ have a similar magnitude
while the inertia frequency $\omega_{I}$ is orders of magnitude smaller
and decreases faster with particle radius. All the relevant values
for the physical parameters appearing in the system are listed in
Table$~$\ref{tab:parameters}. 

\section{Equations of Motion\label{sec:Equations-of-motion}}

From the Hamiltonian in Eq.$~$\eqref{eq:hamiltonian_reduced} we
obtain the equations of motion for the set of operators $\hat{\boldsymbol{\xi}}=(\hat{\mathbf{e}}_{3},\hat{\mathbf{L}},\hat{\mathbf{S}},\hat{a}_{1},\hat{a}_{1}^{\dagger},\hat{a}_{2},\hat{a}_{2}^{\dagger})$
via the Heisenberg equation 
\begin{equation}
\dfrac{\mathrm{d}}{\mathrm{d}t}\hat{\boldsymbol{\xi}}=\dfrac{\mathrm{i}}{\hbar}[\hat{H},\hat{\boldsymbol{\xi}}],\label{eq:Heisenberg_EoM}
\end{equation}
and focus on the classical limit by taking the average value of the
operators over coherent states and disregarding any correlations and noise \citep{gardiner1985input} (see
Appendix$~$\ref{sec:Commutation-Relations}). In the following we
denote $\langle\boldsymbol{\hat{\xi}}\rangle=\boldsymbol{\xi}$. We
also add dissipative terms, which we discuss in detail below. The
coupled equations of motion read 
\begin{align}
\dot{\mathbf{e}}_{3} & =-\omega_{I}\mathbf{e}_{3}\times\mathbf{L},\nonumber \\
\dot{\mathbf{L}} & =2\omega_{D}\left(\mathbf{e}_{3}\cdot\mathbf{S}\right)\left(\mathbf{e}_{3}\times\mathbf{S}\right)-\boldsymbol{\tau}_{\mathrm{G}},\nonumber \\
\dot{\mathbf{S}} & =\mathbf{B}_{{\rm eff}}\times\mathbf{S}-2\omega_{D}\left(\mathbf{e}_{3}\cdot\mathbf{S}\right)\left(\mathbf{e}_{3}\times\mathbf{S}\right)+\boldsymbol{\tau}_{\mathrm{G}},\nonumber \\
\dot{a}_{1} & =\mathrm{i}\Delta_{1}a_{1}-\mathrm{i}gS_{+}a_{2}+\epsilon_{L1}-\dfrac{1}{2}\kappa a_{1},\nonumber \\
\dot{a}_{2} & =\mathrm{i}\Delta_{2}a_{2}-\mathrm{i}gS_{-}a_{1}+\epsilon_{L2}-\dfrac{1}{2}\kappa a_{2},\label{eq:eom}
\end{align}
where $\mathbf{B}_{\mathrm{eff}}=-\omega_{B}\mathbf{e}_{z}+g(a_{1}^{\dagger}a_{2}+a_{2}^{\dagger}a_{1})\mathbf{e}_{x}+\text{i}g(a_{1}^{\dagger}a_{2}-a_{2}^{\dagger}a_{1})\mathbf{e}_{y}$.
The evolution of the total angular momentum $\mathbf{J}=\mathbf{L}+\mathbf{S}$
is given by
\begin{equation}
\dot{\mathbf{J}}=\mathbf{B}_{{\rm eff}}\times\mathbf{S},\label{eq:eomJ}
\end{equation}
and the dynamics is given either in terms of $(\mathbf{S},\mathbf{L})$
or $(\mathbf{S},\mathbf{J})$.

The equations of motion include cavity damping terms $\kappa$, which
we consider to include only radiative losses \citep{buck2003optimal}, i.e. $\kappa=\kappa_\mathrm{rad}$,
although other sources, such as surface roughness, can also be relevant
depending on the size of the particle. Furthermore, we introduced
Gilbert damping to account for intrinsic damping of the spin by adding
the torque $\boldsymbol{\tau}_{\mathrm{G}}$ \citep{gilbert2004phenomenological,keshtgar2017magnetomechanical},
\begin{equation}
\boldsymbol{\tau}_{\mathrm{G}}=\dfrac{\eta_{\mathrm{G}}}{S}\dot{\mathbf{S}}\times\mathbf{S}-\frac{\eta_{\mathrm{G}}}{S}\omega_{I}\mathbf{S}\times(\mathbf{S}\times\mathbf{L}),\label{eq:gilbert_torque}
\end{equation}
where $\eta_{\mathrm{G}}$ is a dimensionless damping parameter (in
the order of $10^{-4}$ for YIG). The second term stems from the fact
that the Gilbert damping is defined in the rotating (body) frame,
describing the relative motion of the magnetization with respect to
the lattice. Thus a coordinate transformation is required to include
damping in the laboratory frame. This additional term gives rise to
the Barnett effect, i.e. magnetization dynamics induced by rotation
\citep{barnett1915magnetization,keshtgar2017magnetomechanical}. Since
Gilbert damping is solely due to internal forces, we subtracted $\boldsymbol{\tau}_{\mathrm{G}}$
in the equation for $\dot{\mathbf{L}}$, such that the total angular
momentum is only affected by the external torque (see Eq.$~$\eqref{eq:eomJ})
\citep{band2018dynamics}. This approximation neglects,
for example, heating effects or damping into other magnon modes.

The dynamics described by Eq.$~$\eqref{eq:eom} conserves two quantities:
\begin{equation}
\begin{aligned}|\mathbf{e}_{3}|= & \sqrt{R_{31}^{2}+R_{32}^{2}+R_{33}^{2}}=1,\\
|\mathbf{S}|= & \sqrt{S_{x}{}^{2}+S_{y}^{2}+S_{z}^{2}}=S.
\end{aligned}
\label{eq:conservation_rules}
\end{equation}
Furthermore, for negligible Gilbert damping (i.e. $\eta_{\mathrm{G}}/S\rightarrow0$)
 \footnote{For YIG particles of $\sim(1\,\mu{\rm m})^{3}$, this approximation
is justified as $\eta_{\mathrm{G}}/S<10^{-14}$.}, the equations of motion yield a third conserved quantity
\begin{equation}
\mathbf{e}_{3}\cdot\mathbf{L}=\text{const}.\label{eq:conservtion_Le3}
\end{equation}
This is obtained from the first two equations of motion for $\boldsymbol{\tau}_{G}=\mathbf{0}$
as it imposes that $\dot{\mathbf{e}}_{3}\cdot\mathbf{L}+\dot{\mathbf{L}}\cdot\mathbf{e}_{3}=\dfrac{\mathrm{d}}{\mathrm{d}t}(\mathbf{e}_{3}\cdot\mathbf{L})=0$.
In this case, if $\mathbf{L}(t=0)=\bm{0}$, at any instant of time
$\mathbf{e}_{3}(t)\cdot\mathbf{L}(t)=0$, and thus, the sphere does
not rotate around its anisotropy axis. The only possible angular motions
are precession and libration, with angular speeds given by the components
of $\mathbf{L}$ perpendicular to $\mathbf{e}_{3}$. 

The equation of motion for $\mathbf{L}$ can be recast in a more familiar
form
\begin{equation}
\hbar\dot{\mathbf{L}}=\bm{\tau}=\hbar\mathbf{e}_{3}\times2\omega_{D}\left(\mathbf{e}_{3}\cdot\mathbf{S}\right)\mathbf{S}=R\mathbf{e}_{3}\times\mathbf{F}.\label{eq:torque}
\end{equation}
The angular torque $\bm{\tau}$ is therefore generated by the force
$\mathbf{F}=\frac{2\hbar\omega_{D}}{R}\left(\mathbf{e}_{3}\cdot\mathbf{S}\right)\mathbf{S}$.
The torque $\bm{\tau}$ can be driven by a laser, which induces, via
optomagnonic coupling, the spin dynamics. In particular, if the system's
initial configuration is the undriven equilibrium point $\bm{S}\parallel\bm{e}_{3}=\bm{e}_{z}$,
the only source of initial motion is the laser drive. Any subsequent
time evolution of the spin, and consequently any angular motion of
the sphere, is generated by a combination of the optically induced
torque and the torque by the external magnetic field. The latter is
responsible for a Larmor precession of the spin around the direction
of the magnetic field. 

From the equations of motion Eq.$~$\eqref{eq:eom} it can be further seen
that in the absence of crystalline anisotropy, namely $\omega_{D}=0$,
the angular motion is completely decoupled from the optomagnonic part
for $\eta_{G}=0$, i.e. in this case one recovers the dynamics for
a pure optomagnonic system as treated for one optical mode in Ref.$~$\citep{violakusminskiyCoupledSpinlightDynamics2016a}. If Gilbert
damping is considered $\eta_{G}\neq0$, then there is a coupling of
the angular motion to the spin due to conservation of angular momentum.

We will now proceed to characterize the dynamics of this driven dissipative
system. We will first obtain the possible steady state configurations
given by Eq.$~$\eqref{eq:eom}, and then study the light-spin-rotation
dynamics for small perturbations around such steady states.

\section{Steady State \label{sec:SteadyState-1}}

We find the
possible steady states of the system by setting $\dfrac{\mathrm{d}\xi}{\mathrm{d}t}=0$
$\forall\,\xi$ in Eq.$~$\eqref{eq:eom}. Denoting by $\bar{\xi}$
the steady state value of the quantity $\xi(t)$, we find
the following possible steady states, as depicted in Fig. \ref{Fig:euler_angles}(c): 
\begin{align}
(\mathrm{I)\;} & \bar{\mathbf{L}}=\mathbf{0},\bar{\mathbf{e}}_{3}\parallel\pm\mathbf{\bar{S}}\nonumber \\
\mathrm{(II)\;} & \bar{\mathbf{L}}=\mathbf{0},\bar{\mathbf{e}}_{3}\perp\bar{\mathbf{S}}\nonumber \\
\mathrm{(III)\;} & \mathbf{\bar{L}\neq0},\bar{\mathbf{e}}_{3}\parallel\pm\bar{\mathbf{L}}\parallel\pm\bar{\mathbf{S}}\label{eq:possible_steadyState}
\end{align}
with $\mathbf{\bar{S}\parallel\pm\bar{\mathbf{B}}_{\mathrm{eff}}}$.
In the case of a non-rotating sphere ($\bar{\mathbf{L}}=\mathbf{0}$), (I)
minimizes the angular-spin interaction energy for an easy-axis anisotropy,
whereas (II) minimizes the interaction energy for a hard-axis anisotropy.
If $\mathbf{\bar{L}\neq0}$, then the sphere rotates around $\mathbf{e}_{3}$
and $\bar{\mathbf{L}}$ must be parallel to $\bar{\mathbf{S}}$, which
implies that $\bar{\mathbf{e}}_{3}\parallel\mathbf{\pm\bar{S}}$.

For the steady states of the spin and the optical fields, we find a solution with a finite average number of circulating photons in both modes ($\vert\bar{a}_{i}\vert^2\neq0$) when both modes are simultaneously driven, $\epsilon_{Li}\neq0$ (Supplement 1, Sec.$~$4).
The light creates a deflection of the magnetization, such that in
the steady-state
\begin{align}
\bar{S}_{x}= & S\cos\phi\sin\theta,\nonumber\\
\bar{S}_{y}= & S\sin\phi\sin\theta,\nonumber\\
\bar{S}_{z}= & S\cos\theta.\label{eq:SpinSteadyState-1}
\end{align}
with the azimuthal angle $\phi$
\begin{equation}
\phi=\arctan\left(\mathrm{i}\dfrac{\bar{a}_{2}^{\ast}\bar{a}_{1}-\bar{a}_{1}^{\ast}\bar{a}_{2}}{\bar{a}_{2}^{\ast}\bar{a}_{1}+\bar{a}_{1}^{\ast}\bar{a}_{2}}\right),\label{eq:azimuthalDeflection}
\end{equation}
and the deflection $\theta$ from the $z$-axis
\begin{equation}
\theta=\arctan\left(\dfrac{2g}{\omega_{B}}|\bar{a}_1||\bar{a}_2|\right),\label{eq:deflectionAngle}
\end{equation}
which depends on the square root of the circulating photon number in each optical mode.

In the limit that the optically induced magnetic field is smaller
than the external magnetic field, i.e. $\omega_{B}\gg4g\vert\bar{a}_{1}\vert\vert\bar{a}_{2}\vert$,
the optical steady state reduces to
\begin{align}
\bar{a}_{1}= & -\frac{\epsilon_{L1}}{\mathrm{i}\left(\Delta_{1}+\frac{2g^{2}\vert\bar{a}_{2}\vert^{2}S}{\omega_{B}}\right)-\frac{1}{2}\kappa},\label{eq:a1_steadyState}
\end{align}
where $\bar{a}_{2}$ is given by 
\begin{equation}
\begin{aligned}\mathrm{i}\Delta_{2}\bar{a}_{2}+\mathrm{i}\frac{2g^{2}S}{\omega_{B}}\frac{\epsilon_{L1}^{2}}{\left(\Delta_{1}+\frac{2g^{2}\vert\bar{a}_{2}\vert^{2}S}{\omega_{B}}\right)^{2}+\frac{1}{4}\kappa^{2}}\bar{a}_{2}\\
-\frac{1}{2}\kappa\bar{a}_{2}+\epsilon_{L2}=0.
\end{aligned}
\label{eq:a2_steadyState}
\end{equation}
Note that it is equivalent to first express $\bar{a}_2$ as a function of $\bar{a}_1$ and then solve the corresponding equation for $\bar{a}_1$. By driving only one optical mode, the steady state of the undriven mode equals zero. As a consequence, the steady state value of the spin is unaffected by the light, as can be seen from Eqs.$~$ \eqref{eq:azimuthalDeflection} and \eqref{eq:deflectionAngle}, and the effect of the driving is purely dynamical. In the following we focus on a pumping scheme where both modes are driven (Fig. \ref{fig:schematic_model}(c)).

For the range of laser powers considered (up to $1\,\mathrm{W}$)
the deflection angle scales linearly with the power and we obtain
$\theta\approx0.2\sqrt{P_{1}P_{2}}{}^{\circ}/W$ for a sphere of
size $R=1\,\mu\mathrm{m}$ (taking $\Delta_{1}=-\omega_{B}$, $\Delta_{2}=0$).
Such small deflection angle justifies the static Mie splitting determined by $M_\mathrm{S}$
along the $z$-axis (see Sec. \ref{sec:Model}) also in the
presence of light. 

The steady states given in Eq.$~$\eqref{eq:possible_steadyState} are not necessarily
stable. In particular for $\mathbf{e}_{3}$ an easy-axis (hard-axis),
the perpendicular (parallel) steady state configuration is always
unstable. Further, we observe in the following that the dynamics do
not fully converge to the steady states in the considered timescales due
to the low dissipation in the system. 

\section{Dynamics \label{sec:Force-and-dynamical}}

\begin{figure}[t]
\includegraphics[width=1\columnwidth]{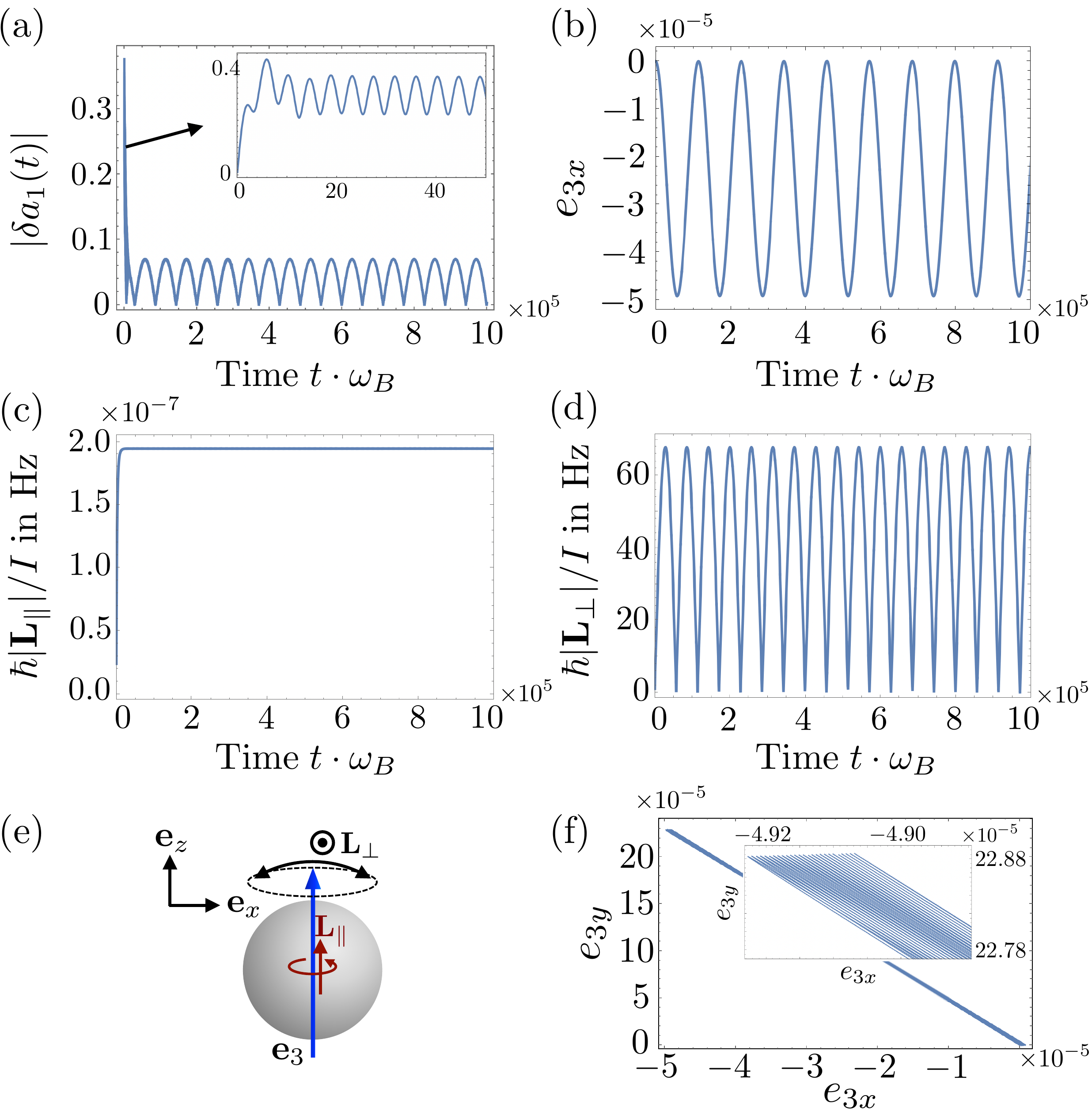} 
\caption{Dynamics of a sphere of size $R=1.5\,\mu\mathrm{m}$ with initial
configuration $\mathbf{S}\parallel\mathbf{e_{3}}\parallel\mathbf{e}_{z}$
and $\mathbf{L}=\bm{0}$. The system is red detuned with $P_{2}=0.02\,\mathrm{W}$
and $P_{1}=P_{2}/2$. (a) Amplitude fluctuation of the cavity mode
1. The inset shows the initial time evolution of the system. After
a transient period, the cavity dynamics is modulated by (b) the dynamics
of the anisotropy axis $\mathbf{e}_{3}$. Angular momentum (c) parallel
$\mathbf{L}_{\parallel}$ and (d) perpendicular $\mathbf{L}_{\perp}$
to $\mathbf{e}_{3}$. As schematically depicted in (e), $\mathbf{L}_{\parallel}$
gives the angular speed of rotation of the sphere around the anisotropy
axis via $\hbar$$\vert\mathbf{L}_{\parallel}\vert/I$ and $\mathbf{L}_{\perp}$
is related to the libration and precession of $\mathbf{e}_{3}$, shown
in the $xy$ plane (f). The time is given in terms of the magnon frequency
$\omega_{B}=10^{10}\,\mathrm{s^{-1}}$. }
\label{fig:a+e3}
\end{figure}

Having characterized the steady states of the system we now turn to
study numerically the dynamics of the system under an optical drive. For concreteness we consider an easy-axis magnet, i.e. $D>0$, and
an initial state in which the spin and the anisotropy axis are aligned
parallel along the external magnetic field, namely $\mathbf{e}_{3}(0)=\mathbf{S}(0)/S=\mathbf{e}_{z}$. For a hard-axis magnet we would retrieve the same results by choosing
an initial configuration in which the anisotropy axis is perpendicular
to the spin and therefore do not explicitly discuss them. For the
optical fields we assume that the initial conditions are given by
Eqs.$~$\eqref{eq:a1_steadyState} and \eqref{eq:a2_steadyState}. We
consider driving laser powers such that the deviation of the steady state with respect to the chosen
initial configuration is sufficiently small.
Thus, we expect the system to oscillate around this configuration
with an amplitude that depends on the driving power. Since we consider a Mie splitting given by the static magnetization along the $z$-axis, our model requires modifications for initial configurations
where the spin is not approximately along the $z$ direction, and it breaks down for pumping powers that exceed a critical value such that
the $z$ component of the spin deviates too far from the $z$-axis. In this case the so-called Suhl instabilities can further occur, opening new dissipation channels \citep{suhl1957theory}. 

We consider two pumping schemes depicted in Fig.$~$\ref{fig:schematic_model}(c):
First, the optical cavity-mode with higher frequency is pumped in resonance, such
that $\Delta_{1}=0$ and $\Delta_{2}=\omega_{B}$, which we refer
to as blue detuning and describes magnon emission; second, the mode
with lower frequency is pumped in resonance, such that $\Delta_{1}=-\omega_{B}$
and $\Delta_{2}=0$, which is related to a magnon absorption and which
we call red detuning. These processes can be interpreted as cavity-enhanced
Stokes and anti-Stokes Brillouin scattering. The Gilbert damping included
in the equations of motion Eq.$~$\eqref{eq:eom} renders the north pole
a stable equilibrium point in the absence of driving. With driving,
for the red detuning scheme, the north pole is stable whereas the
south pole results in runaway solutions which can lead to limit cycles
or chaotic dynamics. For blue detuning, the south pole can become
a stable point depending on the driving power and on the detuning.
For a more detailed discussion see Ref.$~$\citep{violakusminskiyCoupledSpinlightDynamics2016a}.
Note that we choose $P_{1}=P_{2}/2$ for our simulations.  This is
not a restrictive assumption and changing the ratio between
$P_{1}$ and $P_{2}$ does not change the results qualitatively. 

We show the time evolution of $\vert\delta a_{1}(t)\vert=\vert\bar{a}_{1}-a_{1}(t)\vert$,
i.e. the deviations of the amplitude of the higher frequency optical
mode from its steady state value, and of $\mathbf{e}_{3}$ in Figs.$~$\ref{fig:a+e3}(a)
and (b) for the red detuned pumping scheme. The initial configuration
is such that the particle has no initial rotational motion, i.e. $\mathbf{L}(t=0)=\mathbf{0}$.
We shortly discuss the dynamics of an initially rotating particle,
$\mathbf{L}(t=0)\neq\mathbf{0}$, in Appendix \ref{sec:Further-Dynamics}.
This latter case is especially relevant if one starts with an unmagnetized
particle at rest and then magnetizes it, which induces an initial
rotation due to the Einstein-de Haas effect \citep{einstein_de_haas}.

The cavity field dynamics are modulated by the angular motion and,
at short times, the dominant contributions are due to the optomagnonic
coupling and the fast spin dynamics. To gain further insight into
the driven angular motion, we show in Fig.$~$\ref{fig:a+e3}(c) the
dynamics of the absolute value of the components of $\mathbf{L}$
parallel and perpendicular to $\mathbf{e}_{3}$. The parallel component
of the angular momentum $\mathbf{L}_{\parallel}$ gives the angular
speed of rotation of the sphere around its magnetic anisotropy axis via $\hbar$$\vert\mathbf{L}_{\parallel}\vert/I$
in the lab frame, while the perpendicular component $\mathbf{L}_{\perp}$
is related to libration and precession of $\mathbf{e}_{3}$. We notice
that $\vert\mathbf{L}_{\parallel}\vert$ quickly increases but then
stays stationary. This is due to the quick spin dynamics during the
initial evolution from the starting configuration to close-to-steady-state
oscillation. After the transient dynamics the effects of spin damping,
as described by the Gilbert damping term, become negligible and $\vert\mathbf{L}_{\parallel}\vert$
is conserved.

\section{Power Spectrum Analysis\label{sec:Power-Spectrum}}

\begin{figure}[t]
\includegraphics[width=1\columnwidth]{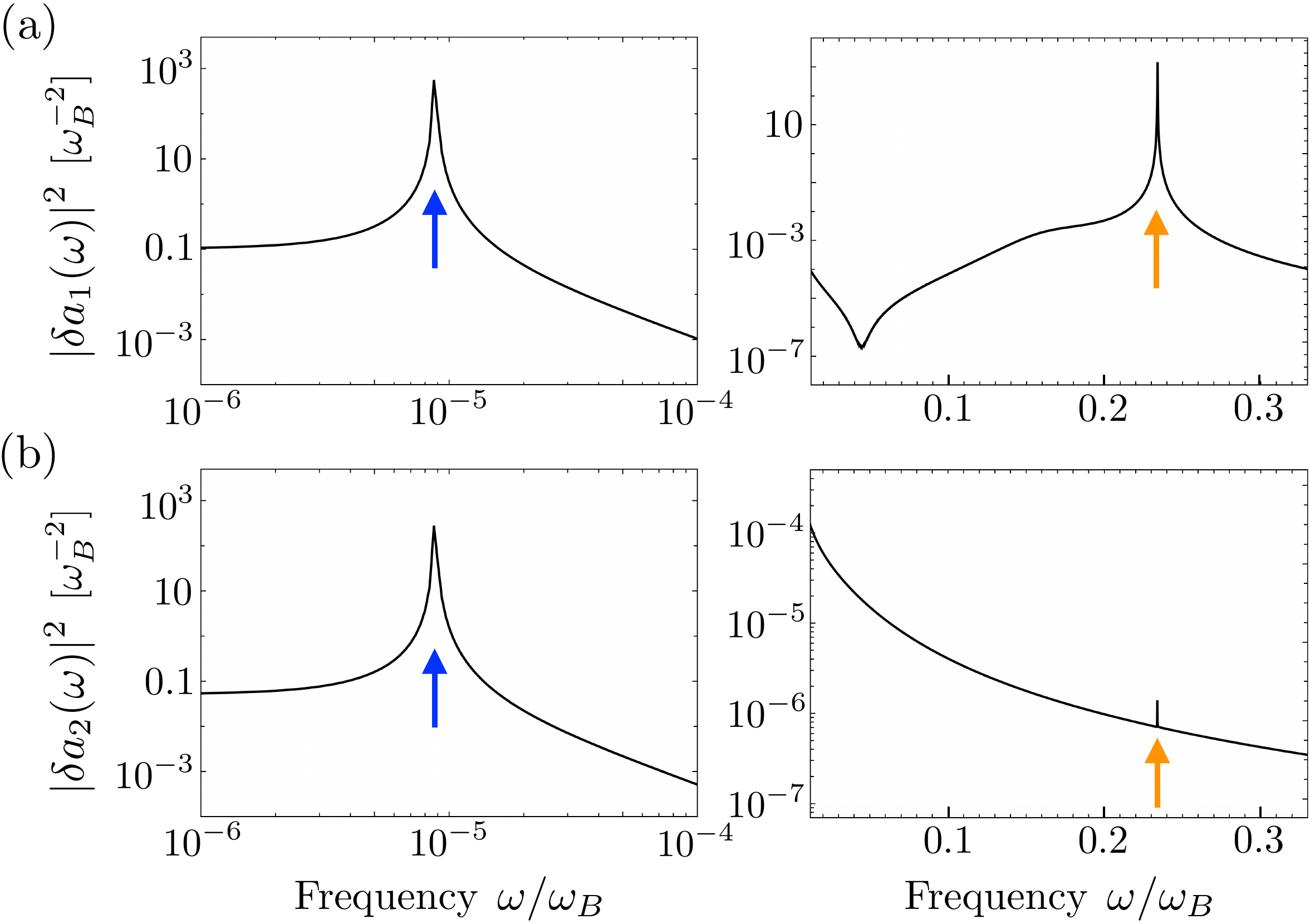}
\caption{Logarithmic power spectrum of the cavity optical modes (a) $\delta a_{1}$
and (b) $\delta a_{2}$ for a red detuned system with $P_{2}=0.02\,\mathrm{W}$
and $P_{1}=P_{2}/2$. The sphere has a radius of $R=1.5\,\mu\mathrm{m}$,
the initial state is $\mathbf{S}\parallel\mathbf{e_{3}}\parallel\mathbf{e}_{z}$,
and $\mathbf{L}=\mathbf{0}$, corresponding to the results presented in Fig.$~$\ref{fig:a+e3}(a).
Both modes yield a peak in the spectrum at a frequency of $\sim90\,\mathrm{kHz}$
(highlighted by the blue arrow) which corresponds to the induced angular
motion. At a frequency of $\sim2.33\,\mathrm{GHz}$ a second peak
can be observed which corresponds to the fast oscillations at small
times (orange arrow), related to the magntization's transient dynamics. Note that the horizontal axis in the right panel is in linear scale.
The frequency is given in terms of the magnon frequency $\omega_{B}=10^{10}\,\mathrm{s^{-1}}$.}
\label{fig:cavity_red}
\end{figure}

As we observed in the previous section, the dynamics of the cavity optical modes exhibit a transitory behavior followed by an evolution modulated
by the anisotropy axis dynamics. Namely, at short times the deviations
$\vert\delta a_{i}(t)\vert$ oscillate with the same frequency as
the spin oscillates and then adapts to the oscillations of the magnetic anisotropy
axis. The frequencies associated with each of these dynamics are imprinted
in the power spectrum of the light, as we show now. 

First, we write the cavity fields as
\begin{equation}
a_{i}(t)=\bar{a}_{i}+\delta a_{i}(t),\label{eq:cavity_field}
\end{equation}
such that $\delta a_{i}(t)$ represents the deviation of $a_{i}$
from its steady state value $\bar{a}_{i}$. We then consider the Fourier
transform of $\delta a_{i}(t)$
\begin{equation}
\mathcal{F}[\delta a_{i}(t)]=\int_{0}^{+\infty}\delta a_{i}(t)\mathrm{e}^{\mathrm{i}\omega t}\mathrm{d}t=\delta a_{i}(\omega),\label{eq:Fourier_trafo}
\end{equation}
from which we compute the power spectrum $|\delta a_{i}(\omega)|^{2}$. 

We show in Fig.$~$\ref{fig:cavity_red} the absolute value of the
fluctuations and the resulting spectrum for an initially non-rotating
sphere ($R=1.5\,\mu\mathrm{m}$) under a red detuned drive. The spectra
for both modes show a clear peak at a frequency of $\sim90\,\mathrm{kHz}$.
This peak corresponds to the oscillation frequency of the cavity field
amplitude after the initial transitory behavior. In fact, this frequency
is the same as the oscillation frequency of the magnetic anisotropy axis  and
is a fingerprint of the modulation of the cavity dynamics by the sphere's
angular motion. Further, the spectra yield a second peak at a frequency
$\sim2.33\,\mathrm{GHz}$, that is related to optomagnonic induced
transient spin oscillations including the effects of its coupling to the
angular motion (a frequency shift). The finite linewidth of the Lorentzian shaped peaks is governed by the low dissipation and the finite simulation time. The dip around $0.4\,\mathrm{GHz}$ in the power spectrum
of mode 1 can be attributed to a Fano resonance \citep{fano1961effects}
emerging from the coupling between spin and angular momentum which,
close to the steady state, act like two coupled harmonic oscillators.
Such Fano resonances have also been observed in optomechanical systems
\citep{wang2020optomechanically,qu2013fano}.

To better understand the origin of the observed resonance frequencies,
we consider now the linearized version of the equations of motion Eq.
\eqref{eq:eom} and derive the susceptibility of the cavity-field
fluctuations. For that, we consider small displacements from
the steady state 
\begin{align*}
\boldsymbol{\xi}(t) & =\bar{\boldsymbol{\xi}}+\delta\boldsymbol{\xi}(t),
\end{align*}
and retain in the equations of motion only linear terms in the fluctuations.
Since the considered optomagnonic effects are weak, to a first approximation
we assume that the steady state of the spin and the anisotropy axis
is sufficiently close to the north pole, i.e. $\bar{S}_{x,y}/S=\bar{e}_{3x,y}\approx0$
and $\bar{S}_{z}/S=\bar{e}_{3z}\approx1$, and take a non-rotating
sphere, $\mathbf{\bar{L}}=\mathbf{0}$. Relegating the detailed calculations
to the Appendix$~$\ref{sec:Resonance-Peak}, we obtain the following
susceptibility for the fluctuations $\delta a_1$ in frequency space: 
\begin{equation}
\chi_{1}(\omega)=\left[-\mathrm{i}\left(\omega+\Delta_{1}-2g^{2}S\vert\bar{a}_{2}\vert^{2}\chi_{+}(\omega)\right)+\frac{\kappa}{2}\right]^{-1},\label{eq:a1(omega)}
\end{equation}
where 
\begin{align*}
\chi_{+}^{-1}(\omega) & =\omega-\tilde{\omega}_{B}-\dfrac{|\bar{B}_{+}|^{2}}{2\omega}+\dfrac{4\omega_{D}^{2}\omega_{I}S^{3}}{2\omega_{D}\omega_{I}S^{2}-\omega^{2}}\\
 & +\dfrac{2g^{2}S|\bar{a}_{1}|^{2}\left(\omega-\Delta_{2}-\mathrm{i}\kappa/2\right)}{\kappa^{2}/4+(\omega-\Delta_{2})^{2}}-\dfrac{\bar{B}_{+}^{2}\bar{B}_{+}^{\ast2}}{4\omega^{2}}\chi_{-}(\omega),\\
\chi_{-}^{-1}(\omega) & =\omega+\tilde{\omega}_{B}-\dfrac{|\bar{B}_{+}|^{2}}{2\omega}-\dfrac{4\omega_{D}^{2}\omega_{I}S^{3}}{2\omega_{D}\omega_{I}S^{2}-\omega^{2}}\\
 & +\dfrac{2g^{2}S|\bar{a}_{1}|^{2}\left(\omega+\Delta_{2}-\mathrm{i}\kappa/2\right)}{\kappa^{2}/4+(\omega+\Delta_{2})^{2}}\\
 & -\dfrac{2g^{2}S|\bar{a}_{2}|^{2}\left(\omega-\Delta_{1}-\mathrm{i}\kappa/2\right)}{\kappa^{2}/4+(\omega-\Delta_{1})^{2}},
\end{align*}
with $\tilde{\omega}_{B}=\omega_{B}+2\omega_{D}S$ and $\bar{B}_{+}=2g\bar{a}_{1}\bar{a}_{2}^{\ast}$.
The peaks of the power spectrum are obtained by the poles of the imaginary part of $\chi_{1}[\omega]$, and are approximately given by the equation 
\begin{align}
(\omega+\Delta_{1})\left(\omega-\tilde{\omega}_{B}-\frac{|\bar{B}_{+}|^{2}}{2\omega}+\dfrac{4\omega_{D}^{2}\omega_{I}S^{3}}{2\omega_{D}\omega_{I}S^{2}-\omega^{2}}\right)\nonumber \\
-2Sg^{2}\vert\bar{a}_{2}\vert^{2}=0.\label{eq:peak_frequencies}
\end{align}
For the values used in Fig.$~$\ref{fig:cavity_red}, we obtain from
Eq.$~$\eqref{eq:peak_frequencies} two resonance frequencies: $\omega_{L}/(2\pi)\approx88.7\,\mathrm{kHz}$
and $\omega_{H}/(2\pi)\approx2.33\,\mathrm{GHz}$, in agreement with
the numerical results shown in Fig.$~$\ref{fig:cavity_red}.

\begin{figure}[t]
\includegraphics[width=1\columnwidth]{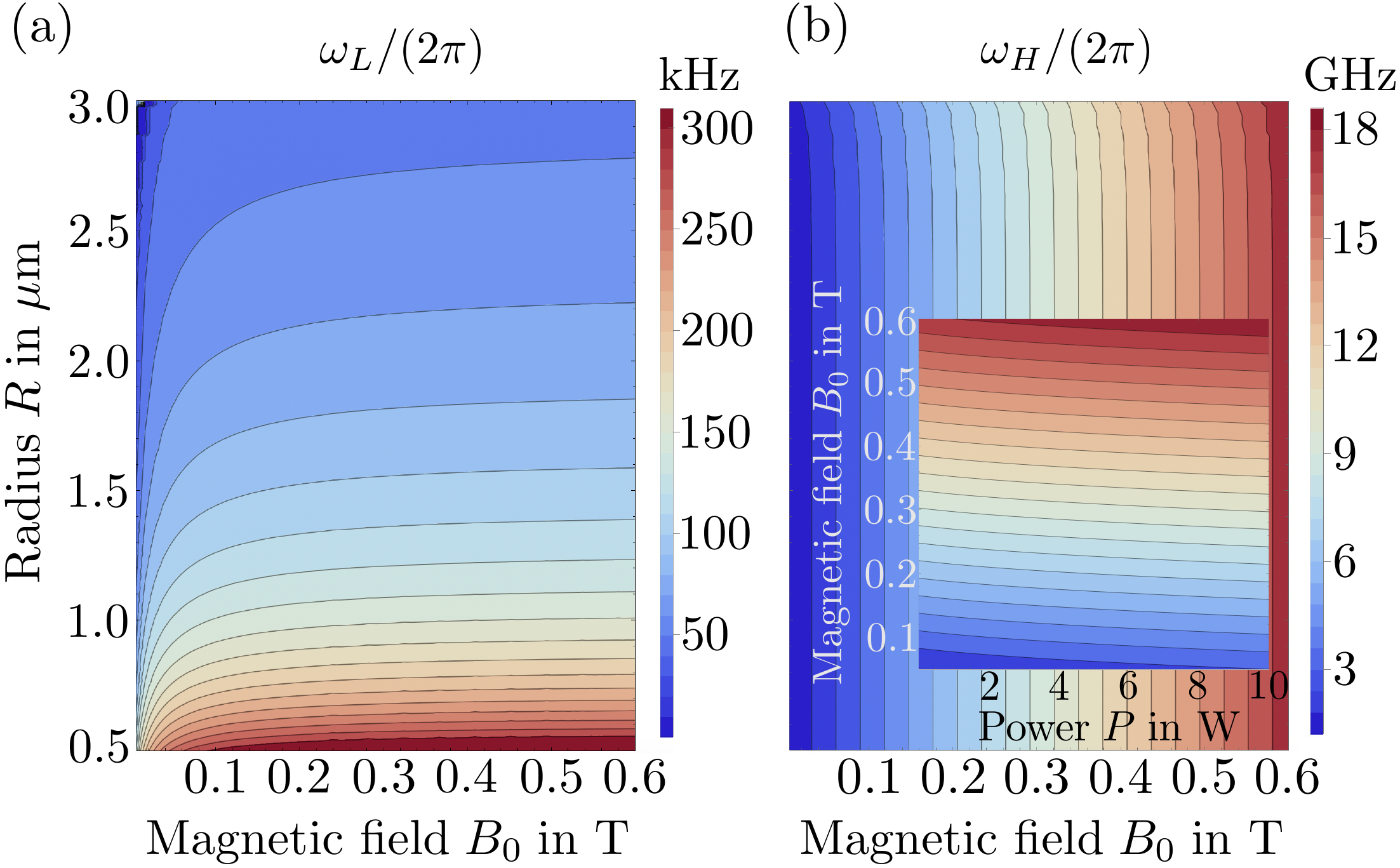}
\caption{Phase diagrams for a red detuned system. Values of the (a) low $\omega_{L}$
and (b) high $\omega_{H}$ resonance frequency depending on the radius
$R$ and the magnetic field $B_0$ for $P_{2}=0.02\,\mathrm{W}$ and
$P_{1}=P_{2}/2$. The inset shows the dependence on $B_0$ and the laser
power $P_{2}$ for an increased optomagnonic coupling $g'=10g$ and
$R=1.5\,\mu\mathrm{{m}}$. The resonance frequencies are obtained
from the imaginary part of the eigenvalues of the linearized system Eq.$~$\eqref{eq:eom_linearized}. }
\label{fig:phase_diagrams}
\end{figure}

The high-frequency peak $\omega_{H}$ can be obtained by discarding the terms $\dfrac{4\omega_{D}^{2}\omega_{I}S^{3}}{2\omega_{D}\omega_{I}S^{2}-\omega^{2}}$
and $|\bar{B}_{+}|^{2}/(2\omega)$ in Eq.$~$\eqref{eq:peak_frequencies}.
This corresponds to neglecting the dynamics of $\mathbf{e}_{3}$ (a
good approximation for describing the dynamics of the system for periods
$\ll1/\omega_{I}$), and to considering the optically induced magnetic
field negligible compared to the applied external magnetic field.
Under these approximations, the high-frequency peak is given by
$\omega_{H}\approx\tilde{\omega}_{B}$. The coupling between spin
and angular motion changes the spin precession frequency by a factor
$2\omega_{D}S$ compared to the uncoupled case \citep{kittel1948theory}. Note that this approximation for the shift in the spin precession frequency is volume independent and only depends on the material properties
like the anisotropy constant or the density. This can also be seen
in Fig.$~$\ref{fig:phase_diagrams} (for radii up to $R\approx2.5\,\mu\mathrm{{m}}$)
where we show the dependence of the resonance frequencies on the particle
radius and the external magnetic field. Regarding its dependence on
the laser power, we observe that in principle $\omega_{H}$ is shifted
to higher frequencies for larger powers. However, for the considered
values of the optomagnonic coupling, this effect is negligible. Nonetheless,
the dependence is shown in the inset of Fig.$~$\ref{fig:phase_diagrams}(b)
for an increased coupling $g'=10g.$ In this case we further find
that an additional peak close to the lower frequency peak appears
in the spectrum for power values $P_{2}>0.1\,\mathrm{{W}}$. The same
can be observed for the unchanged optomagnonic coupling strength for
smaller magnon frequencies. However, this peak is weak in amplitude
and thus we do not further investigate it. 

An initial rotation does not change the power spectrum qualitatively
but decreases the frequency of the lower peak $\omega_L$. Such effects
are typically small for $L\sim S$, which corresponds to frequencies
of $\sim100/(2\pi)\,\mathrm{Hz}$. For $L\gg S$, rotations
yield a more pronounced frequency shift, for instance, for $L=10^{3}S$
($L=10^{4}S$) the position of the lower peak is at $\sim80\,\mathrm{kHz}$
($\sim40\,\mathrm{kHz}$). For large $L$ gyroscopic effects become
more dominant such that the torque due to the fast angular rotation
stabilizes the anisotropy axis and thus decreases its oscillation
frequency. As mentioned before, the discussion presented here is valid
as long as the initial state of the system is close to its steady
state. %
\begin{comment}
The induced torque could be probed by measuring the polarization state
of the transmitted laser beam. The angular momentum transfer renders
the balance of left ($+\hbar$) and right ($-\hbar$) circular components
of the light such that the transmitted beam has a net angular momentum.
The induced torque is determined by difference of the power of left
and right circular components that is measurable with a torque detector,
as demonstrated in \citep{la2004optical}. 
\end{comment}

\section{RF Driven System \label{sec:Driven-System}}

\begin{figure}[t]
\includegraphics[width=1\columnwidth]{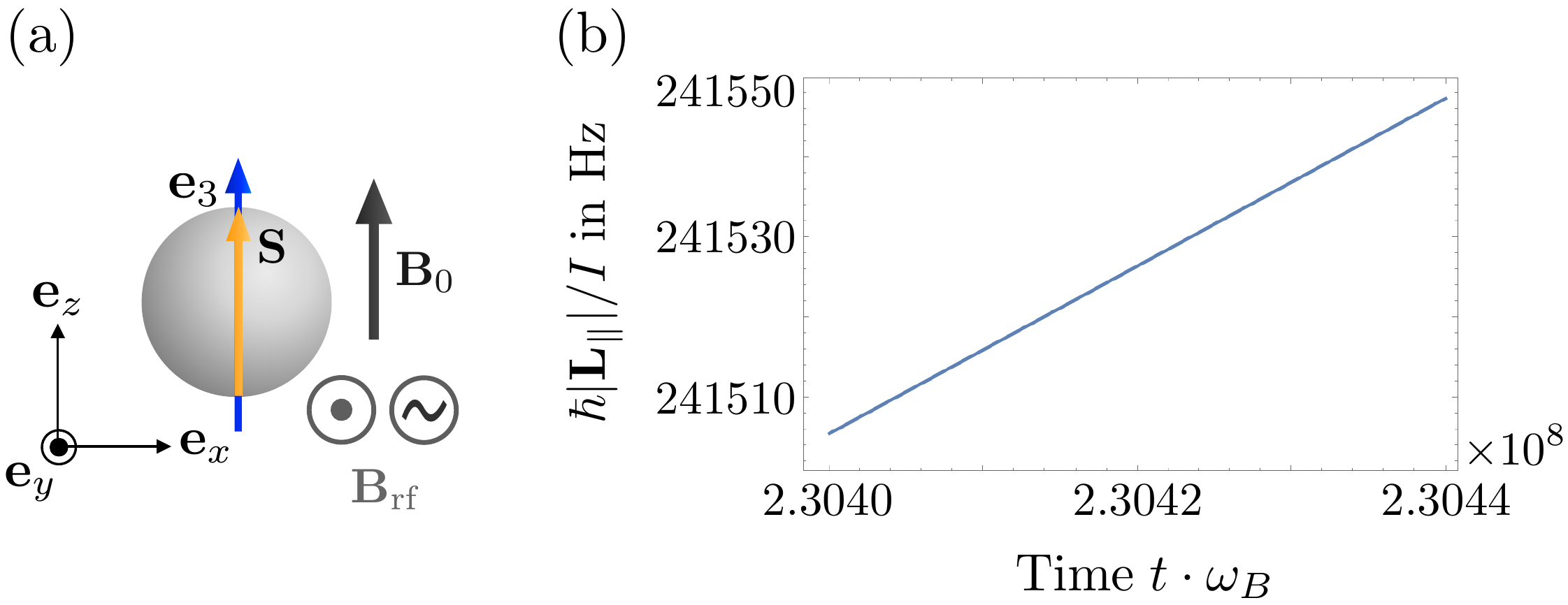}
\caption{(a) Illustration of the driven system. An oscillating magnetic field
is applied perpendicular to the bias field $B_0$. (b) Angular speed of rotation
of the sphere around the anisotropy axis. The system is red detuned
with $R=1.5\,\mu\mathrm{m}$, $P_{2}=0.02\,\mathrm{W}$ and $P_{1}=P_{2}/2$
and initial configuration $\mathbf{S}\parallel\mathbf{e_{3}}\parallel\mathbf{e}_{z}$
and $\mathbf{L}=\bm{0}$. The amplitude of the driving magnetic field
is given by $a_{\mathrm{rf}}=0.01$. The angular speed reached after
$\sim20\,\mathrm{ms}$ is 12 orders of magnitude higher than the corresponding
rotation of the undriven system shown in Fig.$\,$\ref{fig:a+e3}(c).
The time is given in terms of the magnon frequency $\omega_{B}=10^{10}\,\mathrm{s^{-1}}$.}
\label{fig:rotationSpeed_driven}
\end{figure}
To enhance the peak in the power spectrum related to the spin oscillations
as well as the angular speed of rotation of the sphere, we take inspiration
from ferromagnetic resonance experiments where a periodic magnetic
field is applied perpendicular to the bias field, as schematically
depicted in Fig. \ref{fig:rotationSpeed_driven}(a). A resonant drive
corresponds to an oscillating driving field with frequency matching
the magnetization's precession frequency. We thus include in our model
a periodic magnetic field $\mathbf{B}_{\mathrm{rf}}=a_{\mathrm{rf}}B_0\cos(\omega_{\mathrm{rf}}t)\mathbf{e}_{y}$
where $a_{\mathrm{rf}}$ is a dimensionless factor describing the
ratio between the FMR drive amplitude and the bias field strength
(given by $B$). The Hamiltonian Eq.$~$\eqref{eq:hamiltonian_reduced} is
then augmented by the term
\begin{align}
\hat{H}_{\mathrm{\mathrm{rf}}}= & -\hbar a_{\mathrm{rf}}\omega_{B}\hat{S}_{y}\cos(\omega_{\mathrm{rf}}t).\label{eq:Hamiltonian_driven}
\end{align}
The amplitude of the driving field can be related to the excitation
power $P_{\text{mw}}$ by $a_{\mathrm{rf}}\omega_{\mathrm{rf}}=2\sqrt{\kappa_{\text{mw}}P_{\text{mw }}/(\hbar\omega_{\text{mw}})}$
with the microwave frequency $\omega_{\mathrm{mw}}$ and coupling to the antenna
$\kappa_{\mathrm{mw}}$. The value of the latter can be kept close
to the intrinsic magnon dissipation $\kappa_{m}\sim1\,\mathrm{MHz}$
\citep{tabuchi2016quantum,rameshti2021cavity} to enable a maximal
power flow into the magnet. Thus, for $\omega_{\mathrm{mw}}/(2\pi)=10\,\mathrm{GHz}$
this corresponds to $P_{\mathrm{mw}}\sim a_{\mathrm{rf}}^{2}\,\mathrm{pW}$
which is comparable to excitation powers used in Ref.$~$\citep{lachance2017resolving}.

\begin{figure}[t]
\includegraphics[width=1\columnwidth]{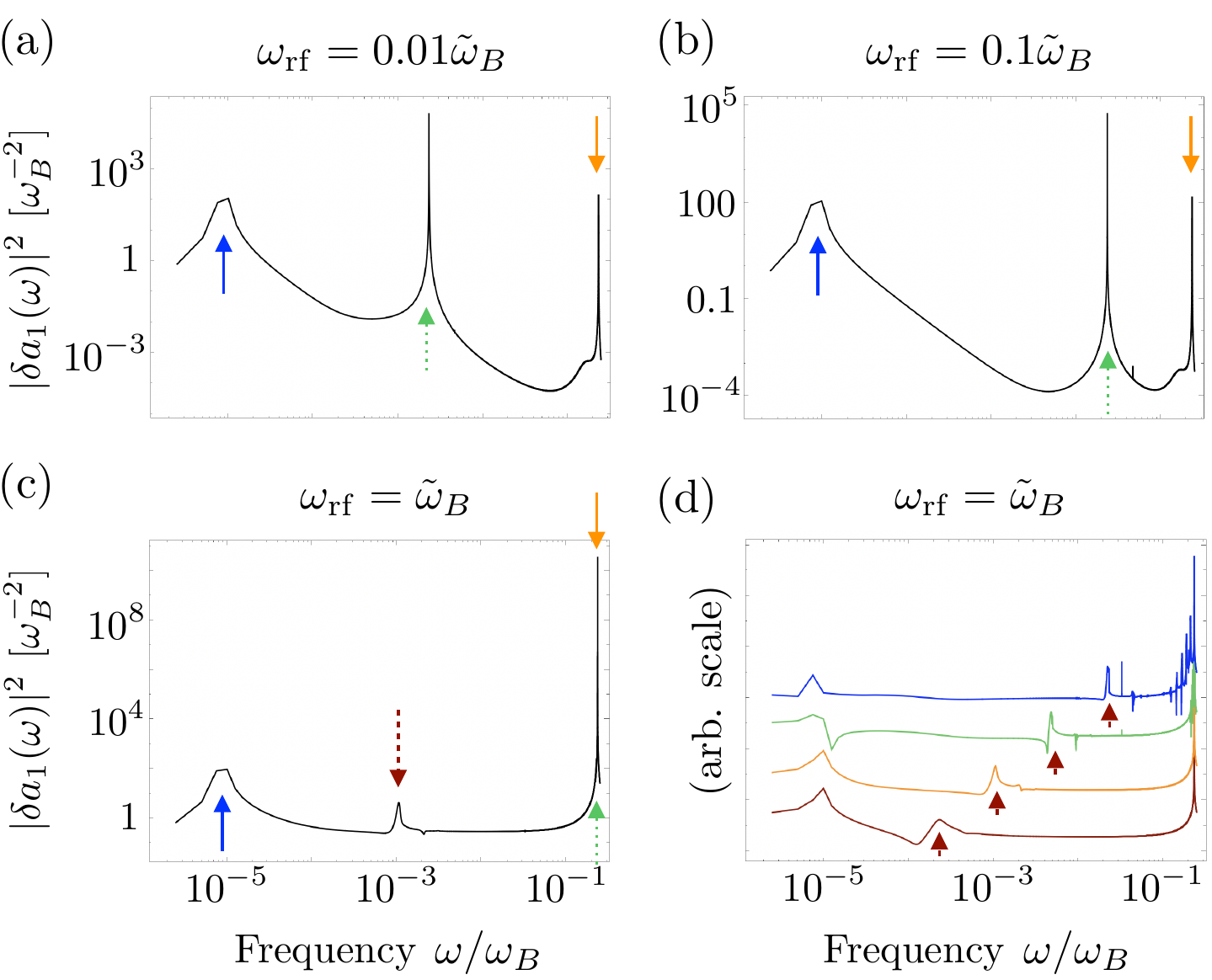}
\caption{Logarithmic power spectrum of the driven system of $\delta a_{1}$
for $a_{\mathrm{rf}}=10^{-3}$ and FMR frequency (a) $\omega_{\mathrm{rf}}=0.01\tilde{\omega}_{B}$,
(b) $\omega_{\mathrm{rf}}=0.1\tilde{\omega}_{B}$, (c) $\omega_{\mathrm{rf}}=\tilde{\omega}_{B}$.
The system is red detuned with $P_{2}=0.02\,\mathrm{W}$, $P_{1}=P_{2}/2$
and $R=1.5\,\mu\mathrm{m}$. The initial state is $\mathbf{S}\parallel\mathbf{e_{3}}\parallel\mathbf{e}_{z}$,
and $\mathbf{L}=\mathbf{0}$. The FMR frequency is imprinted in the spectrum
by a corresponding peak (marked by the green, dotted arrow). For $\omega_{\mathrm{rf}}=\tilde{\omega}_{B}$,
additionally to the peaks at $\sim100\,\mathrm{kHz}$ and $\sim2.33\,\mathrm{GHz}$,
the power spectrum yields a third peak (red, dashed arrow) that depends
on the amplitude of the driving field (d). The driving field amplitude
values are $a_{\mathrm{rf}}=10^{-4},10^{-3},10^{-2},10^{-1}$ from
bottom to top. The frequency is given in terms of the magnon frequency
$\omega_{B}=10^{10}\,\mathrm{s^{-1}}$.}
\label{fig:powerSpectrum_driven}
\end{figure}

This external drive not only enhances the visibility of the peaks
in the cavity spectrum but also introduces new features depending
on the frequency of the drive. Fig.$~$\ref{fig:powerSpectrum_driven}(a)-(c)
shows the power spectrum of cavity mode $1$ for different FMR frequencies $\omega_{\mathrm{rf}}$ which are visible by a corresponding peak.
If we choose the FMR frequency $\omega_{\mathrm{rf}}$ to match the
shifted spin precession frequency $\omega_{H}\approx\tilde{\omega}_{B}$,
the additional field enhances the amplitude of the higher frequency
peak $\omega_H$ and yields a third peak with intermediate frequency between $\omega_L$ and $\omega_H$. This extra resonance made visible by the rf driving field is due to a mismatch between the
optical modes frequency difference and the shifted magnetization precession
frequency due to the coupling to angular motion. This is a consequence of the nonlinearity of the system and is not captured by the analytically calculated susceptibilities, so we have to resort to numerical analysis. The position of the appearing peak depends on the driving
field amplitude (Fig.$~$\ref{fig:powerSpectrum_driven}(d)), namely,
for larger amplitudes this peak shifts to higher frequencies.

The driving magnetic field also affects the angular motion of the
particle. In fact, it increases the angular speed of rotation of the
sphere around the anisotropy axis. This can be seen in Fig. \ref{fig:rotationSpeed_driven}(b)
where we show $\hbar \vert\mathbf{L}_{\parallel}\vert/I$ in the
red detuned scheme for the initial configuration $\mathbf{e}_{3}(0)=\mathbf{S}(0)/S=\mathbf{e}_{z}$,
and $\mathbf{L}(t=0)=\mathbf{0}$. We observe that the induced rotation is
always counterclockwise with respect to $\mathbf{\mathbf{e}_{3}}$.
Thus, an initially clockwise rotating particle will be decelerated.
The effects of rotation are only imprinted in the optical power
spectrum for high rotation speeds as discussed in the previous section. 

\section{Generalizations \label{sec:Generalisations}}

In this section we briefly discuss some possible generalizations
to the Hamiltonian Eq.$~$\eqref{eq:hamiltonian_reduced} of our model. 

\subsection{Cotton-Mouton Effect\label{sec:Cotton-Mouton-Effect}}

So far we considered that only the Faraday effect plays a role in
optomagnonic coupling. Another effect relevant in several optomagnonic
systems is the Cotton-Mouton effect \citep{stancil2009spin,osada2018orbital},
which we now include, discussing its possible impacts on our results. 

The Cotton-Mouton (CM) effect is quadratic in the magnetization and
is taken into account in the full effective permittivity tensor \citep{wettling1975relation} 
\begin{equation}
\varepsilon_{ij}(\mathbf{M})=\varepsilon_{0}\left(\varepsilon\delta_{ij}-\mathrm{i}f\sum_{k}\epsilon_{ijk}M_{k}+\sum_{kl}G_{ijkl}M_{k}M_{l}\right).\label{eq:permettivity_tensor}
\end{equation}
Notice that this includes a term $\propto M_{k}M_{l}$ which was not
included in our previous formalism {[}cf. Eq. \eqref{eq:permittivity_tensor_Faraday}{]}.
For cubic crystals, such as YIG, most of the tensor components $G_{ijkl}$
are zero by symmetry and the non-vanishing ones are \citep{pisarev1971magnetic,wettling1975relation}
$G_{iiii}=G_{11}$, $G_{iijj}=G_{12}$, $G_{ijij}=G_{44}$ for $i\neq j$
and $i,j\in\{1,2,3\}$.

The Hamiltonian obtained by quantizing the CM contribution to the
energy density is given by (for details see Appendix \ref{sec:CM-effect-app}) 
\begin{equation}
\begin{aligned}\hat{H}_{\mathrm{int}}^{\mathrm{CM}}= & \hbar g_{\mathrm{CM}}^{44}\left(\hat{S}_{z}\hat{S}_{+}\hat{a}_{1}^{\dagger}\hat{a}_{2}+\hat{S}_{-}\hat{S}_{z}\hat{a}_{2}^{\dagger}\hat{a}_{1}\right)\\
 & +\hbar g_{\mathrm{CM}}^{12}\hat{S}_{z}^{2}(\hat{a}_{1}^{\dagger}\hat{a}_{1}-\hat{a}_{2}^{\dagger}\hat{a}_{2}),
\end{aligned}
\label{eq:H_Int_CM_final-1}
\end{equation}
where we neglected the constant shift in the photon field. The coupling
constants $g_{\mathrm{CM}}^{44}$ and $g_{\mathrm{CM}}^{12}$ are
given by 
\begin{align}
g_{\mathrm{CM}}^{44} & =\dfrac{\varepsilon_{0}}{2\hbar}\left(\dfrac{M_{\mathrm{S}}}{S}\right)^{2}G_{44}\tilde{g},\nonumber \\
g_{\mathrm{CM}}^{12} & =\dfrac{\omega_{r}}{8\epsilon}\left(\dfrac{M_{\mathrm{S}}}{S}\right)^{2}(G_{11}-G_{12}),\label{eq:gCM}
\end{align}
under the approximation $\int\mathrm{d}\mathbf{r}|E_{h}|^{2}\approx\int\mathrm{d}\mathbf{r}|E_{v}|^{2}\approx\dfrac{\hbar\omega_{r}}{2\varepsilon_{0}\varepsilon}$. 

The equations of motion for the spin and the cavity fields including
the additional interacting term Eq.$~$\eqref{eq:H_Int_CM_final-1} read
\begin{align}
\dot{\mathbf{S}}= & \mathbf{\tilde{B}}_{{\rm eff}}\times\mathbf{S}-2\omega_{D}\left(\mathbf{e}_{3}\cdot\mathbf{S}\right)\left(\mathbf{e}_{3}\times\mathbf{S}\right)+\tau_{\mathrm{G}},\nonumber \\
\dot{a}_{1}= & \mathrm{i}(\Delta_{1}-g_{\mathrm{CM}}^{12}S_{z})a_{1}-\mathrm{i}(g+g_{\mathrm{CM}}^{44}S_{z})S_{+}a_{2}+\epsilon_{L1}-\dfrac{1}{2}\kappa a_{1},\nonumber \\
\dot{a}_{2}= & \mathrm{i}(\Delta_{2}+g_{\mathrm{CM}}^{12}S_{z})a_{2}-\mathrm{i}(g+g_{\mathrm{CM}}^{44}S_{z})S_{-}a_{1}+\epsilon_{L2}-\dfrac{1}{2}\kappa a_{2},\label{eq:eom_CM}
\end{align}
where the effective field is now given by $\tilde{\mathbf{B}}_{\mathrm{eff}}=\mathbf{B}_{\mathrm{eff}}+\mathbf{B}_{\mathrm{opt}}^{\mathrm{CM}}$
with
\begin{align}
\mathbf{B}_{\mathrm{opt}}^{\mathrm{CM}}= & g_{\mathrm{CM}}^{44}\left(\begin{array}{c}
(a_{1}^{\dagger}a_{2}+a_{2}^{\dagger}a_{1})S_{z}\\
\mathrm{i}(a_{1}^{\dagger}a_{2}-a_{2}^{\dagger}a_{1})S_{z}\\
(a_{1}^{\dagger}a_{2}+a_{2}^{\dagger}a_{1})S_{x}+\mathrm{i}(a_{1}^{\dagger}a_{2}-a_{2}^{\dagger}a_{1})S_{y}
\end{array}\right)\nonumber \\
 & -2g_{\mathrm{CM}}^{12}\left(\begin{array}{c}
(a_{1}^{\dagger}a_{1}-a_{2}^{\dagger}a_{2})S_{x}\\
(a_{1}^{\dagger}a_{1}-a_{2}^{\dagger}a_{2})S_{y}\\
0
\end{array}\right).\label{eq:Bopt_CM}
\end{align}

For YIG, $G_{44}M_{\mathrm{S}}^{2}=-1.14\cdot10^{-4}$ and $(G_{11}-G_{12}-2G_{44})M_{\mathrm{S}}^{2}=5.73\cdot10^{-5}$
\citep{stancil2009spin}. For small deflections of the spin from the
$z$ axis, i.e. $S_{z}\sim S$ and $S_{x,y}\sim0$, the additional
contribution to the optically induced magnetic field is of the same
structure and magnitude as the one caused by the Faraday effect, considering
that the value of $g_{\mathrm{CM}}^{44}$ mainly differs from $g$
given by Eq.$~$\eqref{eq:g} by a factor of $S^{-1}$. Therefore, the
inclusion of Cotton-Mouton effect terms do not change qualitatively
the results presented in the main text. 

\subsection{Anisotropy \label{subsec:Cubic-anisotropy}}

Some crystal systems, including YIG, exhibit cubic symmetry for which
the magnetocrystalline anisotropy energy can be written as (for a
homogeneous material) \citep{chikazumi2009physics,cullity2008introduction,dionne2009anisotropy,landau1984electrodynamics}
\begin{align}
E_{c}= & K_{1}V\left[(\mathbf{s}\cdot\mathbf{e}_{1})^{2}(\mathbf{s}\cdot\mathbf{e}_{2})^{2}+(\mathbf{s}\cdot\mathbf{e}_{1})^{2}(\mathbf{s}\cdot\mathbf{e}_{3})^{2}\right.\nonumber \\
+ & \left.(\mathbf{s}\cdot\mathbf{e}_{2})^{2}(\mathbf{s}\cdot\mathbf{e}_{3})^{2}\right]+K_{2}V(\mathbf{s}\cdot\mathbf{e}_{1})^{2}(s\cdot\mathbf{e}_{2})^{2}(s\cdot\mathbf{e}_{3})^{2},\label{eq:cubic_anisotropy}
\end{align}
where $K_{1}$ and $K_{2}$ denote the respective first- and second-order
cubic anisotropy constants and $\mathbf{s}=\mathbf{S}/S$. As the
term $\propto K_{2}$ is of 6th order it can be safely neglected.
If we assume only small deviations of the spin $\ensuremath{\mathbf{s}}$
from one of its equilibrium directions, e.g. $\ensuremath{\mathbf{s}\cdot\mathbf{e}_{3}}\sim1$,
and keep the other terms up to second order, Eq.$~$\eqref{eq:cubic_anisotropy}
can be approximated as
\begin{equation}
\begin{aligned}E_{c}\approx & K_{1}V[(\mathbf{s}\cdot\mathbf{e}_{1})^{2}+(\mathbf{s}\cdot\mathbf{e}_{2})^{2}]\\
= & K_{1}V\vert\mathbf{s}\times\mathbf{e}_{3}\vert^{2},
\end{aligned}
\label{eq:cubic_anisotropy_approx}
\end{equation}
where we have used that $\mathbf{e}_{3}=\mathbf{e}_{1}\times\mathbf{e}_{2}$
and the vector triple product expansion. By expressing $\sin\phi=\vert\mathbf{s}\times\mathbf{e}_{3}\vert$
and $\cos\phi=\mathbf{s}\cdot\mathbf{e}_{3}$, $\phi$ being the angle
between $\ensuremath{\mathbf{s}}$ and $\mathbf{e}_{3}$, the cubic
anisotropy term Eq.$~$\eqref{eq:cubic_anisotropy_approx} can be rewritten
as 
\begin{equation}
E_{c}\approx K_{1}V\left[1-(\mathbf{s}\cdot\mathbf{e}_{3})^{2}\right],\label{eq:cubic_anisotropy_approx_final}
\end{equation}
which has the same form as the one for uniaxial symmetry apart from
a constant shift. Therefore, for small oscillations of the magnetization
around one of the directions which minimizes the anisotropy energy
density, cubic crystals can be treated to a good approximation as
crystals with uniaxial anisotropy, as is often considered for YIG
\citep{streib2018damping,elyasi2020resources,pacewicz2019rigorous,rijnierse1975optical}.
Furthermore, other magnetic oxides, such as those incorporating bismuth,
exhibit large uniaxial anisotropies \citep{hansen1985magnetic}.

In case of a non-spherical particle there is an additional source
of magnetic anisotropy due to its shape. The shape of the sample generates
a demagnetizing field which is not equal in all directions \citep{cullity2008introduction}.
The energy density of such geometric anisotropy is given by $\mu_{0}\bm{M}\tilde{N}\bm{M}/2$,
where $\tilde{N}$ is the demagnetization. For example, for a prolate
spheroid with semi-major axis along the $\mathbf{e}_{z}$ direction
one has $\tilde{N}={\rm diag}\left[N_{T},N_{T},1-2N_{T}\right]$ with
$N_{T}$ the demagnetization coefficient. A non-spherical shape entails
changes not only in the magnetic energy but also in the rotational
part of the Hamiltonian. On the one hand, it gives rise to a more
complex inertia tensor and on the other hand the electromagnetic fields
inside the rotating particle cannot be assumed to be independent of
the particle orientation. These effects can be considered to be small
if the shape does not deviate much from a sphere. 

\section{Conclusion\label{sec:Conclusion}}

We proposed an optical method to probe the coupled spin-mechanics
of levitated magnetic microparticles via optomagnonic effects. We
showed that the spin dynamics can be also driven by the optical drive,
inducing almost lossless angular oscillations through the spin-mechanics
coupling. This coupling between magnetization and angular motion can
be probed via the power spectrum of the cavity modes, which exhibit
two main resonance peaks at $\sim100\,$kHz and $\sim2\,$GHz for micrometer-sized particles and an applied magnetic field of $\sim60\,$mT. These are attributed to angular oscillations and spin dynamics
that are shifted by the magnetocrystalline anisotropy. We found that
due to the damping of the magnetization (Gilbert damping) a rotation
of the microparticle around the magnetic anisotropy direction can
be optically induced for an initially non-rotating particle. However,
the induced angular frequency is low ($\sim0.2\,\mu\mathrm{Hz}$). We showed
that this angular frequency can be increased by adding an oscillatory
rf magnetic field perpendicular to the bias magnetic field. In particular,
if the drive is at resonance with the shifted spin precession frequency,
the aforementioned effects are enhanced and the setup can be used
to induce fast rotations of the particle around its magnetic anisotropy
axis. This resonant condition can be identified by the appearance
of an additional peak in the spectrum. 

In general, the optical quality
factor of the particle and the supported driving power limit the driven
angular motion. In experiments the quality factor will be restricted
by additional sources, like surface roughness, and not only by the
considered radiative losses. In order to improve the quality factor
the system can be modified by placing the particle in an external
optical cavity. 

We have studied the classical dynamics of the levitated
magnetic particle coupled to light, in particular neglecting thermal
and shot noise. These, together with the effect of the trapping potential,
need to be included in order to study the dynamics in the quantum regime. Experimentally, levitated
magnetic particles can provide a unique platform to probe transduction
of angular momentum in the quantum regime, and eventually serve as
ultrasensitive torque sensors \citep{la2004optical,ahn2020ultrasensitive}. 

\section*{Acknowledgements}
%\medskip
%\textbf{Acknowledgments.} 
We acknowledge funding from the Max Planck Society and from the Deutsche Forschungsgemeinschaft (DFG, German Research Foundation) through Project-ID 429529648-TRR 306 QuCoLiMa ("Quantum Cooperativity of Light and Matter"). V.W. thanks K. Kustura, A. E. Rubio L\'opez and C. C. Rusconi for fruitful discussions.

%\bibliography{OpticalSignatures}
\input{OpticalSignatures.bbl}

\appendix

\section{Decay Rate and Q Factor\label{sec:DecayRate}}
In order to calculate the decay rate for the optical modes we need
to find the resonance location, i.e. the eigenmodes that are determined
by the roots of a characteristic equation. This characteristic equation
is obtained by satisfying the boundary conditions at the surface of
the microsphere, which implies that the components of the electromagnetic
fields inside and outside the sphere match. The resulting equation
can be reduced to \citep{buck2003optimal}
\begin{equation}
\frac{j_{l-1}(kR)}{j_{l}(kR)}-\frac{1}{n}\frac{h_{l-1}^{(1)}(kR/n)}{h_{l}^{(1)}(kR/n)}=0\label{eq:characteristicEq_TE}
\end{equation}
for TE modes and to
\begin{equation}
\frac{j_{l-1}(kR)}{j_{l}(kR)}-n\frac{h_{l-1}^{(1)}(kR/n)}{h_{l}^{(1)}(kR/n)}+\frac{n^{2}l}{kR}-\frac{l}{kR}=0\label{eq:characteristicEq_TM}
\end{equation}
for TM modes, where $j_{l}(x)$ and $h_{l}^{(1)}(x)$ are the spherical
Bessel and Hankel function of first kind and $k=2\pi n/\lambda_{0}$
is the wave vector inside the sphere with refractive index $n$ and
vacuum laser wavelength $\lambda_{0}$. We determine the relevant
angular mode number by $l=2\pi nR/\lambda_{0}$. From the complex
roots $k_{\rho}$ we then obtain the resonance frequency $\omega_{r}=\mathrm{Re}\{k_{\rho}c/(nR)\}$
and the decay rate $\kappa=|\mathrm{Im}\{k_{\rho}c/(nR)\}|$. For
large mode numbers $l$ an approximate analytic expression for the
complex resonance frequency was derived \citep{datsyuk1992some,weinstein1969open} 

\begin{equation}
\omega_{\rho}=\dfrac{c}{nR}\left[l+\dfrac{1}{2}-(t_{p}^{0}+\Delta t_{l})\xi\right],\label{eq:resonance_freq_approx}
\end{equation}
where
\[
\Delta t_{l}=n^{1-2b}\dfrac{1+\mathrm{i}\mathrm{e^{-2T_{l}}}}{\xi\sqrt{n^{2}-1}},
\]
\[
T_{l}=\left(l+\dfrac{1}{2}\right)(\eta_{l}-\mathrm{tanh}(\eta_{l})),
\]
\[
\eta_{l}=\mathrm{arccosh}\left[n\left[1-\dfrac{1}{l+\dfrac{1}{2}}\left(t_{p}^{0}\xi+\dfrac{l^{1-2b}}{\sqrt{l^{2}-1}}\right)\right]^{-1}\right],
\]
\[
\xi=\left[\dfrac{1}{2}\left(l+\dfrac{1}{2}\right)\right]^{1/3},
\]
and
\[
b=\begin{cases}
0 & \mathrm{TE\;modes}\\
1 & \mathrm{TM\;modes.}
\end{cases}
\]
Also, $t_{p}^{0}$ denotes the $p$th zero of the Airy function, where
$p$ corresponds to the radial mode number. Note that we only consider
$p=1$ modes \citep{buck2003optimal} and focus on TE modes for our calculations. 

\begin{figure}[t]
\includegraphics[width=1\columnwidth]{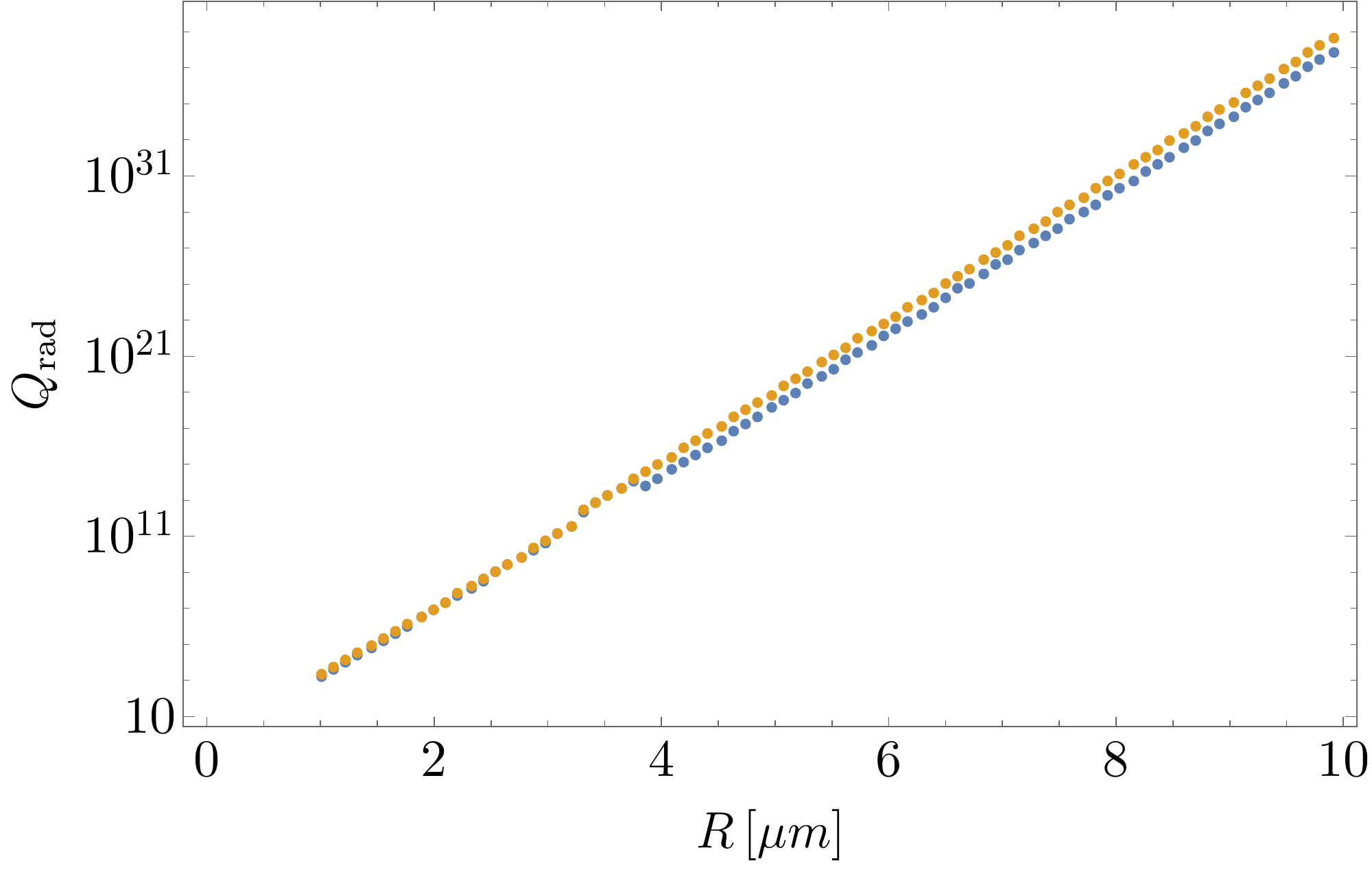}
\caption{Semi-log plot of the radiative Q factor as a function of the size
of the particle $R$ for TE modes (for $\lambda_{0}=1500\,\mathrm{nm}$
and $n=2.2)$. The blue dots depict the solutions obtained from numerically solving the characteristic equation Eq.$~$\eqref{eq:characteristicEq_TM},
whereas the orange ones represent the approximated Q factor Eq.$~$\eqref{eq:q_factor_approx}.
}
\label{fig:q_factor}
\end{figure}

The Q factor due to radiative loss can be obtained through the characteristic
equation as 
\begin{equation}
Q_{\mathrm{rad}}=\dfrac{\mathrm{Re}\{k_{\rho}\}}{2|\mathrm{Im}\{k_{\rho}\}|}.\label{eq:Q}
\end{equation}
Using Eq.$~$\eqref{eq:resonance_freq_approx} to calculate the Q
factor leads to 
\begin{equation}
Q_{\mathrm{rad}}^{\mathrm{approx}}=\dfrac{1}{2}\left(l+\dfrac{1}{2}\right)n^{-(1-2b)}(n^{2}-1)^{1/2}\mathrm{e}^{2T_{l}},\label{eq:q_factor_approx}
\end{equation}
where we used $|(t_{p}^{0}+\Delta t_{l})\xi|\ll l+\dfrac{1}{2}$ \citep{datsyuk1992some}.
In Fig.$~$\ref{fig:q_factor} we show the dependence of the Q factor
on the radius of the particle. As the magnitude of the exact and the
approximated solutions match, we use Eq.$~$\eqref{eq:resonance_freq_approx}
to obtain the values for $\omega_{r}$ and $\kappa_{\mathrm{rad}}$
in our simulations of the trajectories, namely, 
\begin{equation}
\omega_{r}=\dfrac{c}{nR}\left(l+\dfrac{1}{2}\right)\label{eq:omega_r}
\end{equation}
and 
\begin{equation}
\kappa_{\mathrm{rad}}=\dfrac{c}{nR}\left[\dfrac{n^{1-2b}}{\sqrt{n^{2}-1}}\mathrm{e^{-2T_{l}}}\right].\label{eq:kappa}
\end{equation}

\section{Transition Amplitude\label{sec:Transition-Amplitude}}
For one magnon absorption the transition amplitude in Eq.$~$\eqref{eq:g}
can be approximated as \citep{almpanis2020spherical}

\begin{figure}[t]
\includegraphics[width=1\columnwidth]{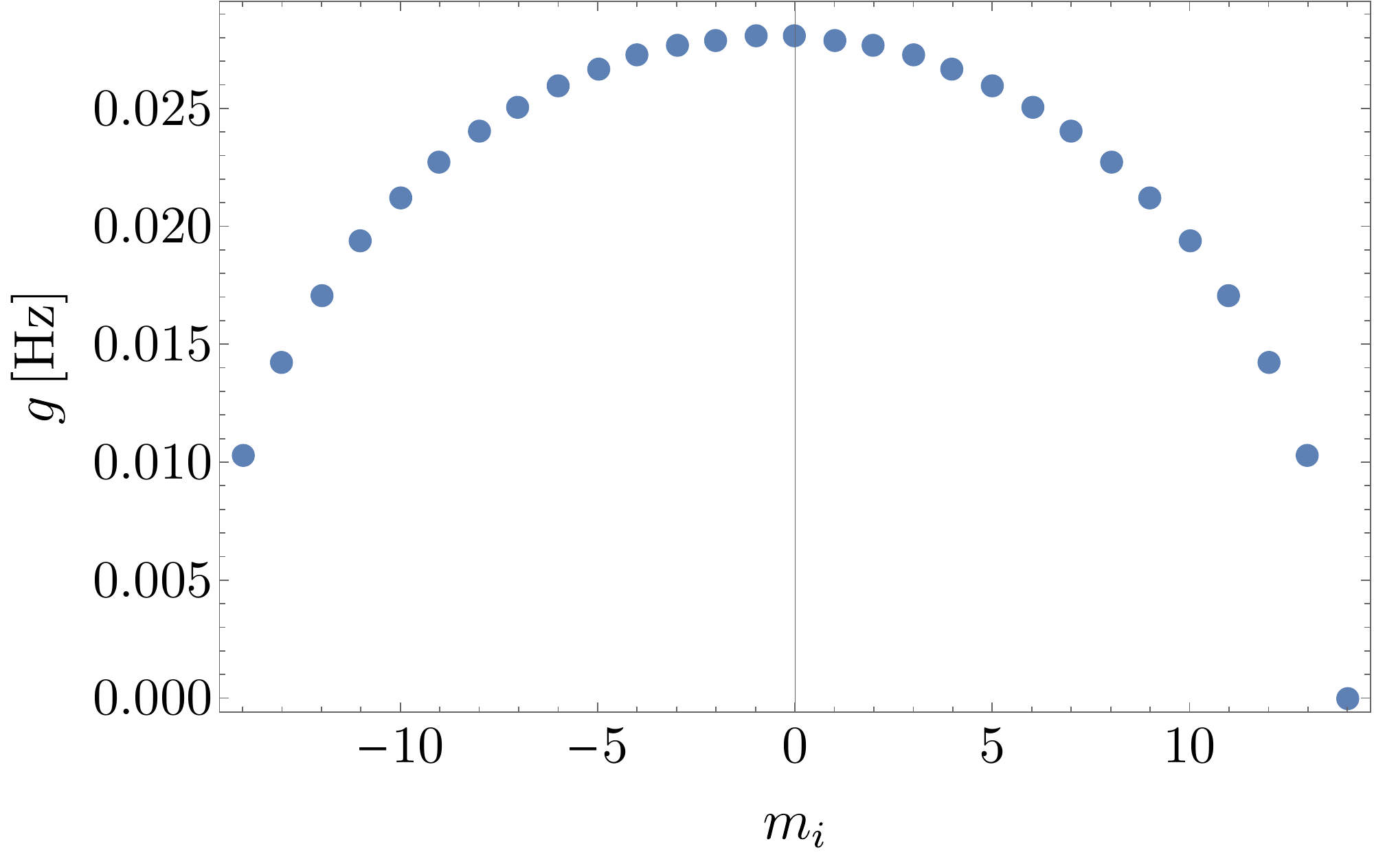}
\caption{Optomagnonic coupling strength $g$ for triple-resonant photon transition
from mode $m_{\mathrm{i}}$ to $m_{\mathrm{i}}+1$ for a sphere of
radius $R=1.5\,\mu\mathrm{m}$. }
\label{fig:g_m}
\end{figure}

\begin{widetext}
\begin{align}
\tilde{g}= & \dfrac{-\hbar\omega_{r}\sqrt{(l-m_{\mathrm{i}})(l+m_{\mathrm{i}}+1)}}{\varepsilon_{0}\varepsilon l(l+1)}\nonumber \\
 & \left[\dfrac{(2l+1)^{2}[j_{l}^{2}(x)-j_{l-1}(x)j_{l+1}(x)]}{(2l+1)^{2}j_{l}^{2}(x)-4l(l+1)j_{l-1}(x)j_{l+1}(x)+(l+1)j_{l-1}^{2}(x)-lj_{l+1}^{2}(x)}\right]_{x=k_{r}R},\label{eq:transitionAmplitude-1}
\end{align}
\label{eq:g_tilde-1}
\end{widetext}with $k_{r}=\omega_{r}n/c$. Note that the absolute
value of the transition amplitude for one magnon emission from mode
$m_{\mathrm{i}}+1$ to $m_{\mathrm{i}}$ is the same as the one for
absorption from $m_{\mathrm{i}}$ to $m_{\mathrm{i}}+1$. We consider
TE modes and choose $m_{\mathrm{i}}=0$ such that the coupling
strength is maximized, which can be seen in Fig.$~$\ref{fig:g_m}
where the optomagnonic coupling is depicted as a function of $m_{\mathrm{i}}.$

\section{Commutation Relations\label{sec:Commutation-Relations}}

The rotation matrix is given by
\begin{align}
R(\hat{\Omega}) & =\begin{pmatrix}\cos\hat{\gamma} & \sin\hat{\gamma} & 0\\
-\sin\hat{\gamma} & \cos\hat{\gamma} & 0\\
0 & 0 & 1
\end{pmatrix}\begin{pmatrix}\cos\hat{\beta} & 0 & -\sin\hat{\beta}\\
0 & 1 & 0\\
\sin\hat{\beta} & 0 & \cos\hat{\beta}
\end{pmatrix}\label{eq:rotationMatrix-1}\\
 & \begin{pmatrix}\cos\hat{\alpha} & \sin\hat{\alpha} & 0\\
-\sin\hat{\alpha} & \cos\hat{\alpha} & 0\\
0 & 0 & 1
\end{pmatrix}\nonumber 
\end{align}
and its elements commute with the spin operator. The anisotropy axis
in the laboratory frame reads
\begin{align}
\mathbf{e}_{3}(\hat{\Omega}) & =R_{31}(\hat{\Omega})\mathbf{e}_{x}+R_{32}(\hat{\Omega})\mathbf{e}_{y}+R_{33}(\hat{\Omega})\mathbf{e}_{z}.\label{eq:e3-1-1}
\end{align}
From the Hamiltonian in Eq.$~$\eqref{eq:hamiltonian_reduced} we
obtain the coupled Heisenberg equations of motion for the set of operators
$\hat{\boldsymbol{\xi}}=(\hat{\mathbf{e}}_{3},\hat{\mathbf{L}},\hat{\mathbf{S}},\hat{a}_{1},\hat{a}_{1}^{\dagger},\hat{a}_{2},\hat{a}_{2}^{\dagger})$
by using the following commutation relations \citep{edmonds1996angular,rusconiMagneticRigidRotor2016}
\begin{align*}
\left[\hat{L}_{j},\hat{L}_{k}\right] & =i\epsilon_{jkl}\hat{L}_{l},\\
\left[\hat{S}_{j},\hat{S}_{k}\right] & =i\epsilon_{jkl}\hat{S}_{l},\\
\left[\hat{L}_{j},R_{kl}(\hat{\Omega})\right] & =i\epsilon_{jlr}R_{kr}(\hat{\Omega}),
\end{align*}
and $\left[\hat{L}_{j},\hat{S}_{k}\right]=\left[\hat{S}_{j},R_{kl}(\hat{\Omega})\right]=\left[R_{jk}(\hat{\Omega}),R_{lr}(\hat{\Omega})\right]=0$.
For the cavity operators the usual bosonic commutation relations hold,
i.e. $[\hat{a}_{i},\hat{a}_{j}^{\dagger}]=\delta_{ij},$ $[\hat{a}_{i},\hat{a}_{i}]=[\hat{a}_{i}^{\dagger},\hat{a}_{i}^{\dagger}]=0$.
Further, we focus on the classical limit in which the operators are
replaced by their expectation values and we neglect any kind of quantum
correlations, i.e. $\langle\hat{\xi}_{i}\hat{\xi}_{j}\rangle\simeq\langle\hat{\xi}_{i}\rangle\langle\hat{\xi}_{j}\rangle$
$\forall\,i,j$. The set of Heisenberg equations of motion $\dfrac{\mathrm{d}\hat{\boldsymbol{\xi}}}{\mathrm{d}t}=\boldsymbol{\Xi}(\hat{\boldsymbol{\xi}})$
is therefore approximated by the closed set of semiclassical equations
$\dfrac{\mathrm{d}\langle\hat{\boldsymbol{\xi}}\rangle}{\mathrm{d}t}=\boldsymbol{\Xi}(\langle\hat{\boldsymbol{\xi}}\rangle)$,
where $\boldsymbol{\Xi}$ is a vector function of $\hat{\boldsymbol{\xi}}$. 

\section{Derivation of the Optomagnonic Steady State }

\label{sec:SteadyState}

In what follows we drop the bar-notation for the steady state. For the steady state of the optical field we need to solve:
\begin{align}
-\mathrm{i}g(S_{+}a_{1}^{\dagger}a_{2}-S_{-}a_{2}^{\dagger}a_{1}) & =0\nonumber \\
\frac{\omega_B}{2}S_{+}+gS_{z}a_{2}^{*}a_{1} & =0\nonumber \\
\mathrm{i}\Delta_{1}a_{1}-\mathrm{i}gS_{+}a_{2}+\epsilon_{L1}-\dfrac{1}{2}\kappa a_{1} & =0\nonumber \\
\mathrm{i}\Delta_{2}a_{2}-\mathrm{i}gS_{-}a_{1}+\epsilon_{L2}-\dfrac{1}{2}\kappa a_{2} & =0\label{eq:steadystate_detail}
\end{align}
 Note, that here we describe the equations for $\mathbf{S}$ in terms
of the ladder operators, such that $\dot{S}_{z}=-\mathrm{i}g(S_{+}a_{1}^{\dagger}a_{2}-S_{-}a_{2}^{\dagger}a_{1})-\dot{L}_{z}$
and $\dot{S}_{+}=-\mathrm{i}\omega_B S_{+}-2\mathrm{i}gS_{z}a_{2}^{\dagger}a_{1}-\dot{L}_{+}$.
From the second equation of Eq.$~$\eqref{eq:steadystate_detail} we get
\begin{align*}
S_{+}=- & \frac{2g}{\omega_B}a_{2}^{*}a_{1}S_{z}.
\end{align*}
As $S_{-}S_{+}=S_{x}^{2}+S_{y}^{2}=S^{2}-S_{z}^{2}$, we multiply
the above equation by $S_{-}=S_{+}^{*}$, such that $S^{2}-S_{z}^{2}=\frac{4g^{2}}{\omega_B^2}\vert a_{1}\vert^{2}\vert a_{2}\vert^{2}S_{z}^{2}$, which results in
\begin{equation}
S_{z}=\frac{S}{\sqrt{1+\frac{4g^{2}}{\omega_B^{2}}\vert a_{1}\vert^{2}\vert a_{2}\vert^{2}}}\label{eq:Sz_steady}
\end{equation}
and
\begin{equation}
S_{+}=-\frac{2ga_{2}^{*}a_{1}S}{\sqrt{\omega_B^{2}+4g^{2}\vert a_{1}\vert^{2}\vert a_{2}\vert^{2}}}.
\label{eq:S+_steady}
\end{equation}
This allows us to rewrite the last two equations in Eq.$~$\eqref{eq:steadystate_detail}
as
\begin{align}
\mathrm{i}\Delta_{1}a_{1}+\mathrm{i}\frac{2g^{2}\vert a_{2}\vert^{2}a_{1}S}{\sqrt{\omega_B^{2}+4g^{2}\vert a_{1}\vert^{2}\vert a_{2}\vert^{2}}}+\epsilon_{L1}-\dfrac{1}{2}\kappa a_{1}= & 0\nonumber\\
\mathrm{i}\Delta_{2}a_{2}+\mathrm{i}\frac{2g^{2}a_{2}\vert a_{1}\vert^{2}S}{\sqrt{\omega_B^{2}+4g^{2}\vert a_{1}\vert^{2}\vert a_{2}\vert^{2}}}+\epsilon_{L2}-\dfrac{1}{2}\kappa a_{2}= & 0, \label{eq:steadustate_optical}
\end{align}
which is a set of coupled equations that involves only the cavity
fields, although being nonlinear. If for instance $\epsilon_{L2}=0$,
then the second equation reads
\[
\mathrm{i}\Delta_{2}a_{2}+\mathrm{i}\frac{2g^{2}a_{2}\vert a_{1}\vert^{2}S}{\sqrt{\omega_B^{2}+4g^{2}\vert a_{1}\vert^{2}\vert a_{2}\vert^{2}}}-\dfrac{1}{2}\kappa a_{2}=0
\]
and multiplying it by $a_{2}^{*}$ results in
\[
\mathrm{i}\Delta_{2}\vert a_{2}\vert^{2}+\mathrm{i}\frac{2g^{2}\vert a_{2}\vert^{2}\vert a_{1}\vert^{2}S}{\sqrt{\omega_B^{2}+4g^{2}\vert a_{1}\vert^{2}\vert a_{2}\vert^{2}}}=\frac{1}{2}\kappa\vert a_{2}\vert^{2}.
\]
Note that the LHS is purely imaginary while the RHS is purely real,
which implies that the only possible solution for this equation is $\vert a_{2}\vert^{2}=0$. From this it follows in
turn that $a_{1}=-\frac{\epsilon_{L1}}{i\Delta_{1}-\frac{1}{2}\kappa}$,
$S_{z}=S$, $S_{+}=S_{-}=S_{x}=S_{y}=0$. In order to get a spin steady state that is affected by light it is necessary to also pump mode 2, i.e.
$\epsilon_{L2}\neq0$.

In the limit $\omega_B\gg4g\vert a_{1}\vert\vert a_{2}\vert$, Eq.$~$\eqref{eq:steadustate_optical} reduces
to
\begin{align}
\mathrm{i}\Delta_{1}a_{1}+\mathrm{i}\frac{2g^{2}\vert a_{2}\vert^{2}a_{1}S}{\omega_B}+\epsilon_{L1}-\dfrac{1}{2}\kappa a_{1}= & 0\nonumber\\
\mathrm{i}\Delta_{2}a_{2}+\mathrm{i}\frac{2g^{2}a_{2}\vert a_{1}\vert^{2}S}{\omega_B}+\epsilon_{L2}-\dfrac{1}{2}\kappa a_{2}= & 0,\label{eq:steadystate_optical_approx}
\end{align}
such that
\begin{equation}
a_{1}=-\frac{\epsilon_{L1}}{\mathrm{i}\left(\Delta_{1}+\frac{2g^{2}\vert a_{2}\vert^{2}S}{\omega_B}\right)-\frac{1}{2}\kappa}\label{eq:appendix_a1_st}
\end{equation}
and
\begin{align}
\mathrm{i}\Delta_{2}a_{2}+\mathrm{i}\frac{2g^{2}S}{\gamma B}\frac{\epsilon_{L1}^{2}}{\left(\Delta_{1}+\frac{2g^{2}\vert a_{2}\vert^{2}S}{\omega_B}\right)^{2}+\frac{1}{4}\kappa^{2}}a_{2}\nonumber\\
-\frac{1}{2}\kappa a_{2}+\epsilon_{L2}=0.
\label{eq:appendix_a2_st}
\end{align}

\section{Further Dynamics\label{sec:Further-Dynamics}}

\begin{figure}[t]
\includegraphics[width=1\columnwidth]{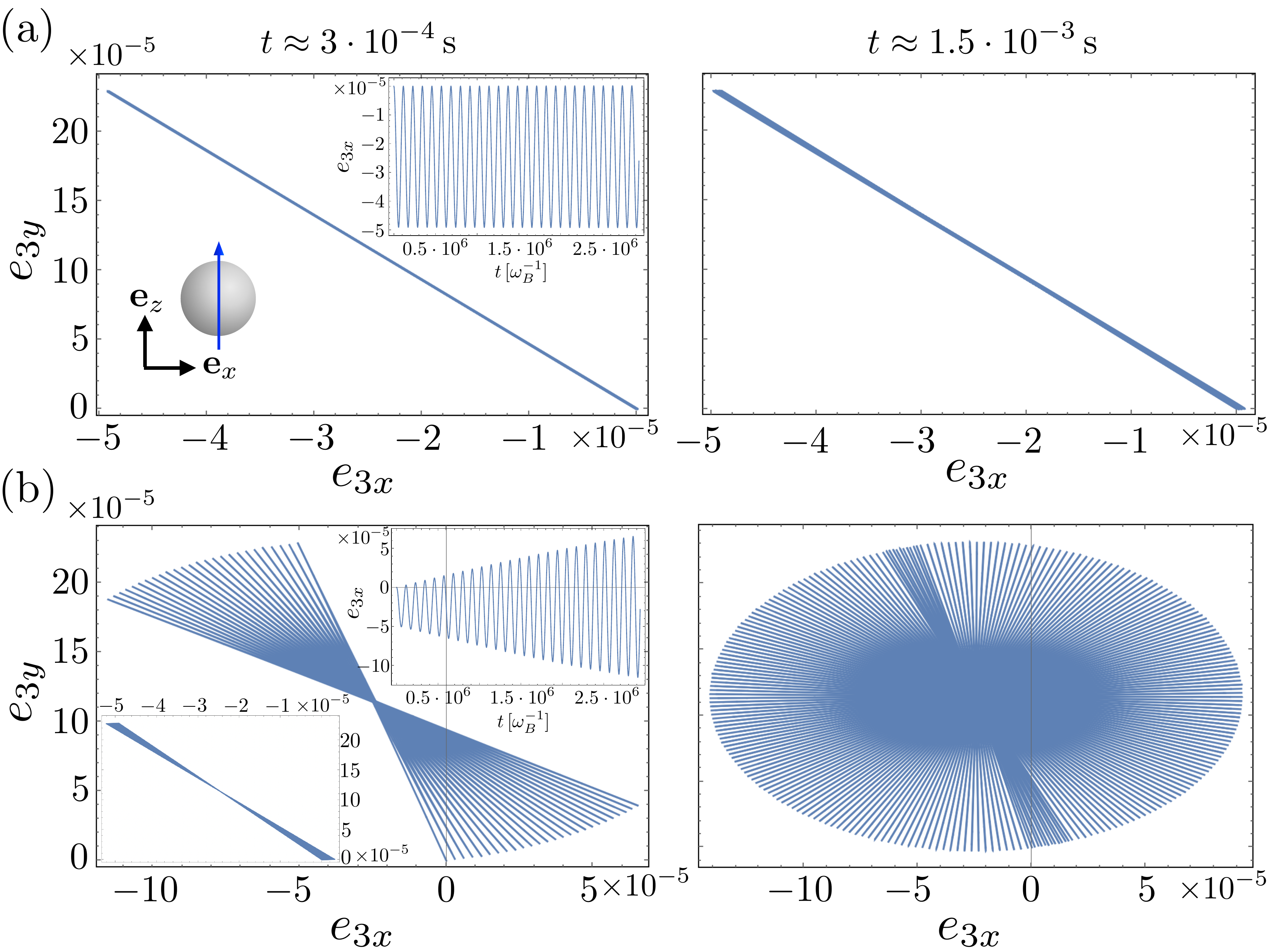}
\caption{Dynamics of $\mathbf{e}_{3}$ in the $xy$ plane for a sphere of size
$R=1.5\,\mu\mathrm{m}$ and a red detuned system with $P_{2}=0.02\,\mathrm{W}$
and $P_{1}=P_{2}/2$. The initial configuration is such that $\mathbf{S}\parallel\mathbf{e_{3}}\parallel\mathbf{e}_{z}$
and $\mathbf{L}(0)=\bm{0}$ and (b)$L_{z}=49/S$ which corresponds
to an initial rotation frequency of $\omega_{\mathrm{rot}}/(2\pi)\approx750\,\mathrm{Hz}$.
The lower inset in (b) shows the dynamics corresponding to $\omega_{\mathrm{rot}}/(2\pi)\approx30\,\mathrm{Hz}$.}
\label{fig:e3_xy}
\end{figure}

\begin{figure}[t]
\includegraphics[width=1\columnwidth]{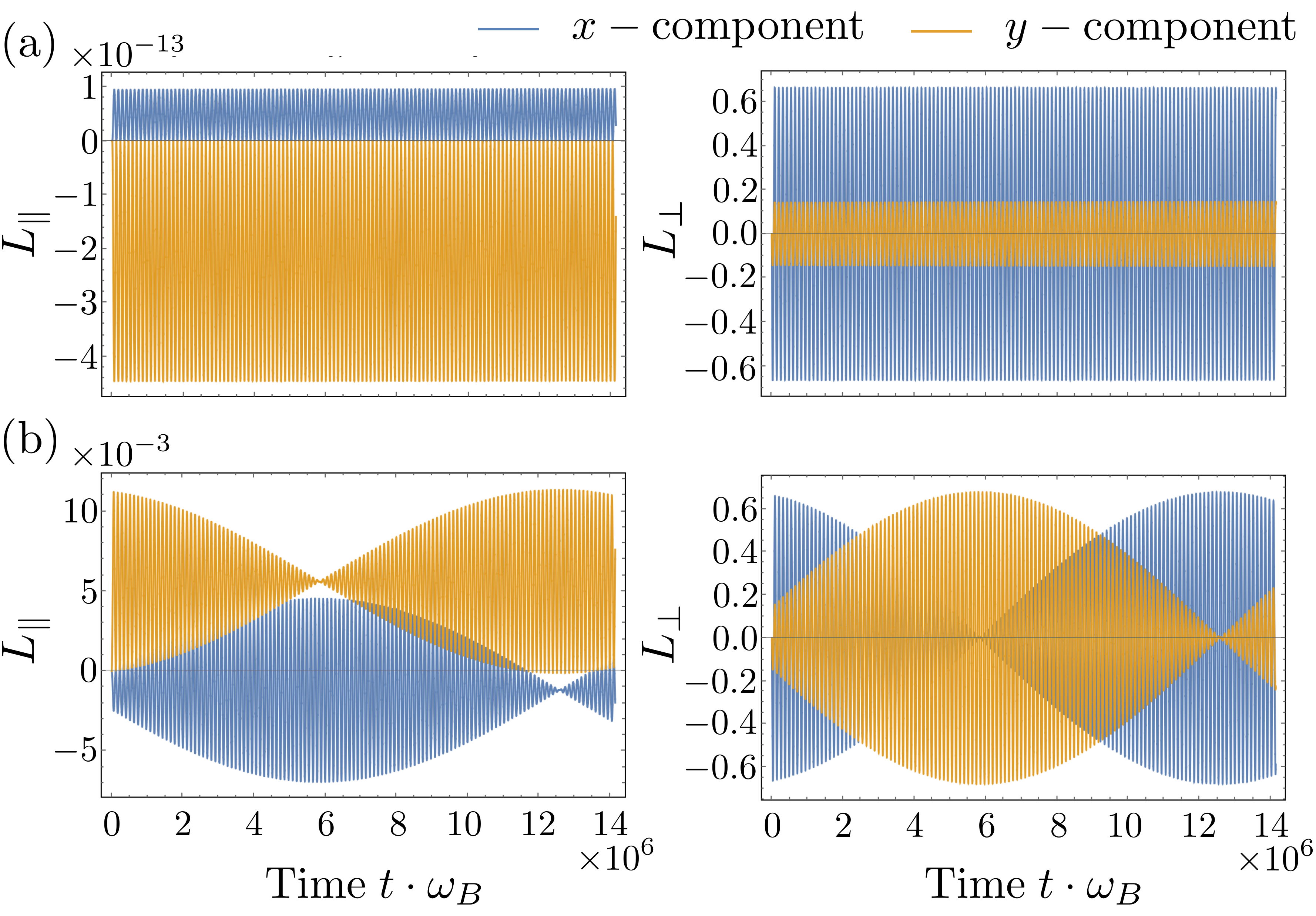}
\caption{Dynamics of the components of the angular momentum parallel $\mathbf{L}_{\parallel}$
and perpendicular $\mathbf{L}_{\perp}$ to $\mathbf{e}_{3}$ for a
sphere of size $R=1.5\,\mu\mathrm{m}$ and a red detuned system with
$P_{2}=0.02\,\mathrm{W}$ and $P_{1}=P_{2}/2$. The initial configuration
is such that $\mathbf{S}\parallel\mathbf{e_{3}}\parallel\mathbf{e}_{z}$
and (a) $\mathbf{L}(0)=\bm{0}$ and (b) $L_{z}=49/S$ which corresponds
to an initial rotation frequency of $\omega_{\mathrm{rot}}/(2\pi)\approx750\,\mathrm{Hz}$. }
\label{fig:angular_motion}
\end{figure}

If the particle is initially unmagnetized and non-rotating, then 
switching on the magnetic field induces a rotational motion,
as a change of magnetization results in mechanical rotation. This is due to conservation of angular momentum, as stated by the Einstein-de Haas
effect \citep{einstein_de_haas}. The relation between magnetization
and mechanical angular momentum is given by 
\begin{equation}
\Delta MV=-\gamma\Delta L.\label{eq:EdH}
\end{equation}
In our case $\Delta M=M_{\mathrm{S}}$, thus, the resulting rotation
frequency is $\omega_{\mathrm{rot}}/(2\pi)\approx30\,\mathrm{Hz}$.
In Fig.$~$\ref{fig:e3_xy} we show the evolution of $\mathbf{e}_{3}$
in the $xy$ plane in the red detuning scheme for $\mathbf{e}_{3}(0)=\mathbf{S}(0)/S=\mathbf{e}_{z}$
and (a) $\mathbf{L}(0)=\bm{0}$, (b) $\mathbf{L}(0)\neq\bm{0}$. It
can be seen that in the case of no initial angular momentum $\mathbf{e}_{3}$
mainly undergoes a librational motion with a very slow precession.
If the sphere rotates initially, then the precessional motion is amplified
and it becomes recognizable that the trajectory is not closed. We
want to comment that we chose a higher initial rotation frequency
than the one due to the Einstein-de Haas effect in order to make the
influence more visible. Higher laser powers do not change the motion
in the $xy$ plane qualitatively, but increase the amplitude of the
oscillations. Further we observed that in this case there is a modulation
of the components of the angular momentum with the initial angular
speed, which can be seen in Fig.$~$\ref{fig:angular_motion}. The
evolution of the absolute values, however, behaves the same as in
the case of $\mathbf{L}(t=0)=\bm{0}$, i.e. the results are the same
as in Fig.$~$\ref{fig:a+e3}(c,d), except that $|\mathbf{L}_{\parallel}|$
takes the constant value $L_{z}(t=0).$ 

\begin{figure}[t]
\includegraphics[width=1\columnwidth]{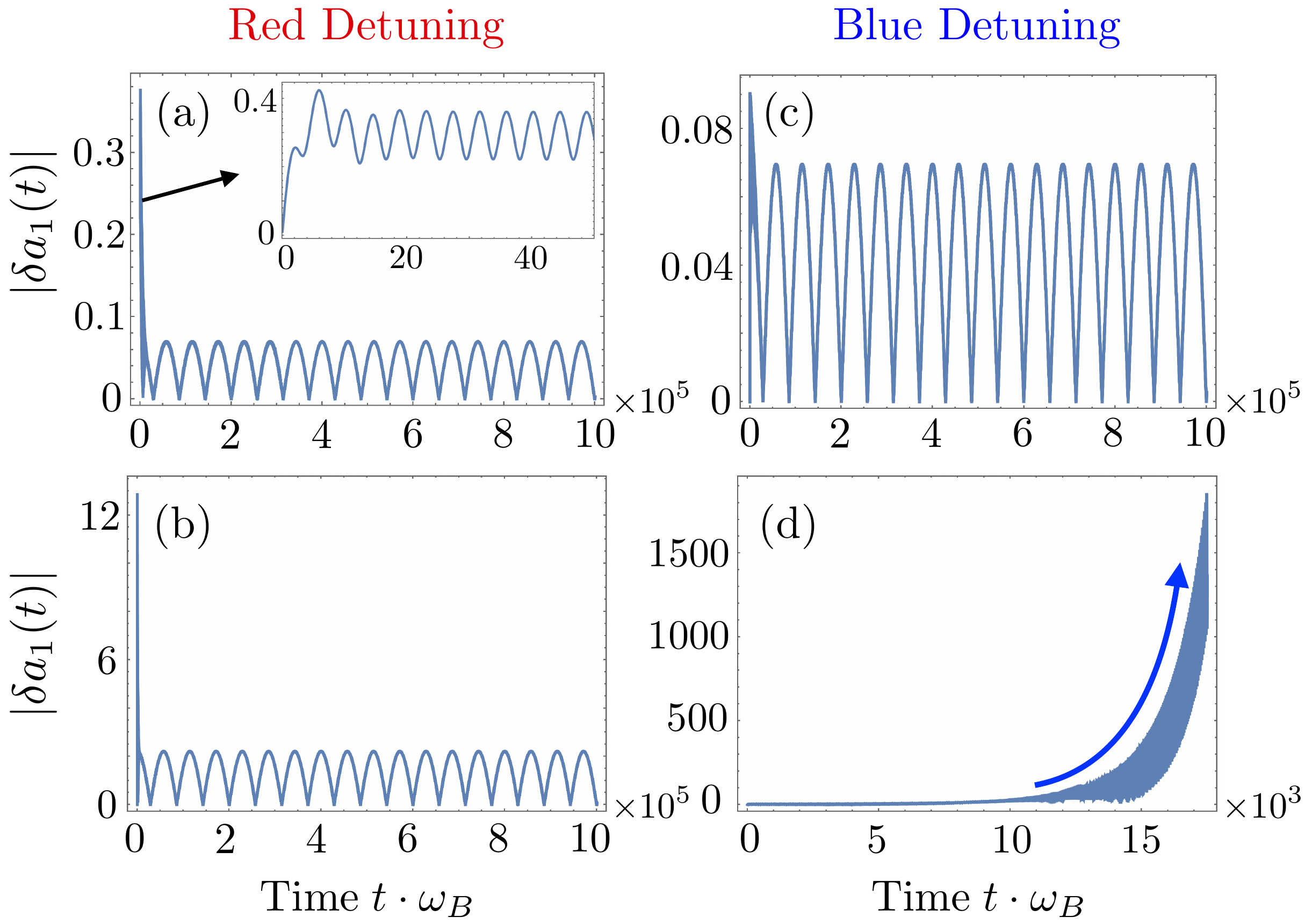}
\caption{Dynamics of the cavity mode $1$ for a sphere of size $R=1.5\,\mu\mathrm{m}$.
The plots are for both red and blue detuning schemes, and for (a,c)
$P_{2}=0.02\,\mathrm{W}$, and (b,d) $P_{2}=0.2\,\mathrm{W}$. For
both cases $P_{1}=P_{2}/2$. The initial configuration is such that
$\mathbf{S}\parallel\mathbf{e_{3}}\parallel\mathbf{e}_{z}$. The inset
of (a) shows the dynamics in the initial times evolution. After a
transient period, the cavity dynamics is modulated by the dynamics
of the anisotropy axis. The blue detuning scheme can lead to instabilities
as shown in (d).}
\label{fig:CavityDynamics01-1}
\end{figure}

The cavity field dynamics is modulated by the angular motion as we
show in Fig.$~$\ref{fig:CavityDynamics01-1} for both blue and red
detuning for two different sets of laser powers. The figure depicts
the time evolution of $\vert\delta a_{1}(t)\vert=\vert\bar{a}_{1}-a_{1}(t)\vert$,
i.e. the deviations of the higher frequency mode from its steady state.
At short times, the dominant contributions are due to the optomagnonic
coupling and the fast spin dynamics. We notice that the blue detuning
pumping scheme leads to instabilities in the system and an exponential
generation of photons that, for the considered parameters, is not
compensated by decay.

\section{Resonance Peak\label{sec:Resonance-Peak}}

In order to obtain the frequencies of the peaks in the power spectra
analytically, we derive the eigenfrequencies of the system by linearizing
the equations of motion Eq.$~$\eqref{eq:eom} (here $\mathbf{l}=\mathbf{L}/S$,
$\mathbf{s}=\mathbf{S}/S$) around the steady state and consider small
oscillations around it, 
\begin{align*}
\boldsymbol{\xi}(t) & =\bar{\boldsymbol{\xi}}+\delta\boldsymbol{\xi}(t).
\end{align*}
 To simplify the calculations, we assume that the steady state for
the spin and the anisotropy axis is sufficiently close to the north
pole, i.e. $\bar{s}_{+}=\bar{e}_{3+}\approx0$ and $\bar{s}_{z}=\bar{e}_{3z}\approx1$.
Further, we take a non-rotating sphere, $\mathbf{\bar{l}}=\mathbf{0}$, and
denote the steady states of the optical field as $\alpha_{1,2}.$
In what follows we adopt the shorthand notation $\delta\xi\equiv\xi$
and $e_{3+}=e_{3x}+\mathrm{i}e_{3y}$, $l_{+}=l_{x}+\mathrm{i}l_{y}$
. Under these assumptions, we can write the linearized equations of
motion as 
\begin{align}
\dot{e}_{3+} & =-\mathrm{i}\omega_{I}Sl_{+},\nonumber \\
\dot{l}_{+} & =2\mathrm{i}\omega_{D}S(s_{+}-e_{3+}),\nonumber \\
\dot{s}_{+} & =-i\tilde{\omega}_{B}s_{+}-i\bar{B}_{+}s_{z}-\mathrm{i}\delta B_{+}+2\mathrm{i}\omega_{D}Se_{3+},\nonumber \\
\dot{s}_{z} & =-\frac{\mathrm{i}}{2}\bar{B}_{+}^{\ast}s_{+}+\frac{\mathrm{i}}{2}\bar{B}_{+}s_{+}^{*}\nonumber \\
\dot{a}_{1} & =\left(i\Delta_{1}-\frac{\kappa}{2}\right)a_{1}-igS\alpha_{2}s_{+}+\delta\varepsilon_{1},\nonumber \\
\dot{a}_{2} & =\left(i\Delta_{2}-\frac{\kappa}{2}\right)a_{2}-igS\alpha_{1}s_{+}^{*}+\delta\varepsilon_{2},\label{eq:eom_linearized}
\end{align}
where we introduced $\tilde{\omega}_{B}=\omega_{B}+2\omega_{D}S$
and $B_{+}=2ga_{1}a_{2}^{\ast}$. Moving to frequency space via $\xi(\omega)=\int_{-\infty}^{\infty}e^{i\omega t}\xi(t)\mathrm{d}t$,
$\xi^{*}(\omega)=\int_{-\infty}^{\infty}e^{i\omega t}\xi^{*}(t)\mathrm{d}t$
leads to a set of algebraic equations
\begin{align}
-\mathrm{i}\omega e_{3+}(\omega) & =-\mathrm{i}\omega_{I}Sl_{+}(\omega),\nonumber \\
-\mathrm{i}\omega l_{+}(\omega) & =2\mathrm{i}\omega_{D}S(s_{+}(\omega)-e_{3+}(\omega)),\nonumber \\
-\mathrm{i}\omega s_{+}(\omega) & =-\mathrm{i}\tilde{\omega}_{B}s_{+}(\omega)-\mathrm{i}\bar{B}_{+}s_{z}(\omega)\nonumber\\&-2\mathrm{i}g\left(\alpha_{2}^{*}a_{1}(\omega)+\alpha_{1}a_{2}^{*}(\omega)\right)+2\mathrm{i}\omega_{D}Se_{3+}(\omega),\nonumber \\
\mathrm{-i}\omega s_{z}(\omega) & =-\frac{\mathrm{i}}{2}\bar{B}_{+}^{\ast}s_{+}(\omega)+\frac{\mathrm{i}}{2}\bar{B}_{+}s_{+}^{*}(\omega),\nonumber \\
-\mathrm{i}\omega a_{1}(\omega) & =\left(i\Delta_{1}-\frac{\kappa}{2}\right)a_{1}(\omega)-igS\alpha_{2}s_{+}(\omega)+\delta\varepsilon_{1}(\omega),\nonumber \\
-\mathrm{i}\omega a_{2}(\omega) & =\left(i\Delta_{2}-\frac{\kappa}{2}\right)a_{2}(\omega)-igS\alpha_{1}s_{+}^{\ast}(\omega)+\delta\varepsilon_{2}(\omega),\label{eq:eom_algebraic}
\end{align}
which we want to solve for $a_{1}(\omega)$. Rearranging and substituting
the equations for $e_{3+}(\omega)$ and $l_{+}(\omega)$ into each
other yields
\[
e_{3+}(\omega)=\dfrac{2\omega_{D}\omega_{I}S^{2}}{2\omega_{D}\omega_{I}S^{2}-\omega^{2}}s_{+}(\omega).
\]

Inserting this expression and the equation for $s_{z}(\omega)$ into
the one for $s_{+}(\omega)$ gives 
\begin{align}
-&\mathrm{i}\left(\omega-\tilde{\omega}_{B}-\frac{|\bar{B}_{+}|^{2}}{2\omega}+\dfrac{4\omega_{D}^{2}\omega_{I}S^{3}}{2\omega_{D}\omega_{I}S^{2}-\omega^{2}}\right)s_{+}(\omega)\nonumber\\&=\frac{\mathrm{i}}{2\omega}\bar{B}_{+}{}^{2}s_{+}^{*}(\omega)-2\mathrm{i}g(\alpha_{2}^{*}a_{1}(\omega)+\alpha_{1}a_{2}^{*}(\omega)).\label{eq:s_plus_unarranged}
\end{align}
For now we neglect the term $\propto s_{+}^{*}(\omega)$ and by defining
\begin{equation}
\chi_{+}(\omega)=\left[\omega-\tilde{\omega}_{B}-\frac{|\bar{B}_{+}|^{2}}{2\omega}+\dfrac{4\omega_{D}^{2}\omega_{I}S^{3}}{2\omega_{D}\omega_{I}S^{2}-\omega^{2}}\right]^{-1},\label{eq:xi_plus}
\end{equation}
we get 
\begin{equation}
s_{+}(\omega)=2g\alpha_{2}^{*}\chi_{+}(\omega)a_{1}(\omega)+2g\alpha_{1}\chi_{+}(\omega)a_{2}^{*}(\omega).\label{eq:s_plus}
\end{equation}
Substituting this expression in the equation for $a_{1}(\omega)$
and rearranging it yields
\begin{align}
&\left[-\mathrm{i}\left(\omega+\Delta_{1}-2g^{2}S\vert\alpha_{2}\vert^{2}\chi_{+}(\omega)\right)+\frac{\kappa}{2}\right]a_{1}(\omega)=\nonumber\\&\qquad\qquad-2\mathrm{i}g^{2}S\alpha_{2}\alpha_{1}\chi_{+}(\omega)a_{2}^{*}(\omega)+\delta\varepsilon_{1}(\omega),
\end{align}
where again, for now, we do not consider the term $\propto a_{2}^{*}(\omega)$,
such that
\begin{equation}
\left[-\mathrm{i}\left(\omega+\Delta_{1}-2Sg^{2}\vert\alpha_{2}\vert^{2}\chi_{+}(\omega)\right)+\frac{\kappa}{2}\right]a_{1}(\omega)=\delta\varepsilon_{1}(\omega).\label{eq:a1}
\end{equation}
The peaks of the power spectrum are then given by the zeroes of the
imaginary part of the function 
\begin{align}
\chi_{1}^{-1}(\omega)  =&-\mathrm{i}\left(\omega+\Delta_{1}-2Sg^{2}\vert\alpha_{2}\vert^{2}\chi_{+}(\omega)\right)+\frac{\kappa}{2}\nonumber\\
  =&-\mathrm{i}(\omega+\Delta_{1})+\frac{\kappa}{2}\nonumber\\&-\dfrac{2\mathrm{i}Sg^{2}\vert\alpha_{2}\vert^{2}}{\omega-\tilde{\omega}_{B}-\frac{|\bar{B}_{+}|^{2}}{2\omega}+\dfrac{4\omega_{D}^{2}\omega_{I}S^{3}}{2\omega_{D}\omega_{I}S^{2}-\omega^{2}}},\label{eq:xi1}
\end{align}
and the resonance frequencies are obtained to a reasonable approximation
by solving
\begin{align}
(\omega+\Delta_{1})&\left(\omega-\tilde{\omega}_{B}-\frac{|\bar{B}_{+}|^{2}}{2\omega}+\dfrac{4\omega_{D}^{2}\omega_{I}S^{3}}{2\omega_{D}\omega_{I}S^{2}-\omega^{2}}\right)\nonumber\\&\hspace{3cm}-2Sg^{2}\vert\alpha_{2}\vert^{2}=0.
\end{align}

We now want to include the terms $\propto s_{+}^{*}(\omega)$ and
$\propto a_{2}^{*}(\omega)$. By inserting 
\begin{align}
e_{3+}^{\ast}(\omega)  =&\dfrac{2\omega_{D}\omega_{I}S^{2}}{2\omega_{D}\omega_{I}S^{2}-\omega^{2}}s_{+}^{\ast}(\omega),\nonumber \\
a_{1}^{\ast}(\omega)  =&\dfrac{-gS\alpha_{2}^{\ast}\left(\omega-\Delta_{1}-\mathrm{i}\dfrac{\kappa}{2}\right)}{\frac{\kappa^{2}}{4}+(\omega-\Delta_{1})^{2}}s_{+}^{\ast}(\omega)\nonumber\\&+\dfrac{\mathrm{i}\left(\omega-\Delta_{1}\right)+\dfrac{\kappa}{2}}{\frac{\kappa^{2}}{4}+(\omega-\Delta_{1})^{2}}\delta\varepsilon_{1}^{\ast}(\omega),\nonumber \\
a_{2}(\omega)  =&\dfrac{gS\alpha_{1}\left(\omega+\Delta_{2}-\mathrm{i}\dfrac{\kappa}{2}\right)}{\frac{\kappa^{2}}{4}+(\omega+\Delta_{2})^{2}}s_{+}^{\ast}(\omega)\nonumber\\&+\dfrac{\mathrm{i}\left(\omega+\Delta_{2}\right)+\dfrac{\kappa}{2}}{\frac{\kappa^{2}}{4}+(\omega+\Delta_{2})^{2}}\delta\varepsilon_{2}(\omega),\nonumber \\
a_{2}^{\ast}(\omega)  =&\dfrac{-gS\alpha_{1}^{\ast}\left(\omega-\Delta_{2}-\mathrm{i}\dfrac{\kappa}{2}\right)}{\frac{\kappa^{2}}{4}+(\omega-\Delta_{2})^{2}}s_{+}(\omega)\nonumber\\&+\dfrac{\mathrm{i}\left(\omega-\Delta_{2}\right)+\dfrac{\kappa}{2}}{\frac{\kappa^{2}}{4}+(\omega-\Delta_{2})^{2}}\delta\varepsilon_{2}^{\ast}(\omega),\label{eq:e3,a1,a2}
\end{align}
into
\begin{align*}
-\mathrm{i}\omega s_{+}^{\ast}(\omega)&=\mathrm{i}\tilde{\omega}_{B}s_{+}^{\ast}(\omega)+\mathrm{i}\bar{B}_{+}^{\ast}s_{z}(\omega)-2\mathrm{i}\omega_{D}Se_{3+}^{\ast}(\omega)\\&+2\mathrm{i}g(\alpha_{2}a_{1}^{\ast}(\omega)+\alpha_{1}^{\ast}a_{2}(\omega))
\end{align*}
we obtain 
\begin{equation}
s_{+}^{\ast}(\omega)=\chi_{-}(\omega)\left[-\dfrac{1}{2\omega}\bar{B}_{+}^{\ast2}s_{+}(\omega)+C_{1}\right],\label{eq:s_minus}
\end{equation}
with 
\begin{align}
\chi_{-}^{-1}(\omega) & =\left[\omega+\tilde{\omega}_{B}-\dfrac{1}{2\omega}|\bar{B}_{+}|^{2}-\dfrac{4\omega_{D}^{2}\omega_{I}S^{3}}{2\omega_{D}\omega_{I}S^{2}-\omega^{2}}\right.\nonumber\\&\left.+\dfrac{2g^{2}S|\alpha_{1}|^{2}\left(\omega+\Delta_{2}-\mathrm{i}\frac{\kappa}{2}\right)}{\frac{\kappa^{2}}{4}+(\omega+\Delta_{2})^{2}}\right.\nonumber\\&\left.-\dfrac{2g^{2}S|\alpha_{2}|^{2}\left(\omega-\Delta_{1}-\mathrm{i}\frac{\kappa}{2}\right)}{\frac{\kappa^{2}}{4}+(\omega-\Delta_{1})^{2}}\right],\label{eq:xi_minus}
\end{align}
and
\begin{align}
C_{1} & =-\dfrac{2g\alpha_{2}\left[\mathrm{i}\left(\omega-\Delta_{1}\right)+\dfrac{\kappa}{2}\right]}{\frac{\kappa^{2}}{4}+(\omega-\Delta_{1})^{2}}\delta\varepsilon_{1}^{\ast}(\omega)\nonumber\\&-\dfrac{2g\alpha_{1}^{\ast}\left[\mathrm{i}\left(\omega+\Delta_{2}\right)+\dfrac{\kappa}{2}\right]}{\frac{\kappa^{2}}{4}+(\omega+\Delta_{2})^{2}}\delta\varepsilon_{2}(\omega).\label{eq:C1}
\end{align}
Substituting the expressions for $s_{+}^{\ast}(\omega)$ and $\alpha_{2}^{\ast}(\omega)$
in Eq.$~$\eqref{eq:s_plus_unarranged} yields after some rearrangements
\begin{align}
s_{+}(\omega)&=\tilde{\chi}_{+}(\omega)\left[2g\alpha_{2}^{\ast}a_{1}(\omega)-\dfrac{\bar{B}_{+}^{2}}{2\omega}\chi_{-}(\omega)C_{1}\right.\nonumber\\&\left.+\dfrac{2g\alpha_{1}\left[\mathrm{i}(\omega-\Delta_{2})+\frac{\kappa}{2}\right]}{\frac{\kappa^{2}}{4}+(\omega-\Delta_{2})^{2}}\delta\varepsilon_{2}^{\ast}(\omega)\right],\label{eq:s_plus_full}
\end{align}
where 
\begin{align}
\tilde{\chi}_{+}^{-1}(\omega)&=\left[\omega-\tilde{\omega}_{B}-\dfrac{1}{2\omega}|\bar{B}_{+}|^{2}+\dfrac{4\omega_{D}^{2}\omega_{I}S^{3}}{2\omega_{D}\omega_{I}S^{2}-\omega^{2}}\right.\nonumber\\&\left.+\dfrac{2g^{2}S|\alpha_{1}|^{2}\left(\omega-\Delta_{2}-\mathrm{i}\frac{\kappa}{2}\right)}{\frac{\kappa^{2}}{4}+(\omega-\Delta_{2})^{2}}-\dfrac{\bar{B}_{+}^{2}\bar{B}_{+}^{\ast2}}{4\omega^{2}}\chi_{-}(\omega)\right].\label{eq:xi_plus_tilde}
\end{align}
With these the equation for $a_{1}(\omega)$ can be written as
\begin{align}
&\left[-\mathrm{i}\left(\omega+\Delta_{1}-2g^{2}S\vert\alpha_{2}\vert^{2}\tilde{\chi}_{+}(\omega)\right)+\frac{\kappa}{2}\right]a_{1}(\omega)=\nonumber\\&\hspace{2cm}-2\mathrm{i}gS\alpha_{2}\chi_{+}(\omega)C_{2}+\delta\varepsilon_{1}(\omega),\label{eq:a1_full}
\end{align}
with
\begin{equation}
C_{2}=-\dfrac{\bar{B}_{+}^{2}}{2\omega}\chi_{-}(\omega)C_{1}+\dfrac{2g\alpha_{1}\left[\mathrm{i}(\omega-\Delta_{2})+\frac{\kappa}{2}\right]}{\frac{\kappa^{2}}{4}+(\omega-\Delta_{2})^{2}}\delta\varepsilon_{2}^{\ast}(\omega).\label{eq:C2}
\end{equation}
Solving Eq.$~$\eqref{eq:a1_full} for $a_{1}(\omega)$ and calculating
the roots of the imaginary part of the denominator yields the more
precise values of the resonance peaks. 

\section{Details on the Cotton-Mouton Effect} \label{sec:CM-effect-app}

The contribution of the Cotton-Mouton (CM) effect to the permittivity tensor $\varepsilon_{ij}^{\mathrm{CM}}=\varepsilon_0\sum_{kl}G_{ij\mu\nu}M_{k}M_{l}$
reads

\begin{widetext}
\begin{equation}
\varepsilon^{\mathrm{CM}}=\varepsilon_{0}\begin{pmatrix}G_{11}M_{x}^{2}+G_{12}(M_{y}^{2}+M_{z}^{2}) & G_{44}(M_{x}M_{y}+M_{y}M_{x}) & G_{44}(M_{x}M_{z}+M_{z}M_{x})\\
G_{44}(M_{x}M_{y}+M_{y}M_{x}) & G_{11}M_{y}^{2}+G_{12}(M_{x}^{2}+M_{z}^{2}) & G_{44}(M_{z}M_{y}+M_{y}M_{z})\\
G_{44}(M_{x}M_{z}+M_{z}M_{x}) & G_{44}(M_{z}M_{y}+M_{y}M_{z}) & G_{11}M_{z}^{2}+G_{12}(M_{x}^{2}+M_{y}^{2})
\end{pmatrix}.\label{eq:CM_tensor}
\end{equation}
\end{widetext}

We follow the same procedure as for the Faraday effect \citep{PhysRevLett.117.133602},
i.e. we consider a single pair of linearly polarized modes with polarization
vector perpendicular to the WGM plane and perpendicular to the sphere
surface, and use a spherical basis with $\mathbf{v}=\mathbf{e_{z}}$,
$\mathbf{h}=\cos\phi\mathbf{e}_{x}+\sin\phi\mathbf{e}_{y}$, and $\mathbf{k}=\mathbf{\mathbf{h}\times\mathbf{v}}=-\sin\phi\mathbf{e}_{x}+\cos\phi\mathbf{e}_{y}$.
The electric field can then be written as $\mathbf{E}=E_{v}\mathbf{v}+E_{h}\mathbf{h}$
and the CM contribution to the electromagnetic energy is
given by

\begin{equation}
\begin{aligned}\bar{U}_{\mathrm{OM}}^{\mathrm{CM}}= & \dfrac{1}{4}\varepsilon_{0}\int\mathrm{d}\mathbf{r}\left[(G_{11}-G_{12})(|E_{v}|^{2}-|E_{h}|^{2})M_{z}^{2}\right.\\
+ & G_{44}(E_{h}^{\ast}E_{v}+E_{v}^{\ast}E_{h})\left(M_{+}\mathrm{e}^{-\mathrm{i}\phi}+M_{-}\mathrm{e}^{\mathrm{i}\phi}\right)M_{z}\\
+ & \left.M_{\mathrm{S}}^{2}(G_{12}|E_{v}|^{2}+G_{11}|E_{h}|^{2})\right].
\end{aligned}
\label{eq:CM_energy}
\end{equation}
Here we only kept the terms which contain one factor of a transverse
magnetization component $M_{\pm}=M_{x}\pm\mathrm{i}M_{y}$ as these
are relevant for one-magnon scattering processes \citep{wettling1975relation},
and used $M_{x}^{2}+M_{y}^{2}+M_{z}^{2}=M_{\mathrm{S}}^{2}$. Quantizing
the electric field and magnetization as before results in
\begin{equation}
\begin{aligned}\hat{H}_{\mathrm{int}}^{\mathrm{CM}}= & \hbar g_{\mathrm{CM}}^{44}\left(\hat{S}_{z}\hat{S}_{+}\hat{a}_{1}^{\dagger}\hat{a}_{2}+\hat{S}_{-}\hat{S}_{z}\hat{a}_{2}^{\dagger}\hat{a}_{1}\right)\\
 & +\hbar g_{\mathrm{CM}}^{12}\hat{S}_{z}^{2}(\hat{a}_{1}^{\dagger}\hat{a}_{1}-\hat{a}_{2}^{\dagger}\hat{a}_{2}),
\end{aligned}
\label{eq:H_Int_CM_final-1-1}
\end{equation}
where we neglected the constant shift in the photon field. The coupling
constants $g_{\mathrm{CM}}^{44}$ and $g_{\mathrm{CM}}^{12}$ are
given by 

\begin{align}
g_{\mathrm{CM}}^{44} & =\dfrac{\varepsilon_{0}}{2\hbar}\left(\dfrac{M_{\mathrm{S}}}{S}\right)^{2}G_{44}\tilde{g},\nonumber \\
g_{\mathrm{CM}}^{12} & =\dfrac{\omega_{r}}{8\epsilon}\left(\dfrac{M_{\mathrm{S}}}{S}\right)^{2}(G_{11}-G_{12}),\label{eq:gCM-1}
\end{align}
under the approximation $\int\mathrm{d}\mathbf{r}|E_{h}|^{2}\approx\int\mathrm{d}\mathbf{r}|E_{v}|^{2}\approx\dfrac{\hbar\omega_{r}}{2\varepsilon_{0}\varepsilon}$.

\end{document}

%% file: OpticalSignatures.bbl

%% file: OpticalSignatures.bbl
\begin{thebibliography}{117}%
\makeatletter
\providecommand \@ifxundefined [1]{%
 \@ifx{#1\undefined}
}%
\providecommand \@ifnum [1]{%
 \ifnum #1\expandafter \@firstoftwo
 \else \expandafter \@secondoftwo
 \fi
}%
\providecommand \@ifx [1]{%
 \ifx #1\expandafter \@firstoftwo
 \else \expandafter \@secondoftwo
 \fi
}%
\providecommand \natexlab [1]{#1}%
\providecommand \enquote  [1]{``#1''}%
\providecommand \bibnamefont  [1]{#1}%
\providecommand \bibfnamefont [1]{#1}%
\providecommand \citenamefont [1]{#1}%
\providecommand \href@noop [0]{\@secondoftwo}%
\providecommand \href [0]{\begingroup \@sanitize@url \@href}%
\providecommand \@href[1]{\@@startlink{#1}\@@href}%
\providecommand \@@href[1]{\endgroup#1\@@endlink}%
\providecommand \@sanitize@url [0]{\catcode `\\12\catcode `\$12\catcode
  `\&12\catcode `\#12\catcode `\^12\catcode `\_12\catcode `\%12\relax}%
\providecommand \@@startlink[1]{}%
\providecommand \@@endlink[0]{}%
\providecommand \url  [0]{\begingroup\@sanitize@url \@url }%
\providecommand \@url [1]{\endgroup\@href {#1}{\urlprefix }}%
\providecommand \urlprefix  [0]{URL }%
\providecommand \Eprint [0]{\href }%
\providecommand \doibase [0]{https://doi.org/}%
\providecommand \selectlanguage [0]{\@gobble}%
\providecommand \bibinfo  [0]{\@secondoftwo}%
\providecommand \bibfield  [0]{\@secondoftwo}%
\providecommand \translation [1]{[#1]}%
\providecommand \BibitemOpen [0]{}%
\providecommand \bibitemStop [0]{}%
\providecommand \bibitemNoStop [0]{.\EOS\space}%
\providecommand \EOS [0]{\spacefactor3000\relax}%
\providecommand \BibitemShut  [1]{\csname bibitem#1\endcsname}%
\let\auto@bib@innerbib\@empty
%</preamble>
\bibitem [{\citenamefont {Ashkin}(2006)}]{ashkin2006optical}%
  \BibitemOpen
  \bibfield  {author} {\bibinfo {author} {\bibfnamefont {A.}~\bibnamefont
  {Ashkin}},\ }\href@noop {} {\emph {\bibinfo {title} {Optical trapping and
  manipulation of neutral particles using lasers: a reprint volume with
  commentaries}}}\ (\bibinfo  {publisher} {World Scientific},\ \bibinfo {year}
  {2006})\BibitemShut {NoStop}%
\bibitem [{\citenamefont {Grier}(2003)}]{grier2003revolution}%
  \BibitemOpen
  \bibfield  {author} {\bibinfo {author} {\bibfnamefont {D.~G.}\ \bibnamefont
  {Grier}},\ }\bibfield  {title} {\bibinfo {title} {A revolution in optical
  manipulation},\ }\href {https://www.nature.com/articles/nature01935}
  {\bibfield  {journal} {\bibinfo  {journal} {Nature}\ }\textbf {\bibinfo
  {volume} {424}},\ \bibinfo {pages} {810} (\bibinfo {year}
  {2003})}\BibitemShut {NoStop}%
\bibitem [{\citenamefont {Hebestreit}\ \emph
  {et~al.}(2018{\natexlab{a}})\citenamefont {Hebestreit}, \citenamefont
  {Frimmer}, \citenamefont {Reimann},\ and\ \citenamefont
  {Novotny}}]{hebestreit2018sensing}%
  \BibitemOpen
  \bibfield  {author} {\bibinfo {author} {\bibfnamefont {E.}~\bibnamefont
  {Hebestreit}}, \bibinfo {author} {\bibfnamefont {M.}~\bibnamefont {Frimmer}},
  \bibinfo {author} {\bibfnamefont {R.}~\bibnamefont {Reimann}},\ and\ \bibinfo
  {author} {\bibfnamefont {L.}~\bibnamefont {Novotny}},\ }\bibfield  {title}
  {\bibinfo {title} {Sensing static forces with free-falling nanoparticles},\
  }\href {https://journals.aps.org/prl/abstract/10.1103/PhysRevLett.121.063602}
  {\bibfield  {journal} {\bibinfo  {journal} {Phys. Rev. Lett.}\ }\textbf
  {\bibinfo {volume} {121}},\ \bibinfo {pages} {063602} (\bibinfo {year}
  {2018}{\natexlab{a}})}\BibitemShut {NoStop}%
\bibitem [{\citenamefont {Ranjit}\ \emph {et~al.}(2015)\citenamefont {Ranjit},
  \citenamefont {Atherton}, \citenamefont {Stutz}, \citenamefont {Cunningham},\
  and\ \citenamefont {Geraci}}]{ranjit2015attonewton}%
  \BibitemOpen
  \bibfield  {author} {\bibinfo {author} {\bibfnamefont {G.}~\bibnamefont
  {Ranjit}}, \bibinfo {author} {\bibfnamefont {D.~P.}\ \bibnamefont
  {Atherton}}, \bibinfo {author} {\bibfnamefont {J.~H.}\ \bibnamefont {Stutz}},
  \bibinfo {author} {\bibfnamefont {M.}~\bibnamefont {Cunningham}},\ and\
  \bibinfo {author} {\bibfnamefont {A.~A.}\ \bibnamefont {Geraci}},\ }\bibfield
   {title} {\bibinfo {title} {Attonewton force detection using microspheres in
  a dual-beam optical trap in high vacuum},\ }\href
  {https://journals.aps.org/pra/abstract/10.1103/PhysRevA.91.051805} {\bibfield
   {journal} {\bibinfo  {journal} {Phys. Rev. A}\ }\textbf {\bibinfo {volume}
  {91}},\ \bibinfo {pages} {051805} (\bibinfo {year} {2015})}\BibitemShut
  {NoStop}%
\bibitem [{\citenamefont {Ranjit}\ \emph {et~al.}(2016)\citenamefont {Ranjit},
  \citenamefont {Cunningham}, \citenamefont {Casey},\ and\ \citenamefont
  {Geraci}}]{ranjit2016zeptonewton}%
  \BibitemOpen
  \bibfield  {author} {\bibinfo {author} {\bibfnamefont {G.}~\bibnamefont
  {Ranjit}}, \bibinfo {author} {\bibfnamefont {M.}~\bibnamefont {Cunningham}},
  \bibinfo {author} {\bibfnamefont {K.}~\bibnamefont {Casey}},\ and\ \bibinfo
  {author} {\bibfnamefont {A.~A.}\ \bibnamefont {Geraci}},\ }\bibfield  {title}
  {\bibinfo {title} {Zeptonewton force sensing with nanospheres in an optical
  lattice},\ }\href
  {https://journals.aps.org/pra/abstract/10.1103/PhysRevA.93.053801} {\bibfield
   {journal} {\bibinfo  {journal} {Phys. Rev. A}\ }\textbf {\bibinfo {volume}
  {93}},\ \bibinfo {pages} {053801} (\bibinfo {year} {2016})}\BibitemShut
  {NoStop}%
\bibitem [{\citenamefont {Geraci}\ \emph {et~al.}(2010)\citenamefont {Geraci},
  \citenamefont {Papp},\ and\ \citenamefont {Kitching}}]{geraci2010short}%
  \BibitemOpen
  \bibfield  {author} {\bibinfo {author} {\bibfnamefont {A.~A.}\ \bibnamefont
  {Geraci}}, \bibinfo {author} {\bibfnamefont {S.~B.}\ \bibnamefont {Papp}},\
  and\ \bibinfo {author} {\bibfnamefont {J.}~\bibnamefont {Kitching}},\
  }\bibfield  {title} {\bibinfo {title} {Short-range force detection using
  optically cooled levitated microspheres},\ }\href
  {https://journals.aps.org/prl/abstract/10.1103/PhysRevLett.105.101101}
  {\bibfield  {journal} {\bibinfo  {journal} {Phys. Rev. Lett.}\ }\textbf
  {\bibinfo {volume} {105}},\ \bibinfo {pages} {101101} (\bibinfo {year}
  {2010})}\BibitemShut {NoStop}%
\bibitem [{\citenamefont {Geraci}\ and\ \citenamefont
  {Goldman}(2015)}]{geraci2015sensing}%
  \BibitemOpen
  \bibfield  {author} {\bibinfo {author} {\bibfnamefont {A.}~\bibnamefont
  {Geraci}}\ and\ \bibinfo {author} {\bibfnamefont {H.}~\bibnamefont
  {Goldman}},\ }\bibfield  {title} {\bibinfo {title} {Sensing short range
  forces with a nanosphere matter-wave interferometer},\ }\href
  {https://journals.aps.org/prd/abstract/10.1103/PhysRevD.92.062002} {\bibfield
   {journal} {\bibinfo  {journal} {Phys. Rev. D}\ }\textbf {\bibinfo {volume}
  {92}},\ \bibinfo {pages} {062002} (\bibinfo {year} {2015})}\BibitemShut
  {NoStop}%
\bibitem [{\citenamefont {Hempston}\ \emph {et~al.}(2017)\citenamefont
  {Hempston}, \citenamefont {Vovrosh}, \citenamefont {Toro{\v{s}}},
  \citenamefont {Winstone}, \citenamefont {Rashid},\ and\ \citenamefont
  {Ulbricht}}]{hempston2017force}%
  \BibitemOpen
  \bibfield  {author} {\bibinfo {author} {\bibfnamefont {D.}~\bibnamefont
  {Hempston}}, \bibinfo {author} {\bibfnamefont {J.}~\bibnamefont {Vovrosh}},
  \bibinfo {author} {\bibfnamefont {M.}~\bibnamefont {Toro{\v{s}}}}, \bibinfo
  {author} {\bibfnamefont {G.}~\bibnamefont {Winstone}}, \bibinfo {author}
  {\bibfnamefont {M.}~\bibnamefont {Rashid}},\ and\ \bibinfo {author}
  {\bibfnamefont {H.}~\bibnamefont {Ulbricht}},\ }\bibfield  {title} {\bibinfo
  {title} {Force sensing with an optically levitated charged nanoparticle},\
  }\href {https://aip.scitation.org/doi/10.1063/1.4993555} {\bibfield
  {journal} {\bibinfo  {journal} {Appl. Phys. Lett.}\ }\textbf {\bibinfo
  {volume} {111}},\ \bibinfo {pages} {133111} (\bibinfo {year}
  {2017})}\BibitemShut {NoStop}%
\bibitem [{\citenamefont {Blakemore}\ \emph {et~al.}(2019)\citenamefont
  {Blakemore}, \citenamefont {Rider}, \citenamefont {Roy}, \citenamefont
  {Wang}, \citenamefont {Kawasaki},\ and\ \citenamefont
  {Gratta}}]{blakemore2019three}%
  \BibitemOpen
  \bibfield  {author} {\bibinfo {author} {\bibfnamefont {C.~P.}\ \bibnamefont
  {Blakemore}}, \bibinfo {author} {\bibfnamefont {A.~D.}\ \bibnamefont
  {Rider}}, \bibinfo {author} {\bibfnamefont {S.}~\bibnamefont {Roy}}, \bibinfo
  {author} {\bibfnamefont {Q.}~\bibnamefont {Wang}}, \bibinfo {author}
  {\bibfnamefont {A.}~\bibnamefont {Kawasaki}},\ and\ \bibinfo {author}
  {\bibfnamefont {G.}~\bibnamefont {Gratta}},\ }\bibfield  {title} {\bibinfo
  {title} {Three-dimensional force-field microscopy with optically levitated
  microspheres},\ }\href
  {https://journals.aps.org/pra/abstract/10.1103/PhysRevA.99.023816} {\bibfield
   {journal} {\bibinfo  {journal} {Phys. Rev. A}\ }\textbf {\bibinfo {volume}
  {99}},\ \bibinfo {pages} {023816} (\bibinfo {year} {2019})}\BibitemShut
  {NoStop}%
\bibitem [{\citenamefont {Monteiro}\ \emph {et~al.}(2020)\citenamefont
  {Monteiro}, \citenamefont {Li}, \citenamefont {Afek}, \citenamefont {Li},
  \citenamefont {Mossman},\ and\ \citenamefont {Moore}}]{monteiro2020force}%
  \BibitemOpen
  \bibfield  {author} {\bibinfo {author} {\bibfnamefont {F.}~\bibnamefont
  {Monteiro}}, \bibinfo {author} {\bibfnamefont {W.}~\bibnamefont {Li}},
  \bibinfo {author} {\bibfnamefont {G.}~\bibnamefont {Afek}}, \bibinfo {author}
  {\bibfnamefont {C.-l.}\ \bibnamefont {Li}}, \bibinfo {author} {\bibfnamefont
  {M.}~\bibnamefont {Mossman}},\ and\ \bibinfo {author} {\bibfnamefont {D.~C.}\
  \bibnamefont {Moore}},\ }\bibfield  {title} {\bibinfo {title} {Force and
  acceleration sensing with optically levitated nanogram masses at microkelvin
  temperatures},\ }\href
  {https://journals.aps.org/pra/abstract/10.1103/PhysRevA.101.053835}
  {\bibfield  {journal} {\bibinfo  {journal} {Phys. Rev. A}\ }\textbf {\bibinfo
  {volume} {101}},\ \bibinfo {pages} {053835} (\bibinfo {year}
  {2020})}\BibitemShut {NoStop}%
\bibitem [{\citenamefont {Monteiro}\ \emph {et~al.}(2017)\citenamefont
  {Monteiro}, \citenamefont {Ghosh}, \citenamefont {Fine},\ and\ \citenamefont
  {Moore}}]{monteiro2017optical}%
  \BibitemOpen
  \bibfield  {author} {\bibinfo {author} {\bibfnamefont {F.}~\bibnamefont
  {Monteiro}}, \bibinfo {author} {\bibfnamefont {S.}~\bibnamefont {Ghosh}},
  \bibinfo {author} {\bibfnamefont {A.~G.}\ \bibnamefont {Fine}},\ and\
  \bibinfo {author} {\bibfnamefont {D.~C.}\ \bibnamefont {Moore}},\ }\bibfield
  {title} {\bibinfo {title} {Optical levitation of 10-ng spheres with nano-g
  acceleration sensitivity},\ }\href
  {https://journals.aps.org/pra/abstract/10.1103/PhysRevA.96.063841} {\bibfield
   {journal} {\bibinfo  {journal} {Phys. Rev. A}\ }\textbf {\bibinfo {volume}
  {96}},\ \bibinfo {pages} {063841} (\bibinfo {year} {2017})}\BibitemShut
  {NoStop}%
\bibitem [{\citenamefont {Bykov}\ \emph {et~al.}(2015)\citenamefont {Bykov},
  \citenamefont {Schmidt}, \citenamefont {Euser},\ and\ \citenamefont
  {Russell}}]{bykov2015flying}%
  \BibitemOpen
  \bibfield  {author} {\bibinfo {author} {\bibfnamefont {D.~S.}\ \bibnamefont
  {Bykov}}, \bibinfo {author} {\bibfnamefont {O.~A.}\ \bibnamefont {Schmidt}},
  \bibinfo {author} {\bibfnamefont {T.~G.}\ \bibnamefont {Euser}},\ and\
  \bibinfo {author} {\bibfnamefont {P.~S.~J.}\ \bibnamefont {Russell}},\
  }\bibfield  {title} {\bibinfo {title} {Flying particle sensors in hollow-core
  photonic crystal fibre},\ }\href
  {https://www.nature.com/articles/nphoton.2015.94} {\bibfield  {journal}
  {\bibinfo  {journal} {Nature Photonics}\ }\textbf {\bibinfo {volume} {9}},\
  \bibinfo {pages} {461} (\bibinfo {year} {2015})}\BibitemShut {NoStop}%
\bibitem [{\citenamefont {Zeltner}\ \emph {et~al.}(2018)\citenamefont
  {Zeltner}, \citenamefont {Pennetta}, \citenamefont {Xie},\ and\ \citenamefont
  {Russell}}]{zeltner2018flyingparticle}%
  \BibitemOpen
  \bibfield  {author} {\bibinfo {author} {\bibfnamefont {R.}~\bibnamefont
  {Zeltner}}, \bibinfo {author} {\bibfnamefont {R.}~\bibnamefont {Pennetta}},
  \bibinfo {author} {\bibfnamefont {S.}~\bibnamefont {Xie}},\ and\ \bibinfo
  {author} {\bibfnamefont {P.~S.}\ \bibnamefont {Russell}},\ }\bibfield
  {title} {\bibinfo {title} {Flying particle microlaser and temperature sensor
  in hollow core photonic crystal fiber},\ }\href
  {https://www.osapublishing.org/ol/abstract.cfm?uri=ol-43-7-1479} {\bibfield
  {journal} {\bibinfo  {journal} {Optics Letters}\ }\textbf {\bibinfo {volume}
  {43}},\ \bibinfo {pages} {1479} (\bibinfo {year} {2018})}\BibitemShut
  {NoStop}%
\bibitem [{\citenamefont {Zhang}\ \emph {et~al.}(2014)\citenamefont {Zhang},
  \citenamefont {Liang}, \citenamefont {Liu}, \citenamefont {Lei},
  \citenamefont {Yang},\ and\ \citenamefont {Yuan}}]{zhang2014novel}%
  \BibitemOpen
  \bibfield  {author} {\bibinfo {author} {\bibfnamefont {Y.}~\bibnamefont
  {Zhang}}, \bibinfo {author} {\bibfnamefont {P.}~\bibnamefont {Liang}},
  \bibinfo {author} {\bibfnamefont {Z.}~\bibnamefont {Liu}}, \bibinfo {author}
  {\bibfnamefont {J.}~\bibnamefont {Lei}}, \bibinfo {author} {\bibfnamefont
  {J.}~\bibnamefont {Yang}},\ and\ \bibinfo {author} {\bibfnamefont
  {L.}~\bibnamefont {Yuan}},\ }\bibfield  {title} {\bibinfo {title} {A novel
  temperature sensor based on optical trapping technology},\ }\href
  {https://ieeexplore.ieee.org/document/6742608} {\bibfield  {journal}
  {\bibinfo  {journal} {Journal of Lightwave Technology}\ }\textbf {\bibinfo
  {volume} {32}},\ \bibinfo {pages} {1394} (\bibinfo {year}
  {2014})}\BibitemShut {NoStop}%
\bibitem [{\citenamefont {Ashkin}\ \emph {et~al.}(1987)\citenamefont {Ashkin},
  \citenamefont {Dziedzic},\ and\ \citenamefont {Yamane}}]{ashkin1987cells}%
  \BibitemOpen
  \bibfield  {author} {\bibinfo {author} {\bibfnamefont {A.}~\bibnamefont
  {Ashkin}}, \bibinfo {author} {\bibfnamefont {J.~M.}\ \bibnamefont
  {Dziedzic}},\ and\ \bibinfo {author} {\bibfnamefont {T.}~\bibnamefont
  {Yamane}},\ }\bibfield  {title} {\bibinfo {title} {Optical trapping and
  manipulation of single cells using infrared laser beams},\ }\href
  {https://www.nature.com/articles/330769a0#citeas} {\bibfield  {journal}
  {\bibinfo  {journal} {Nature}\ }\textbf {\bibinfo {volume} {330}},\ \bibinfo
  {pages} {769} (\bibinfo {year} {1987})}\BibitemShut {NoStop}%
\bibitem [{\citenamefont {Ashkin}\ and\ \citenamefont
  {Dziedzic}(1987)}]{ashkin1987bacteria}%
  \BibitemOpen
  \bibfield  {author} {\bibinfo {author} {\bibfnamefont {A.}~\bibnamefont
  {Ashkin}}\ and\ \bibinfo {author} {\bibfnamefont {J.~M.}\ \bibnamefont
  {Dziedzic}},\ }\bibfield  {title} {\bibinfo {title} {Optical trapping and
  manipulation of viruses and bacteria},\ }\href
  {https://science.sciencemag.org/content/235/4795/1517} {\bibfield  {journal}
  {\bibinfo  {journal} {Science}\ }\textbf {\bibinfo {volume} {235}},\ \bibinfo
  {pages} {1517} (\bibinfo {year} {1987})}\BibitemShut {NoStop}%
\bibitem [{\citenamefont {Lang}\ and\ \citenamefont
  {Block}(2003)}]{lang2003resource}%
  \BibitemOpen
  \bibfield  {author} {\bibinfo {author} {\bibfnamefont {M.~J.}\ \bibnamefont
  {Lang}}\ and\ \bibinfo {author} {\bibfnamefont {S.~M.}\ \bibnamefont
  {Block}},\ }\bibfield  {title} {\bibinfo {title} {Resource letter: Lbot-1:
  Laser-based optical tweezers},\ }\href
  {https://aapt.scitation.org/doi/10.1119/1.1532323} {\bibfield  {journal}
  {\bibinfo  {journal} {American Journal of Physics}\ }\textbf {\bibinfo
  {volume} {71}},\ \bibinfo {pages} {201} (\bibinfo {year} {2003})}\BibitemShut
  {NoStop}%
\bibitem [{\citenamefont {Neuman}\ and\ \citenamefont
  {Block}(2004)}]{neuman2004optical}%
  \BibitemOpen
  \bibfield  {author} {\bibinfo {author} {\bibfnamefont {K.~C.}\ \bibnamefont
  {Neuman}}\ and\ \bibinfo {author} {\bibfnamefont {S.~M.}\ \bibnamefont
  {Block}},\ }\bibfield  {title} {\bibinfo {title} {Optical trapping},\ }\href
  {https://aip.scitation.org/doi/10.1063/1.1785844} {\bibfield  {journal}
  {\bibinfo  {journal} {Review of Scientific Instruments}\ }\textbf {\bibinfo
  {volume} {75}},\ \bibinfo {pages} {2787} (\bibinfo {year}
  {2004})}\BibitemShut {NoStop}%
\bibitem [{\citenamefont {Marag{\`o}}\ \emph {et~al.}(2013)\citenamefont
  {Marag{\`o}}, \citenamefont {Jones}, \citenamefont {Gucciardi}, \citenamefont
  {Volpe},\ and\ \citenamefont {Ferrari}}]{marago2013optical}%
  \BibitemOpen
  \bibfield  {author} {\bibinfo {author} {\bibfnamefont {O.~M.}\ \bibnamefont
  {Marag{\`o}}}, \bibinfo {author} {\bibfnamefont {P.~H.}\ \bibnamefont
  {Jones}}, \bibinfo {author} {\bibfnamefont {P.~G.}\ \bibnamefont
  {Gucciardi}}, \bibinfo {author} {\bibfnamefont {G.}~\bibnamefont {Volpe}},\
  and\ \bibinfo {author} {\bibfnamefont {A.~C.}\ \bibnamefont {Ferrari}},\
  }\bibfield  {title} {\bibinfo {title} {Optical trapping and manipulation of
  nanostructures},\ }\href {https://www.nature.com/articles/nnano.2013.208}
  {\bibfield  {journal} {\bibinfo  {journal} {Nature Nanotechnology}\ }\textbf
  {\bibinfo {volume} {8}},\ \bibinfo {pages} {807} (\bibinfo {year}
  {2013})}\BibitemShut {NoStop}%
\bibitem [{\citenamefont {Gieseler}\ \emph {et~al.}(2013)\citenamefont
  {Gieseler}, \citenamefont {Novotny},\ and\ \citenamefont
  {Quidant}}]{gieseler2013thermal}%
  \BibitemOpen
  \bibfield  {author} {\bibinfo {author} {\bibfnamefont {J.}~\bibnamefont
  {Gieseler}}, \bibinfo {author} {\bibfnamefont {L.}~\bibnamefont {Novotny}},\
  and\ \bibinfo {author} {\bibfnamefont {R.}~\bibnamefont {Quidant}},\
  }\bibfield  {title} {\bibinfo {title} {Thermal nonlinearities in a
  nanomechanical oscillator},\ }\href {https://doi.org/10.1038/nphys2798}
  {\bibfield  {journal} {\bibinfo  {journal} {Nature Physics}\ }\textbf
  {\bibinfo {volume} {9}},\ \bibinfo {pages} {806} (\bibinfo {year}
  {2013})}\BibitemShut {NoStop}%
\bibitem [{\citenamefont {Gieseler}\ \emph {et~al.}(2014)\citenamefont
  {Gieseler}, \citenamefont {Quidant}, \citenamefont {Dellago},\ and\
  \citenamefont {Novotny}}]{gieseler2014dynamic}%
  \BibitemOpen
  \bibfield  {author} {\bibinfo {author} {\bibfnamefont {J.}~\bibnamefont
  {Gieseler}}, \bibinfo {author} {\bibfnamefont {R.}~\bibnamefont {Quidant}},
  \bibinfo {author} {\bibfnamefont {C.}~\bibnamefont {Dellago}},\ and\ \bibinfo
  {author} {\bibfnamefont {L.}~\bibnamefont {Novotny}},\ }\bibfield  {title}
  {\bibinfo {title} {Dynamic relaxation of a levitated nanoparticle from a
  non-equilibrium steady state},\ }\href
  {https://doi.org/10.1038/nnano.2014.40} {\bibfield  {journal} {\bibinfo
  {journal} {Nature Nanotechnology}\ }\textbf {\bibinfo {volume} {9}},\
  \bibinfo {pages} {358} (\bibinfo {year} {2014})}\BibitemShut {NoStop}%
\bibitem [{\citenamefont {Millen}\ \emph {et~al.}(2014)\citenamefont {Millen},
  \citenamefont {Deesuwan}, \citenamefont {Barker},\ and\ \citenamefont
  {Anders}}]{millen2014nanoscale}%
  \BibitemOpen
  \bibfield  {author} {\bibinfo {author} {\bibfnamefont {J.}~\bibnamefont
  {Millen}}, \bibinfo {author} {\bibfnamefont {T.}~\bibnamefont {Deesuwan}},
  \bibinfo {author} {\bibfnamefont {P.}~\bibnamefont {Barker}},\ and\ \bibinfo
  {author} {\bibfnamefont {J.}~\bibnamefont {Anders}},\ }\bibfield  {title}
  {\bibinfo {title} {Nanoscale temperature measurements using non-equilibrium
  brownian dynamics of a levitated nanosphere},\ }\href
  {https://www.nature.com/articles/nnano.2014.82} {\bibfield  {journal}
  {\bibinfo  {journal} {Nature Nanotechnology}\ }\textbf {\bibinfo {volume}
  {9}},\ \bibinfo {pages} {425} (\bibinfo {year} {2014})}\BibitemShut {NoStop}%
\bibitem [{\citenamefont {Ricci}\ \emph {et~al.}(2017)\citenamefont {Ricci},
  \citenamefont {Rica}, \citenamefont {Spasenovi{\'c}}, \citenamefont
  {Gieseler}, \citenamefont {Rondin}, \citenamefont {Novotny},\ and\
  \citenamefont {Quidant}}]{ricci2017optically}%
  \BibitemOpen
  \bibfield  {author} {\bibinfo {author} {\bibfnamefont {F.}~\bibnamefont
  {Ricci}}, \bibinfo {author} {\bibfnamefont {R.~A.}\ \bibnamefont {Rica}},
  \bibinfo {author} {\bibfnamefont {M.}~\bibnamefont {Spasenovi{\'c}}},
  \bibinfo {author} {\bibfnamefont {J.}~\bibnamefont {Gieseler}}, \bibinfo
  {author} {\bibfnamefont {L.}~\bibnamefont {Rondin}}, \bibinfo {author}
  {\bibfnamefont {L.}~\bibnamefont {Novotny}},\ and\ \bibinfo {author}
  {\bibfnamefont {R.}~\bibnamefont {Quidant}},\ }\bibfield  {title} {\bibinfo
  {title} {Optically levitated nanoparticle as a model system for stochastic
  bistable dynamics},\ }\href
  {https://www.nature.com/articles/ncomms15141?origin=ppub} {\bibfield
  {journal} {\bibinfo  {journal} {Nat. Comm.}\ }\textbf {\bibinfo {volume}
  {8}},\ \bibinfo {pages} {1} (\bibinfo {year} {2017})}\BibitemShut {NoStop}%
\bibitem [{\citenamefont {Arita}\ \emph {et~al.}(2013)\citenamefont {Arita},
  \citenamefont {Mazilu},\ and\ \citenamefont
  {Dholakia}}]{arita2013laserinduced}%
  \BibitemOpen
  \bibfield  {author} {\bibinfo {author} {\bibfnamefont {Y.}~\bibnamefont
  {Arita}}, \bibinfo {author} {\bibfnamefont {M.}~\bibnamefont {Mazilu}},\ and\
  \bibinfo {author} {\bibfnamefont {K.}~\bibnamefont {Dholakia}},\ }\bibfield
  {title} {\bibinfo {title} {Laser-induced rotation and cooling of a trapped
  microgyroscope in vacuum},\ }\href {https://doi.org/10.1038/ncomms3374}
  {\bibfield  {journal} {\bibinfo  {journal} {Nat. Comm.}\ }\textbf {\bibinfo
  {volume} {4}},\ \bibinfo {pages} {2374} (\bibinfo {year} {2013})}\BibitemShut
  {NoStop}%
\bibitem [{\citenamefont {Hebestreit}\ \emph
  {et~al.}(2018{\natexlab{b}})\citenamefont {Hebestreit}, \citenamefont
  {Reimann}, \citenamefont {Frimmer},\ and\ \citenamefont
  {Novotny}}]{hebestreit2018measuring}%
  \BibitemOpen
  \bibfield  {author} {\bibinfo {author} {\bibfnamefont {E.}~\bibnamefont
  {Hebestreit}}, \bibinfo {author} {\bibfnamefont {R.}~\bibnamefont {Reimann}},
  \bibinfo {author} {\bibfnamefont {M.}~\bibnamefont {Frimmer}},\ and\ \bibinfo
  {author} {\bibfnamefont {L.}~\bibnamefont {Novotny}},\ }\bibfield  {title}
  {\bibinfo {title} {Measuring the internal temperature of a levitated
  nanoparticle in high vacuum},\ }\href
  {https://journals.aps.org/pra/abstract/10.1103/PhysRevA.97.043803} {\bibfield
   {journal} {\bibinfo  {journal} {Phys. Rev. A}\ }\textbf {\bibinfo {volume}
  {97}},\ \bibinfo {pages} {043803} (\bibinfo {year}
  {2018}{\natexlab{b}})}\BibitemShut {NoStop}%
\bibitem [{\citenamefont {Hoang}\ \emph {et~al.}(2016)\citenamefont {Hoang},
  \citenamefont {Ma}, \citenamefont {Ahn}, \citenamefont {Bang}, \citenamefont
  {Robicheaux}, \citenamefont {Yin},\ and\ \citenamefont
  {Li}}]{hoang2016torsional}%
  \BibitemOpen
  \bibfield  {author} {\bibinfo {author} {\bibfnamefont {T.~M.}\ \bibnamefont
  {Hoang}}, \bibinfo {author} {\bibfnamefont {Y.}~\bibnamefont {Ma}}, \bibinfo
  {author} {\bibfnamefont {J.}~\bibnamefont {Ahn}}, \bibinfo {author}
  {\bibfnamefont {J.}~\bibnamefont {Bang}}, \bibinfo {author} {\bibfnamefont
  {F.}~\bibnamefont {Robicheaux}}, \bibinfo {author} {\bibfnamefont {Z.-Q.}\
  \bibnamefont {Yin}},\ and\ \bibinfo {author} {\bibfnamefont {T.}~\bibnamefont
  {Li}},\ }\bibfield  {title} {\bibinfo {title} {Torsional optomechanics of a
  levitated nonspherical nanoparticle},\ }\href
  {https://doi.org/10.1103/PhysRevLett.117.123604} {\bibfield  {journal}
  {\bibinfo  {journal} {Phys. Rev. Lett.}\ }\textbf {\bibinfo {volume} {117}},\
  \bibinfo {pages} {123604} (\bibinfo {year} {2016})}\BibitemShut {NoStop}%
\bibitem [{\citenamefont {Rashid}\ \emph {et~al.}(2018)\citenamefont {Rashid},
  \citenamefont {Toro\ifmmode~\check{s}\else \v{s}\fi{}}, \citenamefont
  {Setter},\ and\ \citenamefont {Ulbricht}}]{rashid2018precessionmotion}%
  \BibitemOpen
  \bibfield  {author} {\bibinfo {author} {\bibfnamefont {M.}~\bibnamefont
  {Rashid}}, \bibinfo {author} {\bibfnamefont {M.}~\bibnamefont
  {Toro\ifmmode~\check{s}\else \v{s}\fi{}}}, \bibinfo {author} {\bibfnamefont
  {A.}~\bibnamefont {Setter}},\ and\ \bibinfo {author} {\bibfnamefont
  {H.}~\bibnamefont {Ulbricht}},\ }\bibfield  {title} {\bibinfo {title}
  {Precession motion in levitated optomechanics},\ }\href
  {https://doi.org/10.1103/PhysRevLett.121.253601} {\bibfield  {journal}
  {\bibinfo  {journal} {Phys. Rev. Lett.}\ }\textbf {\bibinfo {volume} {121}},\
  \bibinfo {pages} {253601} (\bibinfo {year} {2018})}\BibitemShut {NoStop}%
\bibitem [{\citenamefont {Deli\ifmmode~\acute{c}\else \'{c}\fi{}}\ \emph
  {et~al.}(2019)\citenamefont {Deli\ifmmode~\acute{c}\else \'{c}\fi{}},
  \citenamefont {Reisenbauer}, \citenamefont {Grass}, \citenamefont {Kiesel},
  \citenamefont {Vuleti\ifmmode~\acute{c}\else \'{c}\fi{}},\ and\ \citenamefont
  {Aspelmeyer}}]{delic2019cavity}%
  \BibitemOpen
  \bibfield  {author} {\bibinfo {author} {\bibfnamefont {U.~c.~v.}\
  \bibnamefont {Deli\ifmmode~\acute{c}\else \'{c}\fi{}}}, \bibinfo {author}
  {\bibfnamefont {M.}~\bibnamefont {Reisenbauer}}, \bibinfo {author}
  {\bibfnamefont {D.}~\bibnamefont {Grass}}, \bibinfo {author} {\bibfnamefont
  {N.}~\bibnamefont {Kiesel}}, \bibinfo {author} {\bibfnamefont
  {V.}~\bibnamefont {Vuleti\ifmmode~\acute{c}\else \'{c}\fi{}}},\ and\ \bibinfo
  {author} {\bibfnamefont {M.}~\bibnamefont {Aspelmeyer}},\ }\bibfield  {title}
  {\bibinfo {title} {Cavity cooling of a levitated nanosphere by coherent
  scattering},\ }\href {https://doi.org/10.1103/PhysRevLett.122.123602}
  {\bibfield  {journal} {\bibinfo  {journal} {Phys. Rev. Lett.}\ }\textbf
  {\bibinfo {volume} {122}},\ \bibinfo {pages} {123602} (\bibinfo {year}
  {2019})}\BibitemShut {NoStop}%
\bibitem [{\citenamefont {Windey}\ \emph {et~al.}(2019)\citenamefont {Windey},
  \citenamefont {Gonzalez-Ballestero}, \citenamefont {Maurer}, \citenamefont
  {Novotny}, \citenamefont {Romero-Isart},\ and\ \citenamefont
  {Reimann}}]{windey2019cavitybased}%
  \BibitemOpen
  \bibfield  {author} {\bibinfo {author} {\bibfnamefont {D.}~\bibnamefont
  {Windey}}, \bibinfo {author} {\bibfnamefont {C.}~\bibnamefont
  {Gonzalez-Ballestero}}, \bibinfo {author} {\bibfnamefont {P.}~\bibnamefont
  {Maurer}}, \bibinfo {author} {\bibfnamefont {L.}~\bibnamefont {Novotny}},
  \bibinfo {author} {\bibfnamefont {O.}~\bibnamefont {Romero-Isart}},\ and\
  \bibinfo {author} {\bibfnamefont {R.}~\bibnamefont {Reimann}},\ }\bibfield
  {title} {\bibinfo {title} {Cavity-based 3d cooling of a levitated
  nanoparticle via coherent scattering},\ }\href
  {https://doi.org/10.1103/PhysRevLett.122.123601} {\bibfield  {journal}
  {\bibinfo  {journal} {Phys. Rev. Lett.}\ }\textbf {\bibinfo {volume} {122}},\
  \bibinfo {pages} {123601} (\bibinfo {year} {2019})}\BibitemShut {NoStop}%
\bibitem [{\citenamefont {Tebbenjohanns}\ \emph {et~al.}(2019)\citenamefont
  {Tebbenjohanns}, \citenamefont {Frimmer}, \citenamefont {Militaru},
  \citenamefont {Jain},\ and\ \citenamefont
  {Novotny}}]{tabbenjohanns2019colddamping}%
  \BibitemOpen
  \bibfield  {author} {\bibinfo {author} {\bibfnamefont {F.}~\bibnamefont
  {Tebbenjohanns}}, \bibinfo {author} {\bibfnamefont {M.}~\bibnamefont
  {Frimmer}}, \bibinfo {author} {\bibfnamefont {A.}~\bibnamefont {Militaru}},
  \bibinfo {author} {\bibfnamefont {V.}~\bibnamefont {Jain}},\ and\ \bibinfo
  {author} {\bibfnamefont {L.}~\bibnamefont {Novotny}},\ }\bibfield  {title}
  {\bibinfo {title} {Cold damping of an optically levitated nanoparticle to
  microkelvin temperatures},\ }\href
  {https://doi.org/10.1103/PhysRevLett.122.223601} {\bibfield  {journal}
  {\bibinfo  {journal} {Phys. Rev. Lett.}\ }\textbf {\bibinfo {volume} {122}},\
  \bibinfo {pages} {223601} (\bibinfo {year} {2019})}\BibitemShut {NoStop}%
\bibitem [{\citenamefont {Conangla}\ \emph {et~al.}(2019)\citenamefont
  {Conangla}, \citenamefont {Ricci}, \citenamefont {Cuairan}, \citenamefont
  {Schell}, \citenamefont {Meyer},\ and\ \citenamefont
  {Quidant}}]{conangla2019optimalfeedback}%
  \BibitemOpen
  \bibfield  {author} {\bibinfo {author} {\bibfnamefont {G.~P.}\ \bibnamefont
  {Conangla}}, \bibinfo {author} {\bibfnamefont {F.}~\bibnamefont {Ricci}},
  \bibinfo {author} {\bibfnamefont {M.~T.}\ \bibnamefont {Cuairan}}, \bibinfo
  {author} {\bibfnamefont {A.~W.}\ \bibnamefont {Schell}}, \bibinfo {author}
  {\bibfnamefont {N.}~\bibnamefont {Meyer}},\ and\ \bibinfo {author}
  {\bibfnamefont {R.}~\bibnamefont {Quidant}},\ }\bibfield  {title} {\bibinfo
  {title} {Optimal feedback cooling of a charged levitated nanoparticle with
  adaptive control},\ }\href {https://doi.org/10.1103/PhysRevLett.122.223602}
  {\bibfield  {journal} {\bibinfo  {journal} {Phys. Rev. Lett.}\ }\textbf
  {\bibinfo {volume} {122}},\ \bibinfo {pages} {223602} (\bibinfo {year}
  {2019})}\BibitemShut {NoStop}%
\bibitem [{\citenamefont {Gieseler}\ \emph {et~al.}(2012)\citenamefont
  {Gieseler}, \citenamefont {Deutsch}, \citenamefont {Quidant},\ and\
  \citenamefont {Novotny}}]{gieseler2012subkelvin}%
  \BibitemOpen
  \bibfield  {author} {\bibinfo {author} {\bibfnamefont {J.}~\bibnamefont
  {Gieseler}}, \bibinfo {author} {\bibfnamefont {B.}~\bibnamefont {Deutsch}},
  \bibinfo {author} {\bibfnamefont {R.}~\bibnamefont {Quidant}},\ and\ \bibinfo
  {author} {\bibfnamefont {L.}~\bibnamefont {Novotny}},\ }\bibfield  {title}
  {\bibinfo {title} {Subkelvin parametric feedback cooling of a laser-trapped
  nanoparticle},\ }\href {https://doi.org/10.1103/PhysRevLett.109.103603}
  {\bibfield  {journal} {\bibinfo  {journal} {Phys. Rev. Lett.}\ }\textbf
  {\bibinfo {volume} {109}},\ \bibinfo {pages} {103603} (\bibinfo {year}
  {2012})}\BibitemShut {NoStop}%
\bibitem [{\citenamefont {Li}\ \emph {et~al.}(2011)\citenamefont {Li},
  \citenamefont {Kheifets},\ and\ \citenamefont {Raizen}}]{li2011millikelvin}%
  \BibitemOpen
  \bibfield  {author} {\bibinfo {author} {\bibfnamefont {T.}~\bibnamefont
  {Li}}, \bibinfo {author} {\bibfnamefont {S.}~\bibnamefont {Kheifets}},\ and\
  \bibinfo {author} {\bibfnamefont {M.~G.}\ \bibnamefont {Raizen}},\ }\bibfield
   {title} {\bibinfo {title} {Millikelvin cooling of an optically trapped
  microsphere in vacuum}\ }(\bibinfo  {publisher} {Nature Publishing Group},\
  \bibinfo {year} {2011})\ pp.\ \bibinfo {pages} {527--530}\BibitemShut
  {NoStop}%
\bibitem [{\citenamefont {Reimann}\ \emph {et~al.}(2018)\citenamefont
  {Reimann}, \citenamefont {Doderer}, \citenamefont {Hebestreit}, \citenamefont
  {Diehl}, \citenamefont {Frimmer}, \citenamefont {Windey}, \citenamefont
  {Tebbenjohanns},\ and\ \citenamefont {Novotny}}]{reinmann2018ghzrotation}%
  \BibitemOpen
  \bibfield  {author} {\bibinfo {author} {\bibfnamefont {R.}~\bibnamefont
  {Reimann}}, \bibinfo {author} {\bibfnamefont {M.}~\bibnamefont {Doderer}},
  \bibinfo {author} {\bibfnamefont {E.}~\bibnamefont {Hebestreit}}, \bibinfo
  {author} {\bibfnamefont {R.}~\bibnamefont {Diehl}}, \bibinfo {author}
  {\bibfnamefont {M.}~\bibnamefont {Frimmer}}, \bibinfo {author} {\bibfnamefont
  {D.}~\bibnamefont {Windey}}, \bibinfo {author} {\bibfnamefont
  {F.}~\bibnamefont {Tebbenjohanns}},\ and\ \bibinfo {author} {\bibfnamefont
  {L.}~\bibnamefont {Novotny}},\ }\bibfield  {title} {\bibinfo {title} {Ghz
  rotation of an optically trapped nanoparticle in vacuum},\ }\href
  {https://doi.org/10.1103/PhysRevLett.121.033602} {\bibfield  {journal}
  {\bibinfo  {journal} {Phys. Rev. Lett.}\ }\textbf {\bibinfo {volume} {121}},\
  \bibinfo {pages} {033602} (\bibinfo {year} {2018})}\BibitemShut {NoStop}%
\bibitem [{\citenamefont {Ahn}\ \emph {et~al.}(2018)\citenamefont {Ahn},
  \citenamefont {Xu}, \citenamefont {Bang}, \citenamefont {Deng}, \citenamefont
  {Hoang}, \citenamefont {Han}, \citenamefont {Ma},\ and\ \citenamefont
  {Li}}]{ahn2018opticallylevitated}%
  \BibitemOpen
  \bibfield  {author} {\bibinfo {author} {\bibfnamefont {J.}~\bibnamefont
  {Ahn}}, \bibinfo {author} {\bibfnamefont {Z.}~\bibnamefont {Xu}}, \bibinfo
  {author} {\bibfnamefont {J.}~\bibnamefont {Bang}}, \bibinfo {author}
  {\bibfnamefont {Y.-H.}\ \bibnamefont {Deng}}, \bibinfo {author}
  {\bibfnamefont {T.~M.}\ \bibnamefont {Hoang}}, \bibinfo {author}
  {\bibfnamefont {Q.}~\bibnamefont {Han}}, \bibinfo {author} {\bibfnamefont
  {R.-M.}\ \bibnamefont {Ma}},\ and\ \bibinfo {author} {\bibfnamefont
  {T.}~\bibnamefont {Li}},\ }\bibfield  {title} {\bibinfo {title} {Optically
  levitated nanodumbbell torsion balance and ghz nanomechanical rotor},\ }\href
  {https://doi.org/10.1103/PhysRevLett.121.033603} {\bibfield  {journal}
  {\bibinfo  {journal} {Phys. Rev. Lett.}\ }\textbf {\bibinfo {volume} {121}},\
  \bibinfo {pages} {033603} (\bibinfo {year} {2018})}\BibitemShut {NoStop}%
\bibitem [{\citenamefont {Chang}\ \emph {et~al.}(2010)\citenamefont {Chang},
  \citenamefont {Regal}, \citenamefont {Papp}, \citenamefont {Wilson},
  \citenamefont {Ye}, \citenamefont {Painter}, \citenamefont {Kimble},\ and\
  \citenamefont {Zoller}}]{chang2010cavity}%
  \BibitemOpen
  \bibfield  {author} {\bibinfo {author} {\bibfnamefont {D.~E.}\ \bibnamefont
  {Chang}}, \bibinfo {author} {\bibfnamefont {C.}~\bibnamefont {Regal}},
  \bibinfo {author} {\bibfnamefont {S.}~\bibnamefont {Papp}}, \bibinfo {author}
  {\bibfnamefont {D.}~\bibnamefont {Wilson}}, \bibinfo {author} {\bibfnamefont
  {J.}~\bibnamefont {Ye}}, \bibinfo {author} {\bibfnamefont {O.}~\bibnamefont
  {Painter}}, \bibinfo {author} {\bibfnamefont {H.~J.}\ \bibnamefont
  {Kimble}},\ and\ \bibinfo {author} {\bibfnamefont {P.}~\bibnamefont
  {Zoller}},\ }\bibfield  {title} {\bibinfo {title} {Cavity opto-mechanics
  using an optically levitated nanosphere},\ }\href
  {https://www.pnas.org/content/107/3/1005} {\bibfield  {journal} {\bibinfo
  {journal} {Proceedings of the National Academy of Sciences}\ }\textbf
  {\bibinfo {volume} {107}},\ \bibinfo {pages} {1005} (\bibinfo {year}
  {2010})}\BibitemShut {NoStop}%
\bibitem [{\citenamefont {Romero-Isart}\ \emph {et~al.}(2011)\citenamefont
  {Romero-Isart}, \citenamefont {Pflanzer}, \citenamefont {Juan}, \citenamefont
  {Quidant}, \citenamefont {Kiesel}, \citenamefont {Aspelmeyer},\ and\
  \citenamefont {Cirac}}]{romero2011optically}%
  \BibitemOpen
  \bibfield  {author} {\bibinfo {author} {\bibfnamefont {O.}~\bibnamefont
  {Romero-Isart}}, \bibinfo {author} {\bibfnamefont {A.~C.}\ \bibnamefont
  {Pflanzer}}, \bibinfo {author} {\bibfnamefont {M.~L.}\ \bibnamefont {Juan}},
  \bibinfo {author} {\bibfnamefont {R.}~\bibnamefont {Quidant}}, \bibinfo
  {author} {\bibfnamefont {N.}~\bibnamefont {Kiesel}}, \bibinfo {author}
  {\bibfnamefont {M.}~\bibnamefont {Aspelmeyer}},\ and\ \bibinfo {author}
  {\bibfnamefont {J.~I.}\ \bibnamefont {Cirac}},\ }\bibfield  {title} {\bibinfo
  {title} {Optically levitating dielectrics in the quantum regime: Theory and
  protocols},\ }\href
  {https://journals.aps.org/pra/abstract/10.1103/PhysRevA.83.013803} {\bibfield
   {journal} {\bibinfo  {journal} {Phys. Rev. A}\ }\textbf {\bibinfo {volume}
  {83}},\ \bibinfo {pages} {013803} (\bibinfo {year} {2011})}\BibitemShut
  {NoStop}%
\bibitem [{\citenamefont {Romero-Isart}\ \emph {et~al.}(2010)\citenamefont
  {Romero-Isart}, \citenamefont {Juan}, \citenamefont {Quidant},\ and\
  \citenamefont {Cirac}}]{romero2010toward}%
  \BibitemOpen
  \bibfield  {author} {\bibinfo {author} {\bibfnamefont {O.}~\bibnamefont
  {Romero-Isart}}, \bibinfo {author} {\bibfnamefont {M.~L.}\ \bibnamefont
  {Juan}}, \bibinfo {author} {\bibfnamefont {R.}~\bibnamefont {Quidant}},\ and\
  \bibinfo {author} {\bibfnamefont {J.~I.}\ \bibnamefont {Cirac}},\ }\bibfield
  {title} {\bibinfo {title} {Toward quantum superposition of living
  organisms},\ }\href
  {https://iopscience.iop.org/article/10.1088/1367-2630/12/3/033015/meta}
  {\bibfield  {journal} {\bibinfo  {journal} {New Journal of Physics}\ }\textbf
  {\bibinfo {volume} {12}},\ \bibinfo {pages} {033015} (\bibinfo {year}
  {2010})}\BibitemShut {NoStop}%
\bibitem [{\citenamefont {Bateman}\ \emph {et~al.}(2014)\citenamefont
  {Bateman}, \citenamefont {Nimmrichter}, \citenamefont {Hornberger},\ and\
  \citenamefont {Ulbricht}}]{bateman2014near}%
  \BibitemOpen
  \bibfield  {author} {\bibinfo {author} {\bibfnamefont {J.}~\bibnamefont
  {Bateman}}, \bibinfo {author} {\bibfnamefont {S.}~\bibnamefont
  {Nimmrichter}}, \bibinfo {author} {\bibfnamefont {K.}~\bibnamefont
  {Hornberger}},\ and\ \bibinfo {author} {\bibfnamefont {H.}~\bibnamefont
  {Ulbricht}},\ }\bibfield  {title} {\bibinfo {title} {Near-field
  interferometry of a free-falling nanoparticle from a point-like source},\
  }\href {https://www.nature.com/articles/ncomms5788} {\bibfield  {journal}
  {\bibinfo  {journal} {Nat. Comm.}\ }\textbf {\bibinfo {volume} {5}},\
  \bibinfo {pages} {1} (\bibinfo {year} {2014})}\BibitemShut {NoStop}%
\bibitem [{\citenamefont {Scala}\ \emph {et~al.}(2013)\citenamefont {Scala},
  \citenamefont {Kim}, \citenamefont {Morley}, \citenamefont {Barker},\ and\
  \citenamefont {Bose}}]{scala2013matter}%
  \BibitemOpen
  \bibfield  {author} {\bibinfo {author} {\bibfnamefont {M.}~\bibnamefont
  {Scala}}, \bibinfo {author} {\bibfnamefont {M.}~\bibnamefont {Kim}}, \bibinfo
  {author} {\bibfnamefont {G.}~\bibnamefont {Morley}}, \bibinfo {author}
  {\bibfnamefont {P.}~\bibnamefont {Barker}},\ and\ \bibinfo {author}
  {\bibfnamefont {S.}~\bibnamefont {Bose}},\ }\bibfield  {title} {\bibinfo
  {title} {Matter-wave interferometry of a levitated thermal nano-oscillator
  induced and probed by a spin},\ }\href
  {https://journals.aps.org/prl/abstract/10.1103/PhysRevLett.111.180403}
  {\bibfield  {journal} {\bibinfo  {journal} {Phys. Rev. Lett.}\ }\textbf
  {\bibinfo {volume} {111}},\ \bibinfo {pages} {180403} (\bibinfo {year}
  {2013})}\BibitemShut {NoStop}%
\bibitem [{\citenamefont {Magrini}\ \emph {et~al.}(2021)\citenamefont
  {Magrini}, \citenamefont {Rosenzweig}, \citenamefont {Bach}, \citenamefont
  {Deutschmann-Olek}, \citenamefont {Hofer}, \citenamefont {Hong},
  \citenamefont {Kiesel}, \citenamefont {Kugi},\ and\ \citenamefont
  {Aspelmeyer}}]{magrini2021real}%
  \BibitemOpen
  \bibfield  {author} {\bibinfo {author} {\bibfnamefont {L.}~\bibnamefont
  {Magrini}}, \bibinfo {author} {\bibfnamefont {P.}~\bibnamefont {Rosenzweig}},
  \bibinfo {author} {\bibfnamefont {C.}~\bibnamefont {Bach}}, \bibinfo {author}
  {\bibfnamefont {A.}~\bibnamefont {Deutschmann-Olek}}, \bibinfo {author}
  {\bibfnamefont {S.~G.}\ \bibnamefont {Hofer}}, \bibinfo {author}
  {\bibfnamefont {S.}~\bibnamefont {Hong}}, \bibinfo {author} {\bibfnamefont
  {N.}~\bibnamefont {Kiesel}}, \bibinfo {author} {\bibfnamefont
  {A.}~\bibnamefont {Kugi}},\ and\ \bibinfo {author} {\bibfnamefont
  {M.}~\bibnamefont {Aspelmeyer}},\ }\bibfield  {title} {\bibinfo {title}
  {Real-time optimal quantum control of mechanical motion at room
  temperature},\ }\href {https://www.nature.com/articles/s41586-021-03602-3}
  {\bibfield  {journal} {\bibinfo  {journal} {Nature}\ }\textbf {\bibinfo
  {volume} {595}},\ \bibinfo {pages} {373} (\bibinfo {year}
  {2021})}\BibitemShut {NoStop}%
\bibitem [{\citenamefont {Tebbenjohanns}\ \emph {et~al.}(2021)\citenamefont
  {Tebbenjohanns}, \citenamefont {Mattana}, \citenamefont {Rossi},
  \citenamefont {Frimmer},\ and\ \citenamefont
  {Novotny}}]{tebbenjohanns2021quantum}%
  \BibitemOpen
  \bibfield  {author} {\bibinfo {author} {\bibfnamefont {F.}~\bibnamefont
  {Tebbenjohanns}}, \bibinfo {author} {\bibfnamefont {M.~L.}\ \bibnamefont
  {Mattana}}, \bibinfo {author} {\bibfnamefont {M.}~\bibnamefont {Rossi}},
  \bibinfo {author} {\bibfnamefont {M.}~\bibnamefont {Frimmer}},\ and\ \bibinfo
  {author} {\bibfnamefont {L.}~\bibnamefont {Novotny}},\ }\bibfield  {title}
  {\bibinfo {title} {Quantum control of a nanoparticle optically levitated in
  cryogenic free space},\ }\href
  {https://www.nature.com/articles/s41586-021-03617-w} {\bibfield  {journal}
  {\bibinfo  {journal} {Nature}\ }\textbf {\bibinfo {volume} {595}},\ \bibinfo
  {pages} {378} (\bibinfo {year} {2021})}\BibitemShut {NoStop}%
\bibitem [{\citenamefont {Einstein}\ and\ \citenamefont
  {De~Haas}(1915)}]{einstein_de_haas}%
  \BibitemOpen
  \bibfield  {author} {\bibinfo {author} {\bibfnamefont {A.}~\bibnamefont
  {Einstein}}\ and\ \bibinfo {author} {\bibfnamefont {W.}~\bibnamefont
  {De~Haas}},\ }\bibfield  {title} {\bibinfo {title} {Experimental proof of the
  existence of amperes molecular currents},\ }in\ \href@noop {} {\emph
  {\bibinfo {booktitle} {Proc. KNAW}}},\ Vol.\ \bibinfo {volume} {181}\
  (\bibinfo {year} {1915})\ p.\ \bibinfo {pages} {696}\BibitemShut {NoStop}%
\bibitem [{\citenamefont {Barnett}(1915)}]{barnett1915magnetization}%
  \BibitemOpen
  \bibfield  {author} {\bibinfo {author} {\bibfnamefont {S.~J.}\ \bibnamefont
  {Barnett}},\ }\bibfield  {title} {\bibinfo {title} {Magnetization by
  rotation},\ }\href {https://doi.org/10.1103/PhysRev.6.239} {\bibfield
  {journal} {\bibinfo  {journal} {Phys. Rev.}\ }\textbf {\bibinfo {volume}
  {6}},\ \bibinfo {pages} {239} (\bibinfo {year} {1915})}\BibitemShut {NoStop}%
\bibitem [{\citenamefont {Chikazumi}\ and\ \citenamefont
  {Graham}(2009)}]{chikazumi2009physics}%
  \BibitemOpen
  \bibfield  {author} {\bibinfo {author} {\bibfnamefont {S.}~\bibnamefont
  {Chikazumi}}\ and\ \bibinfo {author} {\bibfnamefont {C.~D.}\ \bibnamefont
  {Graham}},\ }\href@noop {} {\emph {\bibinfo {title} {Physics of
  Ferromagnetism}}},\ \bibinfo {number} {94}\ (\bibinfo  {publisher} {Oxford
  University Press},\ \bibinfo {year} {2009})\BibitemShut {NoStop}%
\bibitem [{\citenamefont {Romero-Isart}\ \emph {et~al.}(2012)\citenamefont
  {Romero-Isart}, \citenamefont {Clemente}, \citenamefont {Navau},
  \citenamefont {Sanchez},\ and\ \citenamefont {Cirac}}]{romero2012quantum}%
  \BibitemOpen
  \bibfield  {author} {\bibinfo {author} {\bibfnamefont {O.}~\bibnamefont
  {Romero-Isart}}, \bibinfo {author} {\bibfnamefont {L.}~\bibnamefont
  {Clemente}}, \bibinfo {author} {\bibfnamefont {C.}~\bibnamefont {Navau}},
  \bibinfo {author} {\bibfnamefont {A.}~\bibnamefont {Sanchez}},\ and\ \bibinfo
  {author} {\bibfnamefont {J.}~\bibnamefont {Cirac}},\ }\bibfield  {title}
  {\bibinfo {title} {Quantum magnetomechanics with levitating superconducting
  microspheres},\ }\href
  {https://journals.aps.org/prl/abstract/10.1103/PhysRevLett.109.147205}
  {\bibfield  {journal} {\bibinfo  {journal} {Phys. Rev. Lett.}\ }\textbf
  {\bibinfo {volume} {109}},\ \bibinfo {pages} {147205} (\bibinfo {year}
  {2012})}\BibitemShut {NoStop}%
\bibitem [{\citenamefont {Wang}\ \emph {et~al.}(2019)\citenamefont {Wang},
  \citenamefont {Lourette}, \citenamefont {O'Kelley}, \citenamefont {Kayci},
  \citenamefont {Band}, \citenamefont {Kimball}, \citenamefont {Sushkov},\ and\
  \citenamefont {Budker}}]{wang2019dynamics}%
  \BibitemOpen
  \bibfield  {author} {\bibinfo {author} {\bibfnamefont {T.}~\bibnamefont
  {Wang}}, \bibinfo {author} {\bibfnamefont {S.}~\bibnamefont {Lourette}},
  \bibinfo {author} {\bibfnamefont {S.~R.}\ \bibnamefont {O'Kelley}}, \bibinfo
  {author} {\bibfnamefont {M.}~\bibnamefont {Kayci}}, \bibinfo {author}
  {\bibfnamefont {Y.}~\bibnamefont {Band}}, \bibinfo {author} {\bibfnamefont
  {D.~F.~J.}\ \bibnamefont {Kimball}}, \bibinfo {author} {\bibfnamefont
  {A.~O.}\ \bibnamefont {Sushkov}},\ and\ \bibinfo {author} {\bibfnamefont
  {D.}~\bibnamefont {Budker}},\ }\bibfield  {title} {\bibinfo {title} {Dynamics
  of a ferromagnetic particle levitated over a superconductor},\ }\href
  {https://doi.org/10.1103/PhysRevApplied.11.044041} {\bibfield  {journal}
  {\bibinfo  {journal} {Phys. Rev. Appl.}\ }\textbf {\bibinfo {volume} {11}},\
  \bibinfo {pages} {044041} (\bibinfo {year} {2019})}\BibitemShut {NoStop}%
\bibitem [{\citenamefont {Timberlake}\ \emph {et~al.}(2019)\citenamefont
  {Timberlake}, \citenamefont {Gasbarri}, \citenamefont {Vinante},
  \citenamefont {Setter},\ and\ \citenamefont
  {Ulbricht}}]{timberlake2019acceleration}%
  \BibitemOpen
  \bibfield  {author} {\bibinfo {author} {\bibfnamefont {C.}~\bibnamefont
  {Timberlake}}, \bibinfo {author} {\bibfnamefont {G.}~\bibnamefont
  {Gasbarri}}, \bibinfo {author} {\bibfnamefont {A.}~\bibnamefont {Vinante}},
  \bibinfo {author} {\bibfnamefont {A.}~\bibnamefont {Setter}},\ and\ \bibinfo
  {author} {\bibfnamefont {H.}~\bibnamefont {Ulbricht}},\ }\bibfield  {title}
  {\bibinfo {title} {Acceleration sensing with magnetically levitated
  oscillators above a superconductor},\ }\href
  {https://aip.scitation.org/doi/10.1063/1.5129145} {\bibfield  {journal}
  {\bibinfo  {journal} {Appl. Phys. Lett.}\ }\textbf {\bibinfo {volume}
  {115}},\ \bibinfo {pages} {224101} (\bibinfo {year} {2019})}\BibitemShut
  {NoStop}%
\bibitem [{\citenamefont {Lewandowski}\ \emph {et~al.}(2021)\citenamefont
  {Lewandowski}, \citenamefont {Knowles}, \citenamefont {Etienne},\ and\
  \citenamefont {D'Urso}}]{lewandowski2021high}%
  \BibitemOpen
  \bibfield  {author} {\bibinfo {author} {\bibfnamefont {C.~W.}\ \bibnamefont
  {Lewandowski}}, \bibinfo {author} {\bibfnamefont {T.~D.}\ \bibnamefont
  {Knowles}}, \bibinfo {author} {\bibfnamefont {Z.~B.}\ \bibnamefont
  {Etienne}},\ and\ \bibinfo {author} {\bibfnamefont {B.}~\bibnamefont
  {D'Urso}},\ }\bibfield  {title} {\bibinfo {title} {High-sensitivity
  accelerometry with a feedback-cooled magnetically levitated microsphere},\
  }\href
  {https://journals.aps.org/prapplied/abstract/10.1103/PhysRevApplied.15.014050}
  {\bibfield  {journal} {\bibinfo  {journal} {Phys. Rev. Appl.}\ }\textbf
  {\bibinfo {volume} {15}},\ \bibinfo {pages} {014050} (\bibinfo {year}
  {2021})}\BibitemShut {NoStop}%
\bibitem [{\citenamefont {Slezak}\ \emph {et~al.}(2018)\citenamefont {Slezak},
  \citenamefont {Lewandowski}, \citenamefont {Hsu},\ and\ \citenamefont
  {D'Urso}}]{slezak2018cooling}%
  \BibitemOpen
  \bibfield  {author} {\bibinfo {author} {\bibfnamefont {B.~R.}\ \bibnamefont
  {Slezak}}, \bibinfo {author} {\bibfnamefont {C.~W.}\ \bibnamefont
  {Lewandowski}}, \bibinfo {author} {\bibfnamefont {J.-F.}\ \bibnamefont
  {Hsu}},\ and\ \bibinfo {author} {\bibfnamefont {B.}~\bibnamefont {D'Urso}},\
  }\bibfield  {title} {\bibinfo {title} {Cooling the motion of a silica
  microsphere in a magneto-gravitational trap in ultra-high vacuum},\ }\href
  {https://iopscience.iop.org/article/10.1088/1367-2630/aacac1/pdf} {\bibfield
  {journal} {\bibinfo  {journal} {New Journal of Physics}\ }\textbf {\bibinfo
  {volume} {20}},\ \bibinfo {pages} {063028} (\bibinfo {year}
  {2018})}\BibitemShut {NoStop}%
\bibitem [{\citenamefont {Hsu}\ \emph {et~al.}(2016)\citenamefont {Hsu},
  \citenamefont {Ji}, \citenamefont {Lewandowski},\ and\ \citenamefont
  {D'Urso}}]{hsu2016cooling}%
  \BibitemOpen
  \bibfield  {author} {\bibinfo {author} {\bibfnamefont {J.-F.}\ \bibnamefont
  {Hsu}}, \bibinfo {author} {\bibfnamefont {P.}~\bibnamefont {Ji}}, \bibinfo
  {author} {\bibfnamefont {C.~W.}\ \bibnamefont {Lewandowski}},\ and\ \bibinfo
  {author} {\bibfnamefont {B.}~\bibnamefont {D'Urso}},\ }\bibfield  {title}
  {\bibinfo {title} {Cooling the motion of diamond nanocrystals in a
  magneto-gravitational trap in high vacuum},\ }\href
  {https://www.nature.com/articles/srep30125.pdf} {\bibfield  {journal}
  {\bibinfo  {journal} {Scientific Reports}\ }\textbf {\bibinfo {volume} {6}},\
  \bibinfo {pages} {1} (\bibinfo {year} {2016})}\BibitemShut {NoStop}%
\bibitem [{\citenamefont {Prat-Camps}\ \emph {et~al.}(2017)\citenamefont
  {Prat-Camps}, \citenamefont {Teo}, \citenamefont {Rusconi}, \citenamefont
  {Wieczorek},\ and\ \citenamefont {Romero-Isart}}]{prat2017ultrasensitive}%
  \BibitemOpen
  \bibfield  {author} {\bibinfo {author} {\bibfnamefont {J.}~\bibnamefont
  {Prat-Camps}}, \bibinfo {author} {\bibfnamefont {C.}~\bibnamefont {Teo}},
  \bibinfo {author} {\bibfnamefont {C.}~\bibnamefont {Rusconi}}, \bibinfo
  {author} {\bibfnamefont {W.}~\bibnamefont {Wieczorek}},\ and\ \bibinfo
  {author} {\bibfnamefont {O.}~\bibnamefont {Romero-Isart}},\ }\bibfield
  {title} {\bibinfo {title} {Ultrasensitive inertial and force sensors with
  diamagnetically levitated magnets},\ }\href
  {https://journals.aps.org/prapplied/abstract/10.1103/PhysRevApplied.8.034002}
  {\bibfield  {journal} {\bibinfo  {journal} {Phys. Rev. Appl.}\ }\textbf
  {\bibinfo {volume} {8}},\ \bibinfo {pages} {034002} (\bibinfo {year}
  {2017})}\BibitemShut {NoStop}%
\bibitem [{\citenamefont {Gieseler}\ \emph {et~al.}(2020)\citenamefont
  {Gieseler}, \citenamefont {Kabcenell}, \citenamefont {Rosenfeld},
  \citenamefont {Schaefer}, \citenamefont {Safira}, \citenamefont {Schuetz},
  \citenamefont {Gonzalez-Ballestero}, \citenamefont {Rusconi}, \citenamefont
  {Romero-Isart},\ and\ \citenamefont {Lukin}}]{gieseler2020single}%
  \BibitemOpen
  \bibfield  {author} {\bibinfo {author} {\bibfnamefont {J.}~\bibnamefont
  {Gieseler}}, \bibinfo {author} {\bibfnamefont {A.}~\bibnamefont {Kabcenell}},
  \bibinfo {author} {\bibfnamefont {E.}~\bibnamefont {Rosenfeld}}, \bibinfo
  {author} {\bibfnamefont {J.}~\bibnamefont {Schaefer}}, \bibinfo {author}
  {\bibfnamefont {A.}~\bibnamefont {Safira}}, \bibinfo {author} {\bibfnamefont
  {M.~J.}\ \bibnamefont {Schuetz}}, \bibinfo {author} {\bibfnamefont
  {C.}~\bibnamefont {Gonzalez-Ballestero}}, \bibinfo {author} {\bibfnamefont
  {C.~C.}\ \bibnamefont {Rusconi}}, \bibinfo {author} {\bibfnamefont
  {O.}~\bibnamefont {Romero-Isart}},\ and\ \bibinfo {author} {\bibfnamefont
  {M.~D.}\ \bibnamefont {Lukin}},\ }\bibfield  {title} {\bibinfo {title}
  {Single-spin magnetomechanics with levitated micromagnets},\ }\href
  {https://journals.aps.org/prl/abstract/10.1103/PhysRevLett.124.163604}
  {\bibfield  {journal} {\bibinfo  {journal} {Phys. Rev. Lett.}\ }\textbf
  {\bibinfo {volume} {124}},\ \bibinfo {pages} {163604} (\bibinfo {year}
  {2020})}\BibitemShut {NoStop}%
\bibitem [{\citenamefont {O'Brien}\ \emph {et~al.}(2019)\citenamefont
  {O'Brien}, \citenamefont {Dunn}, \citenamefont {Downes},\ and\ \citenamefont
  {Twamley}}]{o2019magneto}%
  \BibitemOpen
  \bibfield  {author} {\bibinfo {author} {\bibfnamefont {M.}~\bibnamefont
  {O'Brien}}, \bibinfo {author} {\bibfnamefont {S.}~\bibnamefont {Dunn}},
  \bibinfo {author} {\bibfnamefont {J.}~\bibnamefont {Downes}},\ and\ \bibinfo
  {author} {\bibfnamefont {J.}~\bibnamefont {Twamley}},\ }\bibfield  {title}
  {\bibinfo {title} {Magneto-mechanical trapping of micro-diamonds at low
  pressures},\ }\href {https://aip.scitation.org/doi/10.1063/1.5066065}
  {\bibfield  {journal} {\bibinfo  {journal} {Appl. Phys. Lett.}\ }\textbf
  {\bibinfo {volume} {114}},\ \bibinfo {pages} {053103} (\bibinfo {year}
  {2019})}\BibitemShut {NoStop}%
\bibitem [{\citenamefont {Houlton}\ \emph {et~al.}(2018)\citenamefont
  {Houlton}, \citenamefont {Chen}, \citenamefont {Brubaker}, \citenamefont
  {Bertness},\ and\ \citenamefont {Rogers}}]{houlton2018axisymmetric}%
  \BibitemOpen
  \bibfield  {author} {\bibinfo {author} {\bibfnamefont {J.}~\bibnamefont
  {Houlton}}, \bibinfo {author} {\bibfnamefont {M.}~\bibnamefont {Chen}},
  \bibinfo {author} {\bibfnamefont {M.}~\bibnamefont {Brubaker}}, \bibinfo
  {author} {\bibfnamefont {K.}~\bibnamefont {Bertness}},\ and\ \bibinfo
  {author} {\bibfnamefont {C.}~\bibnamefont {Rogers}},\ }\bibfield  {title}
  {\bibinfo {title} {Axisymmetric scalable magneto-gravitational trap for
  diamagnetic particle levitation},\ }\href
  {https://aip.scitation.org/doi/10.1063/1.5051667} {\bibfield  {journal}
  {\bibinfo  {journal} {Review of Scientific Instruments}\ }\textbf {\bibinfo
  {volume} {89}},\ \bibinfo {pages} {125107} (\bibinfo {year}
  {2018})}\BibitemShut {NoStop}%
\bibitem [{\citenamefont {Johnsson}\ \emph {et~al.}(2016)\citenamefont
  {Johnsson}, \citenamefont {Brennen},\ and\ \citenamefont
  {Twamley}}]{johnsson2016macroscopic}%
  \BibitemOpen
  \bibfield  {author} {\bibinfo {author} {\bibfnamefont {M.~T.}\ \bibnamefont
  {Johnsson}}, \bibinfo {author} {\bibfnamefont {G.~K.}\ \bibnamefont
  {Brennen}},\ and\ \bibinfo {author} {\bibfnamefont {J.}~\bibnamefont
  {Twamley}},\ }\bibfield  {title} {\bibinfo {title} {Macroscopic
  superpositions and gravimetry with quantum magnetomechanics},\ }\href
  {https://www.nature.com/articles/srep37495.pdf} {\bibfield  {journal}
  {\bibinfo  {journal} {Scientific Reports}\ }\textbf {\bibinfo {volume} {6}},\
  \bibinfo {pages} {1} (\bibinfo {year} {2016})}\BibitemShut {NoStop}%
\bibitem [{\citenamefont {Cirio}\ \emph {et~al.}(2012)\citenamefont {Cirio},
  \citenamefont {Brennen},\ and\ \citenamefont {Twamley}}]{cirio2012quantum}%
  \BibitemOpen
  \bibfield  {author} {\bibinfo {author} {\bibfnamefont {M.}~\bibnamefont
  {Cirio}}, \bibinfo {author} {\bibfnamefont {G.}~\bibnamefont {Brennen}},\
  and\ \bibinfo {author} {\bibfnamefont {J.}~\bibnamefont {Twamley}},\
  }\bibfield  {title} {\bibinfo {title} {Quantum magnetomechanics:
  ultrahigh-q-levitated mechanical oscillators},\ }\href
  {https://journals.aps.org/prl/abstract/10.1103/PhysRevLett.109.147206}
  {\bibfield  {journal} {\bibinfo  {journal} {Phys. Rev. Lett.}\ }\textbf
  {\bibinfo {volume} {109}},\ \bibinfo {pages} {147206} (\bibinfo {year}
  {2012})}\BibitemShut {NoStop}%
\bibitem [{\citenamefont {Vinante}\ \emph {et~al.}(2020)\citenamefont
  {Vinante}, \citenamefont {Falferi}, \citenamefont {Gasbarri}, \citenamefont
  {Setter}, \citenamefont {Timberlake},\ and\ \citenamefont
  {Ulbricht}}]{vinante2020ultralow}%
  \BibitemOpen
  \bibfield  {author} {\bibinfo {author} {\bibfnamefont {A.}~\bibnamefont
  {Vinante}}, \bibinfo {author} {\bibfnamefont {P.}~\bibnamefont {Falferi}},
  \bibinfo {author} {\bibfnamefont {G.}~\bibnamefont {Gasbarri}}, \bibinfo
  {author} {\bibfnamefont {A.}~\bibnamefont {Setter}}, \bibinfo {author}
  {\bibfnamefont {C.}~\bibnamefont {Timberlake}},\ and\ \bibinfo {author}
  {\bibfnamefont {H.}~\bibnamefont {Ulbricht}},\ }\bibfield  {title} {\bibinfo
  {title} {Ultralow mechanical damping with meissner-levitated ferromagnetic
  microparticles},\ }\href
  {https://journals.aps.org/prapplied/abstract/10.1103/PhysRevApplied.13.064027}
  {\bibfield  {journal} {\bibinfo  {journal} {Phys. Rev. Appl.}\ }\textbf
  {\bibinfo {volume} {13}},\ \bibinfo {pages} {064027} (\bibinfo {year}
  {2020})}\BibitemShut {NoStop}%
\bibitem [{\citenamefont {Perdriat}\ \emph {et~al.}(2021)\citenamefont
  {Perdriat}, \citenamefont {Pellet-Mary}, \citenamefont {Huillery},
  \citenamefont {Rondin},\ and\ \citenamefont {H{\'e}tet}}]{perdriat2021spin}%
  \BibitemOpen
  \bibfield  {author} {\bibinfo {author} {\bibfnamefont {M.}~\bibnamefont
  {Perdriat}}, \bibinfo {author} {\bibfnamefont {C.}~\bibnamefont
  {Pellet-Mary}}, \bibinfo {author} {\bibfnamefont {P.}~\bibnamefont
  {Huillery}}, \bibinfo {author} {\bibfnamefont {L.}~\bibnamefont {Rondin}},\
  and\ \bibinfo {author} {\bibfnamefont {G.}~\bibnamefont {H{\'e}tet}},\
  }\bibfield  {title} {\bibinfo {title} {Spin-mechanics with nitrogen-vacancy
  centers and trapped particles},\ }\href {https://doi.org/10.3390/mi12060651}
  {\bibfield  {journal} {\bibinfo  {journal} {Micromachines}\ }\textbf
  {\bibinfo {volume} {12}},\ \bibinfo {pages} {651} (\bibinfo {year}
  {2021})}\BibitemShut {NoStop}%
\bibitem [{\citenamefont {Landau}\ \emph {et~al.}(1984)\citenamefont {Landau},
  \citenamefont {Pitaevskii},\ and\ \citenamefont
  {Lifshitz}}]{landau1984electrodynamics}%
  \BibitemOpen
  \bibfield  {author} {\bibinfo {author} {\bibfnamefont {L.}~\bibnamefont
  {Landau}}, \bibinfo {author} {\bibfnamefont {L.}~\bibnamefont {Pitaevskii}},\
  and\ \bibinfo {author} {\bibfnamefont {E.}~\bibnamefont {Lifshitz}},\
  }\href@noop {} {\emph {\bibinfo {title} {Electrodynamics of Continuous Media,
  volume 8 of}}}\ (\bibinfo  {publisher} {Butterworth-Heinemann},\ \bibinfo
  {year} {1984})\BibitemShut {NoStop}%
\bibitem [{\citenamefont {Stancil}\ and\ \citenamefont
  {Prabhakar}(2009)}]{stancil2009spin}%
  \BibitemOpen
  \bibfield  {author} {\bibinfo {author} {\bibfnamefont {D.~D.}\ \bibnamefont
  {Stancil}}\ and\ \bibinfo {author} {\bibfnamefont {A.}~\bibnamefont
  {Prabhakar}},\ }\href@noop {} {\emph {\bibinfo {title} {Spin waves}}},\
  Vol.~\bibinfo {volume} {5}\ (\bibinfo  {publisher} {Springer},\ \bibinfo
  {year} {2009})\BibitemShut {NoStop}%
\bibitem [{\citenamefont {Viola~Kusminskiy}\ \emph {et~al.}(2016)\citenamefont
  {Viola~Kusminskiy}, \citenamefont {Tang},\ and\ \citenamefont
  {Marquardt}}]{violakusminskiyCoupledSpinlightDynamics2016a}%
  \BibitemOpen
  \bibfield  {author} {\bibinfo {author} {\bibfnamefont {S.}~\bibnamefont
  {Viola~Kusminskiy}}, \bibinfo {author} {\bibfnamefont {H.~X.}\ \bibnamefont
  {Tang}},\ and\ \bibinfo {author} {\bibfnamefont {F.}~\bibnamefont
  {Marquardt}},\ }\bibfield  {title} {\bibinfo {title} {Coupled spin-light
  dynamics in cavity optomagnonics},\ }\href
  {https://journals.aps.org/pra/abstract/10.1103/PhysRevA.94.033821} {\bibfield
   {journal} {\bibinfo  {journal} {Phys. Rev. A}\ }\textbf {\bibinfo {volume}
  {94}} (\bibinfo {year} {2016})}\BibitemShut {NoStop}%
\bibitem [{\citenamefont {Liu}\ \emph {et~al.}(2016)\citenamefont {Liu},
  \citenamefont {Zhang}, \citenamefont {Tang},\ and\ \citenamefont
  {Flatt{\'e}}}]{liu2016optomagnonics}%
  \BibitemOpen
  \bibfield  {author} {\bibinfo {author} {\bibfnamefont {T.}~\bibnamefont
  {Liu}}, \bibinfo {author} {\bibfnamefont {X.}~\bibnamefont {Zhang}}, \bibinfo
  {author} {\bibfnamefont {H.~X.}\ \bibnamefont {Tang}},\ and\ \bibinfo
  {author} {\bibfnamefont {M.~E.}\ \bibnamefont {Flatt{\'e}}},\ }\bibfield
  {title} {\bibinfo {title} {Optomagnonics in magnetic solids},\ }\href
  {https://journals.aps.org/prb/abstract/10.1103/PhysRevB.94.060405} {\bibfield
   {journal} {\bibinfo  {journal} {Phys. Rev. B}\ }\textbf {\bibinfo {volume}
  {94}},\ \bibinfo {pages} {060405} (\bibinfo {year} {2016})}\BibitemShut
  {NoStop}%
\bibitem [{\citenamefont {Sharma}\ \emph {et~al.}(2017)\citenamefont {Sharma},
  \citenamefont {Blanter},\ and\ \citenamefont {Bauer}}]{sharma2017light}%
  \BibitemOpen
  \bibfield  {author} {\bibinfo {author} {\bibfnamefont {S.}~\bibnamefont
  {Sharma}}, \bibinfo {author} {\bibfnamefont {Y.~M.}\ \bibnamefont
  {Blanter}},\ and\ \bibinfo {author} {\bibfnamefont {G.~E.}\ \bibnamefont
  {Bauer}},\ }\bibfield  {title} {\bibinfo {title} {Light scattering by magnons
  in whispering gallery mode cavities},\ }\href
  {https://journals.aps.org/prb/abstract/10.1103/PhysRevB.96.094412} {\bibfield
   {journal} {\bibinfo  {journal} {Phys. Rev. B}\ }\textbf {\bibinfo {volume}
  {96}},\ \bibinfo {pages} {094412} (\bibinfo {year} {2017})}\BibitemShut
  {NoStop}%
\bibitem [{\citenamefont {Almpanis}(2018)}]{almpanis2018dielectric}%
  \BibitemOpen
  \bibfield  {author} {\bibinfo {author} {\bibfnamefont {E.}~\bibnamefont
  {Almpanis}},\ }\bibfield  {title} {\bibinfo {title} {Dielectric magnetic
  microparticles as photomagnonic cavities: Enhancing the modulation of
  near-infrared light by spin waves},\ }\href
  {https://doi.org/10.1103/PhysRevB.97.184406} {\bibfield  {journal} {\bibinfo
  {journal} {Phys. Rev. B}\ }\textbf {\bibinfo {volume} {97}},\ \bibinfo
  {pages} {184406} (\bibinfo {year} {2018})}\BibitemShut {NoStop}%
\bibitem [{\citenamefont {Almpanis}\ \emph {et~al.}(2020)\citenamefont
  {Almpanis}, \citenamefont {Zouros}, \citenamefont {Pantazopoulos},
  \citenamefont {Tsakmakidis}, \citenamefont {Papanikolaou},\ and\
  \citenamefont {Stefanou}}]{almpanis2020spherical}%
  \BibitemOpen
  \bibfield  {author} {\bibinfo {author} {\bibfnamefont {E.}~\bibnamefont
  {Almpanis}}, \bibinfo {author} {\bibfnamefont {G.}~\bibnamefont {Zouros}},
  \bibinfo {author} {\bibfnamefont {P.}~\bibnamefont {Pantazopoulos}}, \bibinfo
  {author} {\bibfnamefont {K.}~\bibnamefont {Tsakmakidis}}, \bibinfo {author}
  {\bibfnamefont {N.}~\bibnamefont {Papanikolaou}},\ and\ \bibinfo {author}
  {\bibfnamefont {N.}~\bibnamefont {Stefanou}},\ }\bibfield  {title} {\bibinfo
  {title} {Spherical optomagnonic microresonators: Triple-resonant photon
  transitions between zeeman-split mie modes},\ }\href
  {https://doi.org/10.1103/PhysRevB.101.054412} {\bibfield  {journal} {\bibinfo
   {journal} {Phys. Rev. B}\ }\textbf {\bibinfo {volume} {101}},\ \bibinfo
  {pages} {054412} (\bibinfo {year} {2020})}\BibitemShut {NoStop}%
\bibitem [{\citenamefont {Osada}\ \emph
  {et~al.}(2018{\natexlab{a}})\citenamefont {Osada}, \citenamefont {Gloppe},
  \citenamefont {Nakamura},\ and\ \citenamefont {Usami}}]{osada2018orbital}%
  \BibitemOpen
  \bibfield  {author} {\bibinfo {author} {\bibfnamefont {A.}~\bibnamefont
  {Osada}}, \bibinfo {author} {\bibfnamefont {A.}~\bibnamefont {Gloppe}},
  \bibinfo {author} {\bibfnamefont {Y.}~\bibnamefont {Nakamura}},\ and\
  \bibinfo {author} {\bibfnamefont {K.}~\bibnamefont {Usami}},\ }\bibfield
  {title} {\bibinfo {title} {Orbital angular momentum conservation in brillouin
  light scattering within a ferromagnetic sphere},\ }\href
  {https://iopscience.iop.org/article/10.1088/1367-2630/aae4b1} {\bibfield
  {journal} {\bibinfo  {journal} {New Journal of Physics}\ }\textbf {\bibinfo
  {volume} {20}},\ \bibinfo {pages} {103018} (\bibinfo {year}
  {2018}{\natexlab{a}})}\BibitemShut {NoStop}%
\bibitem [{\citenamefont
  {Kusminskiy}(2019{\natexlab{a}})}]{kusminskiy2019cavity}%
  \BibitemOpen
  \bibfield  {author} {\bibinfo {author} {\bibfnamefont {S.~V.}\ \bibnamefont
  {Kusminskiy}},\ }\href@noop {} {\bibinfo {title} {Cavity optomagnonics}}
  (\bibinfo {year} {2019}{\natexlab{a}}),\ \Eprint
  {https://arxiv.org/abs/1911.11104} {arXiv:1911.11104} \BibitemShut {NoStop}%
\bibitem [{\citenamefont {Haigh}\ \emph {et~al.}(2016)\citenamefont {Haigh},
  \citenamefont {Nunnenkamp}, \citenamefont {Ramsay},\ and\ \citenamefont
  {Ferguson}}]{PhysRevLett.117.133602}%
  \BibitemOpen
  \bibfield  {author} {\bibinfo {author} {\bibfnamefont {J.~A.}\ \bibnamefont
  {Haigh}}, \bibinfo {author} {\bibfnamefont {A.}~\bibnamefont {Nunnenkamp}},
  \bibinfo {author} {\bibfnamefont {A.~J.}\ \bibnamefont {Ramsay}},\ and\
  \bibinfo {author} {\bibfnamefont {A.~J.}\ \bibnamefont {Ferguson}},\
  }\bibfield  {title} {\bibinfo {title} {Triple-resonant brillouin light
  scattering in magneto-optical cavities},\ }\href
  {https://doi.org/10.1103/PhysRevLett.117.133602} {\bibfield  {journal}
  {\bibinfo  {journal} {Phys. Rev. Lett.}\ }\textbf {\bibinfo {volume} {117}},\
  \bibinfo {pages} {133602} (\bibinfo {year} {2016})}\BibitemShut {NoStop}%
\bibitem [{\citenamefont {Haigh}\ \emph {et~al.}(2018)\citenamefont {Haigh},
  \citenamefont {Lambert}, \citenamefont {Sharma}, \citenamefont {Blanter},
  \citenamefont {Bauer},\ and\ \citenamefont {Ramsay}}]{haigh2018selection}%
  \BibitemOpen
  \bibfield  {author} {\bibinfo {author} {\bibfnamefont {J.}~\bibnamefont
  {Haigh}}, \bibinfo {author} {\bibfnamefont {N.}~\bibnamefont {Lambert}},
  \bibinfo {author} {\bibfnamefont {S.}~\bibnamefont {Sharma}}, \bibinfo
  {author} {\bibfnamefont {Y.}~\bibnamefont {Blanter}}, \bibinfo {author}
  {\bibfnamefont {G.}~\bibnamefont {Bauer}},\ and\ \bibinfo {author}
  {\bibfnamefont {A.}~\bibnamefont {Ramsay}},\ }\bibfield  {title} {\bibinfo
  {title} {Selection rules for cavity-enhanced brillouin light scattering from
  magnetostatic modes},\ }\href
  {https://journals.aps.org/prb/abstract/10.1103/PhysRevB.97.214423} {\bibfield
   {journal} {\bibinfo  {journal} {Phys. Rev. B}\ }\textbf {\bibinfo {volume}
  {97}},\ \bibinfo {pages} {214423} (\bibinfo {year} {2018})}\BibitemShut
  {NoStop}%
\bibitem [{\citenamefont {Osada}\ \emph {et~al.}(2016)\citenamefont {Osada},
  \citenamefont {Hisatomi}, \citenamefont {Noguchi}, \citenamefont {Tabuchi},
  \citenamefont {Yamazaki}, \citenamefont {Usami}, \citenamefont {Sadgrove},
  \citenamefont {Yalla}, \citenamefont {Nomura},\ and\ \citenamefont
  {Nakamura}}]{PhysRevLett.116.223601}%
  \BibitemOpen
  \bibfield  {author} {\bibinfo {author} {\bibfnamefont {A.}~\bibnamefont
  {Osada}}, \bibinfo {author} {\bibfnamefont {R.}~\bibnamefont {Hisatomi}},
  \bibinfo {author} {\bibfnamefont {A.}~\bibnamefont {Noguchi}}, \bibinfo
  {author} {\bibfnamefont {Y.}~\bibnamefont {Tabuchi}}, \bibinfo {author}
  {\bibfnamefont {R.}~\bibnamefont {Yamazaki}}, \bibinfo {author}
  {\bibfnamefont {K.}~\bibnamefont {Usami}}, \bibinfo {author} {\bibfnamefont
  {M.}~\bibnamefont {Sadgrove}}, \bibinfo {author} {\bibfnamefont
  {R.}~\bibnamefont {Yalla}}, \bibinfo {author} {\bibfnamefont
  {M.}~\bibnamefont {Nomura}},\ and\ \bibinfo {author} {\bibfnamefont
  {Y.}~\bibnamefont {Nakamura}},\ }\bibfield  {title} {\bibinfo {title} {Cavity
  optomagnonics with spin-orbit coupled photons},\ }\href
  {https://doi.org/10.1103/PhysRevLett.116.223601} {\bibfield  {journal}
  {\bibinfo  {journal} {Phys. Rev. Lett.}\ }\textbf {\bibinfo {volume} {116}},\
  \bibinfo {pages} {223601} (\bibinfo {year} {2016})}\BibitemShut {NoStop}%
\bibitem [{\citenamefont {Osada}\ \emph
  {et~al.}(2018{\natexlab{b}})\citenamefont {Osada}, \citenamefont {Gloppe},
  \citenamefont {Hisatomi}, \citenamefont {Noguchi}, \citenamefont {Yamazaki},
  \citenamefont {Nomura}, \citenamefont {Nakamura},\ and\ \citenamefont
  {Usami}}]{osada2018brillouin}%
  \BibitemOpen
  \bibfield  {author} {\bibinfo {author} {\bibfnamefont {A.}~\bibnamefont
  {Osada}}, \bibinfo {author} {\bibfnamefont {A.}~\bibnamefont {Gloppe}},
  \bibinfo {author} {\bibfnamefont {R.}~\bibnamefont {Hisatomi}}, \bibinfo
  {author} {\bibfnamefont {A.}~\bibnamefont {Noguchi}}, \bibinfo {author}
  {\bibfnamefont {R.}~\bibnamefont {Yamazaki}}, \bibinfo {author}
  {\bibfnamefont {M.}~\bibnamefont {Nomura}}, \bibinfo {author} {\bibfnamefont
  {Y.}~\bibnamefont {Nakamura}},\ and\ \bibinfo {author} {\bibfnamefont
  {K.}~\bibnamefont {Usami}},\ }\bibfield  {title} {\bibinfo {title} {Brillouin
  light scattering by magnetic quasivortices in cavity optomagnonics},\ }\href
  {https://journals.aps.org/prl/abstract/10.1103/PhysRevLett.120.133602}
  {\bibfield  {journal} {\bibinfo  {journal} {Phys. Rev. Lett.}\ }\textbf
  {\bibinfo {volume} {120}},\ \bibinfo {pages} {133602} (\bibinfo {year}
  {2018}{\natexlab{b}})}\BibitemShut {NoStop}%
\bibitem [{\citenamefont {Zhang}\ \emph {et~al.}(2016)\citenamefont {Zhang},
  \citenamefont {Zhu}, \citenamefont {Zou},\ and\ \citenamefont
  {Tang}}]{zhang2016optomagnonic}%
  \BibitemOpen
  \bibfield  {author} {\bibinfo {author} {\bibfnamefont {X.}~\bibnamefont
  {Zhang}}, \bibinfo {author} {\bibfnamefont {N.}~\bibnamefont {Zhu}}, \bibinfo
  {author} {\bibfnamefont {C.-L.}\ \bibnamefont {Zou}},\ and\ \bibinfo {author}
  {\bibfnamefont {H.~X.}\ \bibnamefont {Tang}},\ }\bibfield  {title} {\bibinfo
  {title} {Optomagnonic whispering gallery microresonators},\ }\href
  {https://journals.aps.org/prl/abstract/10.1103/PhysRevLett.117.123605}
  {\bibfield  {journal} {\bibinfo  {journal} {Phys. Rev. Lett.}\ }\textbf
  {\bibinfo {volume} {117}},\ \bibinfo {pages} {123605} (\bibinfo {year}
  {2016})}\BibitemShut {NoStop}%
\bibitem [{\citenamefont {Rameshti}\ \emph {et~al.}(2021)\citenamefont
  {Rameshti}, \citenamefont {Kusminskiy}, \citenamefont {Haigh}, \citenamefont
  {Usami}, \citenamefont {Lachance-Quirion}, \citenamefont {Nakamura},
  \citenamefont {Hu}, \citenamefont {Tang}, \citenamefont {Bauer},\ and\
  \citenamefont {Blanter}}]{rameshti2021cavity}%
  \BibitemOpen
  \bibfield  {author} {\bibinfo {author} {\bibfnamefont {B.~Z.}\ \bibnamefont
  {Rameshti}}, \bibinfo {author} {\bibfnamefont {S.~V.}\ \bibnamefont
  {Kusminskiy}}, \bibinfo {author} {\bibfnamefont {J.~A.}\ \bibnamefont
  {Haigh}}, \bibinfo {author} {\bibfnamefont {K.}~\bibnamefont {Usami}},
  \bibinfo {author} {\bibfnamefont {D.}~\bibnamefont {Lachance-Quirion}},
  \bibinfo {author} {\bibfnamefont {Y.}~\bibnamefont {Nakamura}}, \bibinfo
  {author} {\bibfnamefont {C.-M.}\ \bibnamefont {Hu}}, \bibinfo {author}
  {\bibfnamefont {H.~X.}\ \bibnamefont {Tang}}, \bibinfo {author}
  {\bibfnamefont {G.~E.~W.}\ \bibnamefont {Bauer}},\ and\ \bibinfo {author}
  {\bibfnamefont {Y.~M.}\ \bibnamefont {Blanter}},\ }\href@noop {} {\bibinfo
  {title} {Cavity magnonics}} (\bibinfo {year} {2021}),\ \Eprint
  {https://arxiv.org/abs/2106.09312} {arXiv:2106.09312} \BibitemShut {NoStop}%
\bibitem [{\citenamefont {Heebner}\ \emph {et~al.}(2008)\citenamefont
  {Heebner}, \citenamefont {Grover}, \citenamefont {Ibrahim},\ and\
  \citenamefont {Ibrahim}}]{heebner2008optical}%
  \BibitemOpen
  \bibfield  {author} {\bibinfo {author} {\bibfnamefont {J.}~\bibnamefont
  {Heebner}}, \bibinfo {author} {\bibfnamefont {R.}~\bibnamefont {Grover}},
  \bibinfo {author} {\bibfnamefont {T.}~\bibnamefont {Ibrahim}},\ and\ \bibinfo
  {author} {\bibfnamefont {T.~A.}\ \bibnamefont {Ibrahim}},\ }\href@noop {}
  {\emph {\bibinfo {title} {Optical Microresonators: Theory, Fabrication, and
  Applications}}},\ Vol.\ \bibinfo {volume} {138}\ (\bibinfo  {publisher}
  {Springer Science \& Business Media},\ \bibinfo {year} {2008})\BibitemShut
  {NoStop}%
\bibitem [{\citenamefont {Bohren}\ and\ \citenamefont
  {Huffman}(2008)}]{bohren2008absorption}%
  \BibitemOpen
  \bibfield  {author} {\bibinfo {author} {\bibfnamefont {C.~F.}\ \bibnamefont
  {Bohren}}\ and\ \bibinfo {author} {\bibfnamefont {D.~R.}\ \bibnamefont
  {Huffman}},\ }\href@noop {} {\emph {\bibinfo {title} {Absorption and
  scattering of light by small particles}}}\ (\bibinfo  {publisher} {John Wiley
  \& Sons},\ \bibinfo {year} {2008})\BibitemShut {NoStop}%
\bibitem [{\citenamefont {Kuznetsov}\ \emph {et~al.}(2016)\citenamefont
  {Kuznetsov}, \citenamefont {Miroshnichenko}, \citenamefont {Brongersma},
  \citenamefont {Kivshar},\ and\ \citenamefont
  {Lukyanchuk}}]{kuznetsov2016optically}%
  \BibitemOpen
  \bibfield  {author} {\bibinfo {author} {\bibfnamefont {A.~I.}\ \bibnamefont
  {Kuznetsov}}, \bibinfo {author} {\bibfnamefont {A.~E.}\ \bibnamefont
  {Miroshnichenko}}, \bibinfo {author} {\bibfnamefont {M.~L.}\ \bibnamefont
  {Brongersma}}, \bibinfo {author} {\bibfnamefont {Y.~S.}\ \bibnamefont
  {Kivshar}},\ and\ \bibinfo {author} {\bibfnamefont {B.}~\bibnamefont
  {Lukyanchuk}},\ }\bibfield  {title} {\bibinfo {title} {Optically resonant
  dielectric nanostructures},\ }\href
  {https://science.sciencemag.org/content/354/6314/aag2472} {\bibfield
  {journal} {\bibinfo  {journal} {Science}\ }\textbf {\bibinfo {volume} {354}}
  (\bibinfo {year} {2016})}\BibitemShut {NoStop}%
\bibitem [{\citenamefont {Ford}\ and\ \citenamefont
  {Wener}(1978)}]{ford1978scattering}%
  \BibitemOpen
  \bibfield  {author} {\bibinfo {author} {\bibfnamefont {G.~W.}\ \bibnamefont
  {Ford}}\ and\ \bibinfo {author} {\bibfnamefont {S.~A.}\ \bibnamefont
  {Wener}},\ }\bibfield  {title} {\bibinfo {title} {Scattering and absorption
  of electromagnetic waves by a gyrotropic sphere},\ }\href
  {https://doi.org/10.1103/PhysRevB.18.6752} {\bibfield  {journal} {\bibinfo
  {journal} {Phys. Rev. B}\ }\textbf {\bibinfo {volume} {18}},\ \bibinfo
  {pages} {6752} (\bibinfo {year} {1978})}\BibitemShut {NoStop}%
\bibitem [{\citenamefont {Fleury}\ and\ \citenamefont
  {Loudon}(1968)}]{fleury1968scattering}%
  \BibitemOpen
  \bibfield  {author} {\bibinfo {author} {\bibfnamefont {P.}~\bibnamefont
  {Fleury}}\ and\ \bibinfo {author} {\bibfnamefont {R.}~\bibnamefont
  {Loudon}},\ }\bibfield  {title} {\bibinfo {title} {Scattering of light by
  one-and two-magnon excitations},\ }\href
  {https://journals.aps.org/pr/abstract/10.1103/PhysRev.166.514} {\bibfield
  {journal} {\bibinfo  {journal} {Physical Review}\ }\textbf {\bibinfo {volume}
  {166}},\ \bibinfo {pages} {514} (\bibinfo {year} {1968})}\BibitemShut
  {NoStop}%
\bibitem [{\citenamefont {Nieminen}\ \emph {et~al.}(2001)\citenamefont
  {Nieminen}, \citenamefont {Heckenberg},\ and\ \citenamefont
  {Rubinsztein-Dunlop}}]{nieminen2001optical}%
  \BibitemOpen
  \bibfield  {author} {\bibinfo {author} {\bibfnamefont {T.~A.}\ \bibnamefont
  {Nieminen}}, \bibinfo {author} {\bibfnamefont {N.~R.}\ \bibnamefont
  {Heckenberg}},\ and\ \bibinfo {author} {\bibfnamefont {H.}~\bibnamefont
  {Rubinsztein-Dunlop}},\ }\bibfield  {title} {\bibinfo {title} {Optical
  measurement of microscopic torques},\ }\href
  {https://www.tandfonline.com/doi/abs/10.1080/09500340108230922} {\bibfield
  {journal} {\bibinfo  {journal} {Journal of Modern Optics}\ }\textbf {\bibinfo
  {volume} {48}},\ \bibinfo {pages} {405} (\bibinfo {year} {2001})}\BibitemShut
  {NoStop}%
\bibitem [{\citenamefont {Friese}\ \emph {et~al.}(1998)\citenamefont {Friese},
  \citenamefont {Nieminen}, \citenamefont {Heckenberg},\ and\ \citenamefont
  {Rubinsztein-Dunlop}}]{friese1998optical}%
  \BibitemOpen
  \bibfield  {author} {\bibinfo {author} {\bibfnamefont {M.~E.}\ \bibnamefont
  {Friese}}, \bibinfo {author} {\bibfnamefont {T.~A.}\ \bibnamefont
  {Nieminen}}, \bibinfo {author} {\bibfnamefont {N.~R.}\ \bibnamefont
  {Heckenberg}},\ and\ \bibinfo {author} {\bibfnamefont {H.}~\bibnamefont
  {Rubinsztein-Dunlop}},\ }\bibfield  {title} {\bibinfo {title} {Optical
  alignment and spinning of laser-trapped microscopic particles},\ }\href
  {https://www.nature.com/articles/28566} {\bibfield  {journal} {\bibinfo
  {journal} {Nature}\ }\textbf {\bibinfo {volume} {394}},\ \bibinfo {pages}
  {348} (\bibinfo {year} {1998})}\BibitemShut {NoStop}%
\bibitem [{\citenamefont {Simpson}\ \emph {et~al.}(2007)\citenamefont
  {Simpson}, \citenamefont {Benito},\ and\ \citenamefont
  {Hanna}}]{simpson2007polarization}%
  \BibitemOpen
  \bibfield  {author} {\bibinfo {author} {\bibfnamefont {S.~H.}\ \bibnamefont
  {Simpson}}, \bibinfo {author} {\bibfnamefont {D.~C.}\ \bibnamefont
  {Benito}},\ and\ \bibinfo {author} {\bibfnamefont {S.}~\bibnamefont
  {Hanna}},\ }\bibfield  {title} {\bibinfo {title} {Polarization-induced torque
  in optical traps},\ }\href {https://doi.org/10.1103/PhysRevA.76.043408}
  {\bibfield  {journal} {\bibinfo  {journal} {Phys. Rev. A}\ }\textbf {\bibinfo
  {volume} {76}},\ \bibinfo {pages} {043408} (\bibinfo {year}
  {2007})}\BibitemShut {NoStop}%
\bibitem [{\citenamefont {Donato}\ \emph {et~al.}(2016)\citenamefont {Donato},
  \citenamefont {Mazzulla}, \citenamefont {Pagliusi}, \citenamefont
  {Magazz{\`u}}, \citenamefont {Hernandez}, \citenamefont {Provenzano},
  \citenamefont {Gucciardi}, \citenamefont {Marag{\`o}},\ and\ \citenamefont
  {Cipparrone}}]{donato2016light}%
  \BibitemOpen
  \bibfield  {author} {\bibinfo {author} {\bibfnamefont {M.}~\bibnamefont
  {Donato}}, \bibinfo {author} {\bibfnamefont {A.}~\bibnamefont {Mazzulla}},
  \bibinfo {author} {\bibfnamefont {P.}~\bibnamefont {Pagliusi}}, \bibinfo
  {author} {\bibfnamefont {A.}~\bibnamefont {Magazz{\`u}}}, \bibinfo {author}
  {\bibfnamefont {R.}~\bibnamefont {Hernandez}}, \bibinfo {author}
  {\bibfnamefont {C.}~\bibnamefont {Provenzano}}, \bibinfo {author}
  {\bibfnamefont {P.}~\bibnamefont {Gucciardi}}, \bibinfo {author}
  {\bibfnamefont {O.}~\bibnamefont {Marag{\`o}}},\ and\ \bibinfo {author}
  {\bibfnamefont {G.}~\bibnamefont {Cipparrone}},\ }\bibfield  {title}
  {\bibinfo {title} {Light-induced rotations of chiral birefringent
  microparticles in optical tweezers},\ }\href
  {https://www.nature.com/articles/srep31977} {\bibfield  {journal} {\bibinfo
  {journal} {Scientific Reports}\ }\textbf {\bibinfo {volume} {6}},\ \bibinfo
  {pages} {1} (\bibinfo {year} {2016})}\BibitemShut {NoStop}%
\bibitem [{\citenamefont {Gilbert}(2004)}]{gilbert2004phenomenological}%
  \BibitemOpen
  \bibfield  {author} {\bibinfo {author} {\bibfnamefont {T.~L.}\ \bibnamefont
  {Gilbert}},\ }\bibfield  {title} {\bibinfo {title} {A phenomenological theory
  of damping in ferromagnetic materials},\ }\href
  {https://ieeexplore.ieee.org/document/1353448} {\bibfield  {journal}
  {\bibinfo  {journal} {IEEE Transactions on Magnetics}\ }\textbf {\bibinfo
  {volume} {40}},\ \bibinfo {pages} {3443} (\bibinfo {year}
  {2004})}\BibitemShut {NoStop}%
\bibitem [{\citenamefont {Antonov}\ \emph {et~al.}(1969)\citenamefont
  {Antonov}, \citenamefont {Agranovskaya}, \citenamefont {Petrova},\ and\
  \citenamefont {Titova}}]{antonov1969optical}%
  \BibitemOpen
  \bibfield  {author} {\bibinfo {author} {\bibfnamefont {A.}~\bibnamefont
  {Antonov}}, \bibinfo {author} {\bibfnamefont {A.}~\bibnamefont
  {Agranovskaya}}, \bibinfo {author} {\bibfnamefont {G.}~\bibnamefont
  {Petrova}},\ and\ \bibinfo {author} {\bibfnamefont {A.}~\bibnamefont
  {Titova}},\ }\bibfield  {title} {\bibinfo {title} {Optical properties of
  yttrium-iron garnet},\ }\href
  {https://link.springer.com/content/pdf/10.1007/BF00607387.pdf} {\bibfield
  {journal} {\bibinfo  {journal} {Journal of Applied Spectroscopy}\ }\textbf
  {\bibinfo {volume} {11}},\ \bibinfo {pages} {1225} (\bibinfo {year}
  {1969})}\BibitemShut {NoStop}%
\bibitem [{\citenamefont {Gatteschi}\ \emph {et~al.}(2006)\citenamefont
  {Gatteschi}, \citenamefont {Sessoli},\ and\ \citenamefont
  {Villain}}]{gatteschi2006molecular}%
  \BibitemOpen
  \bibfield  {author} {\bibinfo {author} {\bibfnamefont {D.}~\bibnamefont
  {Gatteschi}}, \bibinfo {author} {\bibfnamefont {R.}~\bibnamefont {Sessoli}},\
  and\ \bibinfo {author} {\bibfnamefont {J.}~\bibnamefont {Villain}},\
  }\href@noop {} {\emph {\bibinfo {title} {Molecular Nanomagnets}}},\
  Vol.~\bibinfo {volume} {1}\ (\bibinfo  {publisher} {Oxford University
  Press},\ \bibinfo {year} {2006})\BibitemShut {NoStop}%
\bibitem [{\citenamefont {Rusconi}\ \emph {et~al.}(2017)\citenamefont
  {Rusconi}, \citenamefont {P{\"o}chhacker}, \citenamefont {Kustura},
  \citenamefont {Cirac},\ and\ \citenamefont
  {Romero-Isart}}]{rusconi2017quantum}%
  \BibitemOpen
  \bibfield  {author} {\bibinfo {author} {\bibfnamefont {C.~C.}\ \bibnamefont
  {Rusconi}}, \bibinfo {author} {\bibfnamefont {V.}~\bibnamefont
  {P{\"o}chhacker}}, \bibinfo {author} {\bibfnamefont {K.}~\bibnamefont
  {Kustura}}, \bibinfo {author} {\bibfnamefont {J.~I.}\ \bibnamefont {Cirac}},\
  and\ \bibinfo {author} {\bibfnamefont {O.}~\bibnamefont {Romero-Isart}},\
  }\bibfield  {title} {\bibinfo {title} {Quantum spin stabilized magnetic
  levitation},\ }\href
  {https://journals.aps.org/prl/abstract/10.1103/PhysRevLett.119.167202}
  {\bibfield  {journal} {\bibinfo  {journal} {Phys. Rev. Lett.}\ }\textbf
  {\bibinfo {volume} {119}},\ \bibinfo {pages} {167202} (\bibinfo {year}
  {2017})}\BibitemShut {NoStop}%
\bibitem [{\citenamefont
  {Kusminskiy}(2019{\natexlab{b}})}]{kusminskiy2019quantum}%
  \BibitemOpen
  \bibfield  {author} {\bibinfo {author} {\bibfnamefont {S.~V.}\ \bibnamefont
  {Kusminskiy}},\ }\href@noop {} {\emph {\bibinfo {title} {Quantum Magnetism,
  Spin Waves, and Optical Cavities}}}\ (\bibinfo  {publisher} {Springer},\
  \bibinfo {year} {2019})\BibitemShut {NoStop}%
\bibitem [{\citenamefont {Wu}\ and\ \citenamefont
  {Hoffmann}(2013)}]{wu2013recent}%
  \BibitemOpen
  \bibfield  {author} {\bibinfo {author} {\bibfnamefont {M.}~\bibnamefont
  {Wu}}\ and\ \bibinfo {author} {\bibfnamefont {A.}~\bibnamefont {Hoffmann}},\
  }\href@noop {} {\emph {\bibinfo {title} {Recent advances in magnetic
  insulators-from spintronics to microwave applications}}}\ (\bibinfo
  {publisher} {Academic Press},\ \bibinfo {year} {2013})\BibitemShut {NoStop}%
\bibitem [{Note1()}]{Note1}%
  \BibitemOpen
  \bibinfo {note} {For instance, one could consider individual driving terms at
  different frequencies for each mode, which would require additional unitary
  transformations to a rotating frame in which $\protect \hat {H}$ is
  time-independent.}\BibitemShut {Stop}%
\bibitem [{\citenamefont {Gardiner}\ and\ \citenamefont
  {Collett}(1985)}]{gardiner1985input}%
  \BibitemOpen
  \bibfield  {author} {\bibinfo {author} {\bibfnamefont {C.~W.}\ \bibnamefont
  {Gardiner}}\ and\ \bibinfo {author} {\bibfnamefont {M.~J.}\ \bibnamefont
  {Collett}},\ }\bibfield  {title} {\bibinfo {title} {Input and output in
  damped quantum systems: Quantum stochastic differential equations and the
  master equation},\ }\href {https://doi.org/10.1103/PhysRevA.31.3761}
  {\bibfield  {journal} {\bibinfo  {journal} {Phys. Rev. A}\ }\textbf {\bibinfo
  {volume} {31}},\ \bibinfo {pages} {3761} (\bibinfo {year}
  {1985})}\BibitemShut {NoStop}%
\bibitem [{\citenamefont {Buck}\ and\ \citenamefont
  {Kimble}(2003)}]{buck2003optimal}%
  \BibitemOpen
  \bibfield  {author} {\bibinfo {author} {\bibfnamefont {J.}~\bibnamefont
  {Buck}}\ and\ \bibinfo {author} {\bibfnamefont {H.}~\bibnamefont {Kimble}},\
  }\bibfield  {title} {\bibinfo {title} {Optimal sizes of dielectric
  microspheres for cavity qed with strong coupling},\ }\href
  {https://doi.org/10.1103/PhysRevA.67.033806} {\bibfield  {journal} {\bibinfo
  {journal} {Phys. Rev. A}\ }\textbf {\bibinfo {volume} {67}},\ \bibinfo
  {pages} {033806} (\bibinfo {year} {2003})}\BibitemShut {NoStop}%
\bibitem [{\citenamefont {Keshtgar}\ \emph {et~al.}(2017)\citenamefont
  {Keshtgar}, \citenamefont {Streib}, \citenamefont {Kamra}, \citenamefont
  {Blanter},\ and\ \citenamefont {Bauer}}]{keshtgar2017magnetomechanical}%
  \BibitemOpen
  \bibfield  {author} {\bibinfo {author} {\bibfnamefont {H.}~\bibnamefont
  {Keshtgar}}, \bibinfo {author} {\bibfnamefont {S.}~\bibnamefont {Streib}},
  \bibinfo {author} {\bibfnamefont {A.}~\bibnamefont {Kamra}}, \bibinfo
  {author} {\bibfnamefont {Y.~M.}\ \bibnamefont {Blanter}},\ and\ \bibinfo
  {author} {\bibfnamefont {G.~E.}\ \bibnamefont {Bauer}},\ }\bibfield  {title}
  {\bibinfo {title} {Magnetomechanical coupling and ferromagnetic resonance in
  magnetic nanoparticles},\ }\href {https://doi.org/10.1103/PhysRevB.95.134447}
  {\bibfield  {journal} {\bibinfo  {journal} {Phys. Rev. B}\ }\textbf {\bibinfo
  {volume} {95}},\ \bibinfo {pages} {134447} (\bibinfo {year}
  {2017})}\BibitemShut {NoStop}%
\bibitem [{\citenamefont {Band}\ \emph {et~al.}(2018)\citenamefont {Band},
  \citenamefont {Avishai},\ and\ \citenamefont {Shnirman}}]{band2018dynamics}%
  \BibitemOpen
  \bibfield  {author} {\bibinfo {author} {\bibfnamefont {Y.}~\bibnamefont
  {Band}}, \bibinfo {author} {\bibfnamefont {Y.}~\bibnamefont {Avishai}},\ and\
  \bibinfo {author} {\bibfnamefont {A.}~\bibnamefont {Shnirman}},\ }\bibfield
  {title} {\bibinfo {title} {Dynamics of a magnetic needle magnetometer:
  Sensitivity to landau-lifshitz-gilbert damping},\ }\href
  {https://doi.org/10.1103/PhysRevLett.121.160801} {\bibfield  {journal}
  {\bibinfo  {journal} {Phys. Rev. Lett.}\ }\textbf {\bibinfo {volume} {121}},\
  \bibinfo {pages} {160801} (\bibinfo {year} {2018})}\BibitemShut {NoStop}%
\bibitem [{Note2()}]{Note2}%
  \BibitemOpen
  \bibinfo {note} {For YIG particles of $\sim (1\protect \tmspace +\thinmuskip
  {.1667em}\mu {\protect \rm m})^{3}$, this approximation is justified as $\eta
  _{\protect \mathrm {G}}/S<10^{-14}$.}\BibitemShut {Stop}%
\bibitem [{\citenamefont {Suhl}(1957)}]{suhl1957theory}%
  \BibitemOpen
  \bibfield  {author} {\bibinfo {author} {\bibfnamefont {H.}~\bibnamefont
  {Suhl}},\ }\bibfield  {title} {\bibinfo {title} {The theory of ferromagnetic
  resonance at high signal powers},\ }\href
  {https://www.sciencedirect.com/science/article/pii/002236975790010}
  {\bibfield  {journal} {\bibinfo  {journal} {Journal of Physics and Chemistry
  of Solids}\ }\textbf {\bibinfo {volume} {1}},\ \bibinfo {pages} {209}
  (\bibinfo {year} {1957})}\BibitemShut {NoStop}%
\bibitem [{\citenamefont {Fano}(1961)}]{fano1961effects}%
  \BibitemOpen
  \bibfield  {author} {\bibinfo {author} {\bibfnamefont {U.}~\bibnamefont
  {Fano}},\ }\bibfield  {title} {\bibinfo {title} {Effects of configuration
  interaction on intensities and phase shifts},\ }\href
  {https://doi.org/10.1103/PhysRev.124.1866} {\bibfield  {journal} {\bibinfo
  {journal} {Phys. Rev.}\ }\textbf {\bibinfo {volume} {124}},\ \bibinfo {pages}
  {1866} (\bibinfo {year} {1961})}\BibitemShut {NoStop}%
\bibitem [{\citenamefont {Wang}\ \emph {et~al.}(2020)\citenamefont {Wang},
  \citenamefont {Si}, \citenamefont {Lu},\ and\ \citenamefont
  {Wu}}]{wang2020optomechanically}%
  \BibitemOpen
  \bibfield  {author} {\bibinfo {author} {\bibfnamefont {X.-Y.}\ \bibnamefont
  {Wang}}, \bibinfo {author} {\bibfnamefont {L.-G.}\ \bibnamefont {Si}},
  \bibinfo {author} {\bibfnamefont {X.-H.}\ \bibnamefont {Lu}},\ and\ \bibinfo
  {author} {\bibfnamefont {Y.}~\bibnamefont {Wu}},\ }\bibfield  {title}
  {\bibinfo {title} {Optomechanically tuned fano resonance and slow light in a
  quadratically coupled optomechanical system with membranes},\ }\href
  {https://iopscience.iop.org/article/10.1088/1361-6455/abb013} {\bibfield
  {journal} {\bibinfo  {journal} {Journal of Physics B: Atomic, Molecular and
  Optical Physics}\ }\textbf {\bibinfo {volume} {53}},\ \bibinfo {pages}
  {235402} (\bibinfo {year} {2020})}\BibitemShut {NoStop}%
\bibitem [{\citenamefont {Qu}\ and\ \citenamefont
  {Agarwal}(2013)}]{qu2013fano}%
  \BibitemOpen
  \bibfield  {author} {\bibinfo {author} {\bibfnamefont {K.}~\bibnamefont
  {Qu}}\ and\ \bibinfo {author} {\bibfnamefont {G.}~\bibnamefont {Agarwal}},\
  }\bibfield  {title} {\bibinfo {title} {Fano resonances and their control in
  optomechanics},\ }\href {https://doi.org/10.1103/PhysRevA.87.063813}
  {\bibfield  {journal} {\bibinfo  {journal} {Phys. Rev. A}\ }\textbf {\bibinfo
  {volume} {87}},\ \bibinfo {pages} {063813} (\bibinfo {year}
  {2013})}\BibitemShut {NoStop}%
\bibitem [{\citenamefont {Kittel}(1948)}]{kittel1948theory}%
  \BibitemOpen
  \bibfield  {author} {\bibinfo {author} {\bibfnamefont {C.}~\bibnamefont
  {Kittel}},\ }\bibfield  {title} {\bibinfo {title} {On the theory of
  ferromagnetic resonance absorption},\ }\href
  {https://journals.aps.org/pr/pdf/10.1103/PhysRev.73.155} {\bibfield
  {journal} {\bibinfo  {journal} {Physical Review}\ }\textbf {\bibinfo {volume}
  {73}},\ \bibinfo {pages} {155} (\bibinfo {year} {1948})}\BibitemShut
  {NoStop}%
\bibitem [{\citenamefont {Tabuchi}\ \emph {et~al.}(2016)\citenamefont
  {Tabuchi}, \citenamefont {Ishino}, \citenamefont {Noguchi}, \citenamefont
  {Ishikawa}, \citenamefont {Yamazaki}, \citenamefont {Usami},\ and\
  \citenamefont {Nakamura}}]{tabuchi2016quantum}%
  \BibitemOpen
  \bibfield  {author} {\bibinfo {author} {\bibfnamefont {Y.}~\bibnamefont
  {Tabuchi}}, \bibinfo {author} {\bibfnamefont {S.}~\bibnamefont {Ishino}},
  \bibinfo {author} {\bibfnamefont {A.}~\bibnamefont {Noguchi}}, \bibinfo
  {author} {\bibfnamefont {T.}~\bibnamefont {Ishikawa}}, \bibinfo {author}
  {\bibfnamefont {R.}~\bibnamefont {Yamazaki}}, \bibinfo {author}
  {\bibfnamefont {K.}~\bibnamefont {Usami}},\ and\ \bibinfo {author}
  {\bibfnamefont {Y.}~\bibnamefont {Nakamura}},\ }\bibfield  {title} {\bibinfo
  {title} {Quantum magnonics: The magnon meets the superconducting qubit},\
  }\href {https://www.sciencedirect.com/science/article/pii/S1631070516300603}
  {\bibfield  {journal} {\bibinfo  {journal} {Comptes Rendus Physique}\
  }\textbf {\bibinfo {volume} {17}},\ \bibinfo {pages} {729} (\bibinfo {year}
  {2016})}\BibitemShut {NoStop}%
\bibitem [{\citenamefont {Lachance-Quirion}\ \emph {et~al.}(2017)\citenamefont
  {Lachance-Quirion}, \citenamefont {Tabuchi}, \citenamefont {Ishino},
  \citenamefont {Noguchi}, \citenamefont {Ishikawa}, \citenamefont {Yamazaki},\
  and\ \citenamefont {Nakamura}}]{lachance2017resolving}%
  \BibitemOpen
  \bibfield  {author} {\bibinfo {author} {\bibfnamefont {D.}~\bibnamefont
  {Lachance-Quirion}}, \bibinfo {author} {\bibfnamefont {Y.}~\bibnamefont
  {Tabuchi}}, \bibinfo {author} {\bibfnamefont {S.}~\bibnamefont {Ishino}},
  \bibinfo {author} {\bibfnamefont {A.}~\bibnamefont {Noguchi}}, \bibinfo
  {author} {\bibfnamefont {T.}~\bibnamefont {Ishikawa}}, \bibinfo {author}
  {\bibfnamefont {R.}~\bibnamefont {Yamazaki}},\ and\ \bibinfo {author}
  {\bibfnamefont {Y.}~\bibnamefont {Nakamura}},\ }\bibfield  {title} {\bibinfo
  {title} {Resolving quanta of collective spin excitations in a
  millimeter-sized ferromagnet},\ }\href
  {https://advances.sciencemag.org/content/3/7/e1603150} {\bibfield  {journal}
  {\bibinfo  {journal} {Science Advances}\ }\textbf {\bibinfo {volume} {3}},\
  \bibinfo {pages} {e1603150} (\bibinfo {year} {2017})}\BibitemShut {NoStop}%
\bibitem [{\citenamefont {Wettling}\ \emph {et~al.}(1975)\citenamefont
  {Wettling}, \citenamefont {Cottam},\ and\ \citenamefont
  {Sandercock}}]{wettling1975relation}%
  \BibitemOpen
  \bibfield  {author} {\bibinfo {author} {\bibfnamefont {W.}~\bibnamefont
  {Wettling}}, \bibinfo {author} {\bibfnamefont {M.}~\bibnamefont {Cottam}},\
  and\ \bibinfo {author} {\bibfnamefont {J.}~\bibnamefont {Sandercock}},\
  }\bibfield  {title} {\bibinfo {title} {The relation between one-magnon light
  scattering and the complex magneto-optic effects in yig},\ }\href
  {https://iopscience.iop.org/article/10.1088/0022-3719/8/2/014} {\bibfield
  {journal} {\bibinfo  {journal} {Journal of Physics C: Solid State Physics}\
  }\textbf {\bibinfo {volume} {8}},\ \bibinfo {pages} {211} (\bibinfo {year}
  {1975})}\BibitemShut {NoStop}%
\bibitem [{\citenamefont {Pisarev}\ \emph {et~al.}(1971)\citenamefont
  {Pisarev}, \citenamefont {Sinii}, \citenamefont {Kolpakova},\ and\
  \citenamefont {Yakovlev}}]{pisarev1971magnetic}%
  \BibitemOpen
  \bibfield  {author} {\bibinfo {author} {\bibfnamefont {R.}~\bibnamefont
  {Pisarev}}, \bibinfo {author} {\bibfnamefont {I.}~\bibnamefont {Sinii}},
  \bibinfo {author} {\bibfnamefont {N.}~\bibnamefont {Kolpakova}},\ and\
  \bibinfo {author} {\bibfnamefont {Y.~M.}\ \bibnamefont {Yakovlev}},\
  }\bibfield  {title} {\bibinfo {title} {Magnetic birefringence of light in
  iron garnets},\ }\href@noop {} {\bibfield  {journal} {\bibinfo  {journal}
  {Sov. Phys. JETP}\ }\textbf {\bibinfo {volume} {33}},\ \bibinfo {pages}
  {1175} (\bibinfo {year} {1971})}\BibitemShut {NoStop}%
\bibitem [{\citenamefont {Cullity}\ and\ \citenamefont
  {Graham}(2008)}]{cullity2008introduction}%
  \BibitemOpen
  \bibfield  {author} {\bibinfo {author} {\bibfnamefont {B.~D.}\ \bibnamefont
  {Cullity}}\ and\ \bibinfo {author} {\bibfnamefont {C.~D.}\ \bibnamefont
  {Graham}},\ }\href@noop {} {\emph {\bibinfo {title} {Introduction to magnetic
  materials, 2nd Edition}}}\ (\bibinfo  {publisher} {Wiley-IEEE Press},\
  \bibinfo {year} {2008})\BibitemShut {NoStop}%
\bibitem [{\citenamefont {Dionne}(2009)}]{dionne2009anisotropy}%
  \BibitemOpen
  \bibfield  {author} {\bibinfo {author} {\bibfnamefont {G.~F.}\ \bibnamefont
  {Dionne}},\ }\bibfield  {title} {\bibinfo {title} {Anisotropy and
  magnetoelastic properties},\ }in\ \href
  {https://doi.org/10.1007/978-1-4419-0054-8_5} {\emph {\bibinfo {booktitle}
  {Magnetic Oxides}}}\ (\bibinfo  {publisher} {Springer},\ \bibinfo {year}
  {2009})\ pp.\ \bibinfo {pages} {201--271}\BibitemShut {NoStop}%
\bibitem [{\citenamefont {Streib}\ \emph {et~al.}(2018)\citenamefont {Streib},
  \citenamefont {Keshtgar},\ and\ \citenamefont {Bauer}}]{streib2018damping}%
  \BibitemOpen
  \bibfield  {author} {\bibinfo {author} {\bibfnamefont {S.}~\bibnamefont
  {Streib}}, \bibinfo {author} {\bibfnamefont {H.}~\bibnamefont {Keshtgar}},\
  and\ \bibinfo {author} {\bibfnamefont {G.~E.}\ \bibnamefont {Bauer}},\
  }\bibfield  {title} {\bibinfo {title} {Damping of magnetization dynamics by
  phonon pumping},\ }\href
  {https://journals.aps.org/prl/abstract/10.1103/PhysRevLett.121.027202}
  {\bibfield  {journal} {\bibinfo  {journal} {Phys. Rev. Lett.}\ }\textbf
  {\bibinfo {volume} {121}},\ \bibinfo {pages} {027202} (\bibinfo {year}
  {2018})}\BibitemShut {NoStop}%
\bibitem [{\citenamefont {Elyasi}\ \emph {et~al.}(2020)\citenamefont {Elyasi},
  \citenamefont {Blanter},\ and\ \citenamefont {Bauer}}]{elyasi2020resources}%
  \BibitemOpen
  \bibfield  {author} {\bibinfo {author} {\bibfnamefont {M.}~\bibnamefont
  {Elyasi}}, \bibinfo {author} {\bibfnamefont {Y.~M.}\ \bibnamefont
  {Blanter}},\ and\ \bibinfo {author} {\bibfnamefont {G.~E.}\ \bibnamefont
  {Bauer}},\ }\bibfield  {title} {\bibinfo {title} {Resources of nonlinear
  cavity magnonics for quantum information},\ }\href
  {https://journals.aps.org/prb/abstract/10.1103/PhysRevB.101.054402}
  {\bibfield  {journal} {\bibinfo  {journal} {Phys. Rev. B}\ }\textbf {\bibinfo
  {volume} {101}},\ \bibinfo {pages} {054402} (\bibinfo {year}
  {2020})}\BibitemShut {NoStop}%
\bibitem [{\citenamefont {Pacewicz}\ \emph {et~al.}(2019)\citenamefont
  {Pacewicz}, \citenamefont {Krupka}, \citenamefont {Salski}, \citenamefont
  {Aleshkevych},\ and\ \citenamefont {Kopyt}}]{pacewicz2019rigorous}%
  \BibitemOpen
  \bibfield  {author} {\bibinfo {author} {\bibfnamefont {A.}~\bibnamefont
  {Pacewicz}}, \bibinfo {author} {\bibfnamefont {J.}~\bibnamefont {Krupka}},
  \bibinfo {author} {\bibfnamefont {B.}~\bibnamefont {Salski}}, \bibinfo
  {author} {\bibfnamefont {P.}~\bibnamefont {Aleshkevych}},\ and\ \bibinfo
  {author} {\bibfnamefont {P.}~\bibnamefont {Kopyt}},\ }\bibfield  {title}
  {\bibinfo {title} {Rigorous broadband study of the intrinsic ferromagnetic
  linewidth of monocrystalline garnet spheres},\ }\href
  {https://www.nature.com/articles/s41598-019-45699-7} {\bibfield  {journal}
  {\bibinfo  {journal} {Scientific Reports}\ }\textbf {\bibinfo {volume} {9}},\
  \bibinfo {pages} {1} (\bibinfo {year} {2019})}\BibitemShut {NoStop}%
\bibitem [{\citenamefont {Rijnierse}\ \emph {et~al.}(1975)\citenamefont
  {Rijnierse}, \citenamefont {Logmans}, \citenamefont {Metselaar},\ and\
  \citenamefont {Stacy}}]{rijnierse1975optical}%
  \BibitemOpen
  \bibfield  {author} {\bibinfo {author} {\bibfnamefont {P.}~\bibnamefont
  {Rijnierse}}, \bibinfo {author} {\bibfnamefont {H.}~\bibnamefont {Logmans}},
  \bibinfo {author} {\bibfnamefont {R.}~\bibnamefont {Metselaar}},\ and\
  \bibinfo {author} {\bibfnamefont {W.}~\bibnamefont {Stacy}},\ }\bibfield
  {title} {\bibinfo {title} {Optical measurement of magnetic anisotropy in thin
  garnet films},\ }\href {https://link.springer.com/article/10.1007/BF00896031}
  {\bibfield  {journal} {\bibinfo  {journal} {Applied Physics}\ }\textbf
  {\bibinfo {volume} {8}},\ \bibinfo {pages} {143} (\bibinfo {year}
  {1975})}\BibitemShut {NoStop}%
\bibitem [{\citenamefont {Hansen}\ \emph {et~al.}(1985)\citenamefont {Hansen},
  \citenamefont {Klages}, \citenamefont {Schuldt},\ and\ \citenamefont
  {Witter}}]{hansen1985magnetic}%
  \BibitemOpen
  \bibfield  {author} {\bibinfo {author} {\bibfnamefont {P.}~\bibnamefont
  {Hansen}}, \bibinfo {author} {\bibfnamefont {C.-P.}\ \bibnamefont {Klages}},
  \bibinfo {author} {\bibfnamefont {J.}~\bibnamefont {Schuldt}},\ and\ \bibinfo
  {author} {\bibfnamefont {K.}~\bibnamefont {Witter}},\ }\bibfield  {title}
  {\bibinfo {title} {Magnetic and magneto-optical properties of
  bismuth-substituted lutetium iron garnet films},\ }\href
  {https://journals.aps.org/prb/pdf/10.1103/PhysRevB.31.5858} {\bibfield
  {journal} {\bibinfo  {journal} {Phys. Rev. B}\ }\textbf {\bibinfo {volume}
  {31}},\ \bibinfo {pages} {5858} (\bibinfo {year} {1985})}\BibitemShut
  {NoStop}%
\bibitem [{\citenamefont {La~Porta}\ and\ \citenamefont
  {Wang}(2004)}]{la2004optical}%
  \BibitemOpen
  \bibfield  {author} {\bibinfo {author} {\bibfnamefont {A.}~\bibnamefont
  {La~Porta}}\ and\ \bibinfo {author} {\bibfnamefont {M.~D.}\ \bibnamefont
  {Wang}},\ }\bibfield  {title} {\bibinfo {title} {Optical torque wrench:
  Angular trapping, rotation, and torque detection of quartz microparticles},\
  }\href {https://doi.org/10.1103/PhysRevLett.92.190801} {\bibfield  {journal}
  {\bibinfo  {journal} {Phys. Rev. Lett.}\ }\textbf {\bibinfo {volume} {92}},\
  \bibinfo {pages} {190801} (\bibinfo {year} {2004})}\BibitemShut {NoStop}%
\bibitem [{\citenamefont {Ahn}\ \emph {et~al.}(2020)\citenamefont {Ahn},
  \citenamefont {Xu}, \citenamefont {Bang}, \citenamefont {Ju}, \citenamefont
  {Gao},\ and\ \citenamefont {Li}}]{ahn2020ultrasensitive}%
  \BibitemOpen
  \bibfield  {author} {\bibinfo {author} {\bibfnamefont {J.}~\bibnamefont
  {Ahn}}, \bibinfo {author} {\bibfnamefont {Z.}~\bibnamefont {Xu}}, \bibinfo
  {author} {\bibfnamefont {J.}~\bibnamefont {Bang}}, \bibinfo {author}
  {\bibfnamefont {P.}~\bibnamefont {Ju}}, \bibinfo {author} {\bibfnamefont
  {X.}~\bibnamefont {Gao}},\ and\ \bibinfo {author} {\bibfnamefont
  {T.}~\bibnamefont {Li}},\ }\bibfield  {title} {\bibinfo {title}
  {Ultrasensitive torque detection with an optically levitated nanorotor},\
  }\href {https://www.nature.com/articles/s41565-019-0605-9} {\bibfield
  {journal} {\bibinfo  {journal} {Nature Nanotechnology}\ }\textbf {\bibinfo
  {volume} {15}},\ \bibinfo {pages} {89} (\bibinfo {year} {2020})}\BibitemShut
  {NoStop}%
\bibitem [{\citenamefont {Datsyuk}(1992)}]{datsyuk1992some}%
  \BibitemOpen
  \bibfield  {author} {\bibinfo {author} {\bibfnamefont {V.}~\bibnamefont
  {Datsyuk}},\ }\bibfield  {title} {\bibinfo {title} {Some characteristics of
  resonant electromagnetic modes in a dielectric sphere},\ }\href
  {https://link.springer.com/article/10.1007/BF00331893} {\bibfield  {journal}
  {\bibinfo  {journal} {Appl. Phys. B}\ }\textbf {\bibinfo {volume} {54}},\
  \bibinfo {pages} {184} (\bibinfo {year} {1992})}\BibitemShut {NoStop}%
\bibitem [{\citenamefont {Weinstein}(1969)}]{weinstein1969open}%
  \BibitemOpen
  \bibfield  {author} {\bibinfo {author} {\bibfnamefont {L.~A.}\ \bibnamefont
  {Weinstein}},\ }\href@noop {} {\emph {\bibinfo {title} {Open resonators and
  open waveguides}}}\ (\bibinfo  {publisher} {Golem Press},\ \bibinfo {year}
  {1969})\BibitemShut {NoStop}%
\bibitem [{\citenamefont {Edmonds}(1996)}]{edmonds1996angular}%
  \BibitemOpen
  \bibfield  {author} {\bibinfo {author} {\bibfnamefont {A.~R.}\ \bibnamefont
  {Edmonds}},\ }\href@noop {} {\emph {\bibinfo {title} {Angular momentum in
  quantum mechanics}}},\ Vol.~\bibinfo {volume} {4}\ (\bibinfo  {publisher}
  {Princeton university press},\ \bibinfo {year} {1996})\BibitemShut {NoStop}%
\bibitem [{\citenamefont {Rusconi}\ and\ \citenamefont
  {{Romero-Isart}}(2016)}]{rusconiMagneticRigidRotor2016}%
  \BibitemOpen
  \bibfield  {author} {\bibinfo {author} {\bibfnamefont {C.~C.}\ \bibnamefont
  {Rusconi}}\ and\ \bibinfo {author} {\bibfnamefont {O.}~\bibnamefont
  {{Romero-Isart}}},\ }\bibfield  {title} {\bibinfo {title} {Magnetic rigid
  rotor in the quantum regime: {{Theoretical}} toolbox},\ }\href
  {https://doi.org/10.1103/PhysRevB.93.054427} {\bibfield  {journal} {\bibinfo
  {journal} {Phys. Rev. B}\ }\textbf {\bibinfo {volume} {93}},\ \bibinfo
  {pages} {054427} (\bibinfo {year} {2016})}\BibitemShut {NoStop}%
\end{thebibliography}
